\documentclass[12pt,letterpaper]{article}
\usepackage[T1]{fontenc}
\usepackage{latexml}   
\iflatexml
  \usepackage{amssymb}   
\else
  \usepackage[charter,cal=cmcal]{mathdesign}
  \DeclareFontShape{T1}{mdbch}{m}{up}{<->ssub*mdbch/m/n}{}
\fi
\usepackage{amsmath,amsthm,mathtools,graphicx,booktabs,array,enumitem,microtype,setspace,needspace,xcolor}
\usepackage[margin=1in]{geometry}
\usepackage[round,authoryear]{natbib}
\usepackage[colorlinks=true,linkcolor=blue,citecolor=blue,urlcolor=blue]{hyperref}
\setlist[enumerate]{label=(\alph*),leftmargin=2em,itemsep=3pt,topsep=4pt}
\setlist[itemize]{label=$\bullet$,leftmargin=2em,itemsep=3pt,topsep=4pt}
\numberwithin{equation}{section}
\newtheorem{theorem}{Theorem}

\newtheorem{lemma}{Lemma}[section]
\theoremstyle{definition}
\newtheorem{definition}{Definition}
\theoremstyle{plain}
\newenvironment{condition}[2]{\par\Needspace{7\baselineskip}\medskip\noindent\textbf{Assumption~#1\if\relax\detokenize{#2}\relax\else\space(#2)\fi.}\ }{\par\medskip}
\newcommand{\E}{\mathbb E}
\newcommand{\PP}{\mathbb P}
\newcommand{\R}{\mathbb R}
\newcommand{\ind}{\mathbf 1}
\newcommand{\indep}{\mathrel{\perp\!\!\!\perp}}
\newcommand{\Var}{\operatorname{Var}}
\newcommand{\Cov}{\operatorname{Cov}}
\newcommand{\supp}{\operatorname{supp}}
\newcommand{\pto}{\mathrel{\to_p}}
\newcommand{\dto}{\mathrel{\Rightarrow}}
\newcommand{\norm}[1]{\left\lVert#1\right\rVert}

\DeclareMathOperator*{\argmax}{arg\,max}
\definecolor{songblue}{RGB}{51,51,179}
\hypersetup{pdftitle={Estimating a Rational Inattention Discrete Choice Model: A Semiparametric Approach},pdfauthor={Hanqi Wang}}
\begin{document}
\raggedbottom
\hypersetup{pageanchor=false}
\begin{titlepage}
\renewcommand{\thefootnote}{}
\centering
\vspace*{.1in}
{\Large\textsc{Estimating a Rational Inattention Discrete Choice Model: A Semiparametric Approach}\par}
\vspace{.25in}
{Hanqi Wang\par}
{\itshape Vancouver School of Economics, University of British Columbia\par}
\vspace{.20in}
{November 3, 2026\par}
\smallskip
{\href{https://hanqiwang-econ.github.io/files/Wang_Hanqi_JMP.pdf}{Click here for the latest version}\par}
\vspace{.25in}
\begin{minipage}{.88\textwidth}
\small
\textsc{Abstract.} People's choices depend on both the payoffs of the available options and what they learn about those payoffs before deciding. This paper shows how choice data identify both unobserved payoff heterogeneity and the cost of information acquisition in a rational inattention discrete-choice model. I establish conditions for semiparametric point identification using exogenous payoff and information-cost shifters, without imposing parametric forms on the taste-shock distribution or on how the unit cost of information varies with observed characteristics. In the model with binary choice, I construct an unobserved choice threshold whose cumulative distribution function is the choice probability as a function of the payoff shifter. Moments of this threshold identify the payoff coefficients and the unit cost of information at each value of the information-cost shifter. A deconvolution argument then identifies the taste-shock distribution. I then extend identification to the model with multinomial choice. To estimate the model, I propose a sieve minimum-distance estimator based on observed market shares and establish its consistency. For inference, I construct confidence regions by test inversion because binding parameter constraints can make the estimator's limiting distribution nonnormal. In an application to Amazon Mechanical Turk, model counterfactuals show that lower unit costs of information raise the average task-taking probability and reduce wage elasticity.
\end{minipage}
\footnotetext[1]{I am deeply grateful to Kyungchul Song, Paul Schrimpf, and Vitor Farinha Luz for their invaluable guidance, mentorship, and helpful comments. I thank Raffaele Saggio for his advice on the empirical application. I also thank Giovanni Gallipoli; Ying Gao; Hiro Kasahara; Li, Hao; Wei Li; Vadim Marmer; Jesse Perla; Mike Peters; and participants at the UBC Econometrics Lunch Seminar, my job market practice talk at UBC, the 2026 North American Summer Meeting of the Econometric Society, and the 2026 Canadian Economics Association Annual Conference for helpful comments and suggestions. An earlier version of this paper circulated under the title ``Estimating a Rational Inattention Discrete Choice Model with Costly-to-Learn Individual Heterogeneity.''}
\vfill
\thispagestyle{empty}
\end{titlepage}
\hypersetup{pageanchor=true}
\section{Introduction}\label{s:introduction}
Following \citet{mcfadden}, econometricians use random utility models to explain heterogeneity in observed choices. These models typically assume that individuals know their taste shocks, the idiosyncratic payoff components unobserved by the econometrician. In real-world situations, however, individuals may be uncertain about these components, and learning about them may require time and effort. To account for the cost of learning, \citet{mm} apply the rational inattention (RI) framework pioneered by \citet{sims} to discrete choice. In this model, rationally inattentive individuals optimally choose what to learn by trading off the expected payoff gains from learning against the cost of acquiring information. This paper studies how the observed choices of these individuals can identify the taste-shock distribution and the unit cost of information as a function of observed characteristics.

Existing identification analyses of RI models have relied on parametric specifications \citep{liao,joo,brownjeon}. For example, \citet{brownjeon} establish identification of an insurance-demand model using a Cardell distribution \citep{cardell} for taste shocks and a parametric specification of how the unit cost of information varies with observed characteristics. Their distributional assumptions yield closed-form choice probabilities, which they use together with the exogeneity of observed characteristics to derive identification conditions. Less is known about identification without these parametric restrictions.

Without imposing parametric forms on the taste-shock distribution or on the unit cost of information as a function of observed characteristics, this paper establishes conditions under which exogenous variation in choice data identifies both. For choices between two alternatives, the required variation comes from two observed characteristics: one changes payoffs and the other changes the unit cost of information. An observed utility shifter enters the payoff component known to the individual linearly. An observed information-cost shifter changes the unit cost of information without changing that known payoff component. The taste shocks are independent of both shifters. With this variation, I extend \citeauthor{lewbel}'s (\citeyear{lewbel}) method of recovering payoff parameters from moments of an unobserved choice threshold to the RI model. I construct a similar threshold, which I call a switch point; when the utility shifter is a wage, it can be interpreted as an implied reservation wage. The switch point's cumulative distribution function gives the probability of choosing one alternative over a reference alternative, which I call alternative 0. This probability is a function of the utility shifter, holding the information-cost shifter fixed. The moments of this switch point identify the payoff coefficients and the unit cost of information at each value of the information-cost shifter. A deconvolution argument then recovers the taste-shock distribution.

\Needspace{3\baselineskip}
Extending the argument for two alternatives establishes identification of the RI model with multiple alternatives. At positive information costs, payoff variation, such as lower wages, can leave only alternative 0 and one other alternative with positive choice probabilities. The individual's problem is then a binary RI problem. Applying the same switch-point moment identities to each other alternative paired with alternative 0 identifies the payoff coefficients and the unit cost of information common to the alternatives. I then use the probability of choosing alternative 0 and multivariate deconvolution to recover the joint distribution of taste shocks.

This paper proposes a sieve minimum-distance estimator using market shares and develops a test-inversion approach for inference that remains asymptotically valid when parameter constraints make the estimator's limiting distribution nonnormal. The estimator uses finite-dimensional approximations to the taste-shock distribution and the unit cost of information as a function of observed characteristics. The RI model gives an equation that the true parameters and population choice probabilities must satisfy. I estimate the parameters by minimizing the average squared difference between the equation's two sides after replacing population choice probabilities with observed shares. I establish consistency even though the criterion can be discontinuous at zero information cost. Inference must accommodate estimates on the boundary of the parameter space: information costs cannot be negative, so estimated costs can be zero. Similarly, when the taste-shock distribution is approximated by nonnegative probability weights on a grid extending beyond its true support, some estimated weights can be zero. When these constraints bind, the estimator's limiting distribution need not be normal. I therefore construct confidence regions by test inversion and obtain intervals for functions of the model parameters by taking their ranges over parameter values that the test does not reject.

The empirical application uses data collected by \citet{dube} from Amazon Mechanical Turk to study how workers' labour supply decisions vary with the difficulty of acquiring information about jobs. On this online platform, workers are paid to complete short tasks, such as transcribing conversations, answering survey questions, and entering data. Estimated information costs are lowest and close to zero when descriptions specify both what workers must do and how much work is required. The wage elasticity of the average task-taking probability, calculated over a doubling of posted rewards, falls from 0.287 at estimated information costs to 0.068 when costs are halved and 0.037 under full information. At the original rewards, the average task-taking probability rises from 2.62\% at estimated costs to 8.90\% when costs are halved and 12.87\% under full information. These counterfactual results suggest that lower unit costs of information can increase workers' willingness to take jobs while reducing their responsiveness to higher wages in this online labour market.

\par\medskip
\Needspace{6\baselineskip}
\noindent\textbf{Related Literature}\enspace
This paper contributes to three strands of literature. First, this paper contributes to RI theory by applying the framework to observational choice data when individuals are initially uncertain about their own idiosyncratic payoff components, which the researcher does not observe. \citet{sims} shows how slow responses to economic shocks can arise from limited information processing even when actions can be adjusted immediately. Individuals optimally choose the information on which to base their decisions, subject to a constraint measured by Shannon mutual information. \citet{mm} apply this framework to discrete choice under Shannon information costs and derive a modified multinomial logit formula for choice probabilities conditional on realized payoffs. The alternative-specific weights in this formula depend on prior beliefs\footnote{The probability distribution that describes an individual's beliefs about possible payoffs before acquiring information is called a prior belief. A prior is correct when it matches the true payoff distribution. Subjective priors reflect individuals' own assessments and may differ from that distribution.} and information costs. \citet{fosgerau} show that RI models with more general information costs can generate the choice probabilities of additive random utility models such as nested logit. \citet{cdl2022} derive restrictions on observed choices that distinguish Shannon costs from more general information costs. These restrictions concern how learning changes with prior beliefs and with different descriptions of the same payoff uncertainty.

Second, this paper contributes to the econometric literature on RI discrete-choice models. Empirical RI research has studied models with known or parametrically specified distributions of idiosyncratic taste shocks. In contrast, I establish identification without imposing a parametric form on this distribution or on the unit cost of information as a function of observed characteristics. The most closely related work is \citet{brownjeon}, who identify and estimate an insurance-demand model using Cardell distributions and a parametric specification of information costs. In their application, consumers acquire costly information about individual-specific expected out-of-pocket costs and other plan characteristics observed by the econometrician. Therefore, in their framework, the data identify choice probabilities conditional on these characteristics and other observed covariates. By contrast, I study identification when agents acquire costly information about idiosyncratic payoff components that are unobserved by the econometrician. \citet{joo} develops a parametric RI demand model without specifying consumers' subjective priors, showing that a prior exists under which the model's choice probabilities are optimal. He uses retail promotions that affect information about products but do not enter consumption utility to test the model against full-information demand. \citet{liao} studies identification of prior beliefs and payoff parameters in an RI discrete-choice model where individuals are uncertain about market-level payoff shocks unobserved by the researcher. Unlike in my model, individuals initially know their private taste shocks, and the researcher knows their distribution. Liao also uses exogenous variation in the available alternatives to test full-information choice and shows how costly information acquisition can explain violations of the full-information model's predictions. More generally, \citet{gualdani} leave information structures unrestricted within a model with parameterized payoffs and prior beliefs. Using Bayes correlated equilibrium, they derive sharp identified sets for these parameters and bounds on counterfactual predictions.

Third, this paper contributes to the semiparametric discrete-choice literature by establishing identification in an RI model. \citet{lewbel} recovers moments of latent choice thresholds from discrete choices, providing the starting point for my identification argument. \citet{fox2012} identify the random-coefficient distribution in mixed logit using local variation in product characteristics. \citet{wang2023} develops identification and sieve estimation of mixed-logit demand with a nonparametric random-coefficient distribution. \citet{iariawang2025} establish real analyticity of mixed-logit and mixed-probit demand in mean utility indices for arbitrary mixing distributions, and show how this property extends identification beyond an observed open set. At positive information costs, my identification argument likewise uses analyticity and continuity to recover the choice-probability function beyond an observed interval on which both alternatives have positive choice probability, without requiring full support of the utility shifter.

The remaining discussion provides a broader overview of RI research and related literatures. Revealed-preference tests and experiments examine costly information acquisition in settings where the researcher knows individuals' prior beliefs and the realized payoffs of the alternatives, while individuals are initially uncertain about those payoffs. Using choice probabilities conditional on realized payoffs, \citet{caplindean} give necessary and sufficient conditions for observed choices to be consistent with optimal learning under some information-cost function, and derive bounds on relative information costs. \citet{deanneligh} use laboratory experiments to show that participants' choices are broadly consistent with optimal information acquisition under an RI model with no specified functional form for information costs. They also estimate several information-cost specifications, including one based on generalized entropy.

RI provides a theory of how consideration sets, the subsets of available alternatives from which individuals actually choose, form endogenously. \citet{cdl} give necessary and sufficient conditions for optimal choice under Shannon information costs, including conditions determining which alternatives receive zero choice probability. Other consideration-set models study what choices reveal about preferences when individuals overlook some available alternatives. \citet{masatlioglu} assume that individuals choose their preferred considered alternative and that removing an unconsidered alternative leaves the consideration set unchanged. If removing an unchosen alternative changes the choice, the removed alternative must have been considered and ranked below the originally chosen alternative, revealing a preference comparison. Allowing consideration to be random, \citet{manzini} assume independent consideration with probabilities fixed across choice sets and uniquely identify both the preference ranking and consideration probabilities from choice probabilities observed across choice sets. \citet{cattaneo} allow dependent random consideration, requiring that removing an option outside a consideration set cannot reduce the probability of considering exactly that set. They characterize the preference rankings consistent with observed choices and construct confidence sets for those rankings.

RI research also examines how information acquisition affects market outcomes and welfare. \citet{afrouzi} shows that firms facing fewer competitors pay less attention to aggregate conditions, consistent with survey evidence. In his calibrated model, this reduced attention makes monetary shocks have larger effects on output. With complete markets and Shannon information costs, \citet{angeletossastry} establish that competitive equilibrium is efficient. They also give conditions under which other information costs create welfare gains from simplifying prices or markets.

Finally, empirical and theoretical studies examine how attention and existing information affect responses to macroeconomic news and information treatments. \citet{songstern} use firms' annual reports to measure attention to macroeconomic conditions and find that attentive firms' market values rise more after monetary easing and fall less after tightening. \citet{weber} find smaller changes in inflation expectations after information treatments when inflation is high, with evidence that more households and firms already know the information provided. In a dynamic RI model, \citet{mw2026} show that stronger incentives to be informed can reduce information-treatment effects through learning before an experiment, but can increase them through learning during the experiment when individuals acquire no information beforehand.

\subsection*{Notation}\label{s:notation}
I use $a\coloneqq b$ to define $a$ by $b$, and $F_V$ for the distribution of a random variable or vector $V$. For a distribution $H$, the subscripts in $\E_H$, $\PP_H$, $\Var_H$, and $\Cov_H$ specify the distribution used. In particular, if $V\sim H$, then
\[
\E_H[f(V)]\coloneqq\int f(v)\,dH(v),\qquad\text{and}\qquad
\PP_H\{V\in B\}\coloneqq\int\ind\{v\in B\}\,dH(v),
\]
where $\ind\{\cdot\}$ is the indicator function. In Section~\ref{s:model}, $F$ is the joint distribution of the signal $T$ and payoff vector $U$, and $F_T$ is its marginal distribution for $T$. Thus $\E_F[U_j\mid T]$ conditions under the joint distribution, whereas $\E_{F_T}$ averages over signals. The operators $\E_G$ and $\PP_G$ use the taste-shock distribution $G$, and $H_1\otimes H_2$ denotes the product of two distributions. Unless otherwise indicated, expectations, probabilities, variances, and covariances are evaluated at the true parameter values.

I use the same symbol for a distribution and its cumulative distribution function. In particular, $G(B)$ is the probability of a set $B$ under the taste-shock distribution, while $G(t)\coloneqq\PP_G\{\Omega\leq t\}$. For vectors, this inequality is componentwise.

\Needspace{8\baselineskip}
\section{The Rational Inattention Model}\label{s:model}
\subsection{The Rational Inattention Problem}
For ease of exposition, I begin with an RI discrete-choice model with two alternatives, 0 and 1, and then extend the analysis to multiple alternatives. The individual is initially uncertain about a component of alternative 1's payoff and can acquire information before choosing.

\begingroup\emergencystretch=2em Each market $m$, such as a job posting, has observed characteristics $Z_m\in\R$ and $X_m\in\R^{d_x}$, where $d_x$ is a fixed positive integer. Write $\mathcal X\coloneqq\supp(X_m)$. Let \mbox{$\lambda:\mathcal X\to[0,\infty)$} give the unit cost of information as a function of the information-cost shifter $X_m$. For individual $i$ in this market, I specify the payoffs and unit cost of information as
\begin{equation}\label{e:payoffs}
U_{i0m}=0,\qquad
U_{i1m}=\alpha+\delta Z_m+\Omega_{im},
\qquad\text{and}\qquad \lambda_m\coloneqq\lambda(X_m).
\end{equation}
The utility shifter $Z_m$ enters the known payoff component $\alpha+\delta Z_m$ linearly.\par\endgroup

The individual knows the taste-shock distribution $G$ but does not know her own taste shock $\Omega_{im}$ before acquiring information. She also knows $(Z_m,X_m)$, the payoff component $\alpha+\delta Z_m$, and the unit cost $\lambda(X_m)$. The taste shocks are independent and identically distributed across individuals $i$ and markets $m$. I also assume that the taste shock $\Omega_{im}$ is independent of the market characteristics $(Z_m,X_m)$:
\[
\Omega_{im}\indep(Z_m,X_m).
\]
The analyst observes market characteristics and market shares but observes neither the taste shocks nor the information individuals acquire. I write $\theta\coloneqq(\alpha,\delta,\lambda,G)$ for a parameter value.\footnote{Additional covariates can be considered by conditioning the binary identification argument in Subsection~\ref{s:id} on utility characteristics $H_U$ and cost characteristics $H_\lambda$. For the specification $\alpha(H_U)+\delta Z+\Omega$ with cost $\lambda(X,H_\lambda)$, the recovery formulas apply at conditioning values where the stated exclusion, support, and low-cost conditions hold. This gives a partially linear payoff specification with a flexible intercept. A common slope or taste-shock distribution across conditioning values is an additional restriction when imposed.}

The individual can learn about her uncertain payoff by acquiring a \emph{signal} $T$, the information produced by a learning procedure. For example, a worker considering a posting offering four dollars to ``Transcribe this MP3 of a 5 minute interview'' may be uncertain how demanding accurate transcription would be for her. Personal difficulty, which contributes to $\Omega_{im}$, depends on how well the recording's pronunciation and vocabulary match her experience. The signal can be an internal assessment formed by reading the requirements, recalling comparable work, and considering how that experience applies to the assignment.

An \emph{information structure} specifies how likely the individual is to receive each signal at each possible payoff, and the individual chooses this structure before receiving the signal \citep{mm}. Fix a market with characteristics $(z,x)$. I suppress the individual and market indices and write $U\coloneqq(U_0,U_1)$. Let $P_U$ denote the prior distribution of $U$ implied by the known payoff component $\alpha+\delta z$ and the taste-shock distribution $G$. Following \citet{mm}, I represent the information structure by a joint distribution $F\coloneqq F_{T,U}$ of the signal $T$ and payoffs $U$, with marginal distributions $F_T$ and $F_U$. The individual chooses $F$ subject to $F_U=P_U$, thereby choosing how the signal depends on the payoffs while leaving their distribution unchanged.\footnote{For a fixed distribution of signals recommending which alternative to choose, the individual solves an entropy-regularized optimal transport problem with the payoff distribution fixed. \citet{engh2026} shows that the RI problem is obtained by also choosing the signal distribution to maximize the value of this transport problem.} In the worker example, $F$ specifies how likely each assessment is at each possible level of personal difficulty. If the distribution of assessments is the same at every payoff, the signal conveys no information.

After receiving the signal $T=t$, the individual chooses the alternative with the highest expected payoff given that signal. She evaluates payoffs under the distribution implied by $F$ after observing $t$, called her posterior distribution. She therefore maximizes
\begin{equation}\label{e:posteriorchoice}
\E_F[U_j\mid T=t].
\end{equation}

Before receiving the signal, the individual chooses the information structure to trade off expected payoff gains against information costs. Following \citet{mm}, I take these costs to be proportional to Shannon mutual information. Her problem is
\begin{equation}\label{e:RIproblem}
\max_{F:\,F_U=P_U}
\left\{
\E_{F_T}\!\left[\max_{j\in\{0,1\}}\E_F[U_j\mid T]\right]
-\lambda(x) I_F(U;T)
\right\},
\end{equation}
where
\begin{equation}\label{e:mutual}
I_F(U;T)\coloneqq\int\log\!\left(\frac{dF}{d(F_T\otimes P_U)}\right)dF.
\end{equation}
The first term in \eqref{e:RIproblem} is the expected payoff from choosing after receiving the signal; the second is the cost of the information obtained. The mutual information $I_F(U;T)$\footnote{\samepage When payoffs take finitely many possible values, mutual information equals prior entropy minus expected posterior entropy.
\[
I_F(U;T)=H(P_U)-\int H(F_{U\mid T=t})\,dF_T(t).
\]
Here, entropy is defined by
\[
H(Q)\coloneqq-\sum_u Q(\{u\})\log Q(\{u\}),
\]
with $0\log 0\coloneqq0$. The first term measures uncertainty about payoffs before receiving the signal. The second measures the uncertainty remaining after receiving the signal, averaged over the possible signals. Their difference measures how much the individual learns. An uninformative signal leaves the payoff distribution unchanged, so mutual information is zero.} defined in \eqref{e:mutual} measures the amount of information the signal $T$ conveys about the payoffs $U$. It equals zero when the signal and payoffs are independent. Multiplying this amount by the unit cost of information $\lambda(x)$ gives the information cost $\lambda(x)I_F(U;T)$.

\Needspace{5\baselineskip}
The unit cost of information $\lambda(x)\geq0$ measures the payoff the individual gives up for each unit of information acquired. A lower $\lambda(x)$ makes any given amount of information cheaper. When information is clearly organized and easy to interpret, learning which option is better may require very little cognitive effort. At positive cost, the individual may leave some uncertainty unresolved because learning more would not justify its cost. At zero cost, she can make a full-information choice by learning which alternative yields the higher payoff.

\Needspace{8\baselineskip}
\subsection{Choice Probabilities of Rationally Inattentive Agents}\label{s:modelprobabilities}
At positive information cost, choice probabilities can be computed without solving directly for the information structure. I write $b(z,x;\theta)$ for the \emph{ex ante choice probability}: the probability of choosing alternative 1 conditional on market characteristics $(z,x)$, before receiving the signal. For $\lambda(x)>0$, \citet{mm} characterize it as the maximizer of the objective below. A nondegenerate $G$ makes this objective strictly concave in $q$, so the maximizer is unique:
\begin{equation}\label{e:shareproblem}
b(z,x;\theta)\coloneqq\argmax_{q\in[0,1]}
\int\log\!\left(1-q+q e^{(\alpha+\delta z+\omega)/\lambda(x)}\right)dG(\omega).
\end{equation}

The ex ante probability $b(z,x;\theta)$ averages conditional choice probabilities over the taste shocks. I call $b(z,x;\theta)$ \emph{interior} when $0<b(z,x;\theta)<1$, so both alternatives have positive choice probability. At positive cost and an interior $b(z,x;\theta)$, \citet{mm} give the probability of choosing alternative 1 conditional on $\Omega=\omega$ as
\begin{equation}\label{e:conditionalRI}
p_\theta(\omega;z,x)\coloneqq
\frac{b(z,x;\theta)e^{(\alpha+\delta z+\omega)/\lambda(x)}}
{1-b(z,x;\theta)+b(z,x;\theta)e^{(\alpha+\delta z+\omega)/\lambda(x)}}.
\end{equation}
\Needspace{6\baselineskip}
Averaging $p_\theta(\omega;z,x)$ over the taste-shock distribution $G$ gives the ex ante choice probability:
\begin{equation}\label{e:fixedpoint}
b(z,x;\theta)=\int p_\theta(\omega;z,x)\,dG(\omega).
\end{equation}
Since the ex ante choice probability also appears in \eqref{e:conditionalRI},
equation \eqref{e:fixedpoint} is a fixed-point restriction: the average
of the conditional probabilities must equal the ex ante probability used in their formula. For $0<b(z,x;\theta)<1$, \eqref{e:fixedpoint} is equivalent to the first-order condition for \eqref{e:shareproblem}.

\Needspace{6\baselineskip}
When information is free, the individual chooses alternative 1 when its payoff exceeds zero:
\begin{equation}\label{e:fullinfo}
b(z,x;\theta)=\PP_G\{\alpha+\delta z+\Omega>0\}.
\end{equation}
For identification, I use a right-continuous version of the zero-cost choice-probability function.\footnote{With a nonzero slope and a continuous distribution of $Z$ conditional on $X$, exact payoff ties have probability zero across markets. Choosing either alternative at a tie therefore gives the same choice probabilities for almost every market.}

\subsection{Observations}\label{s:observations}
The analyst observes market characteristics and the share of individuals who choose alternative 1 in each of $M$ markets. Market $m$ has a deterministic number $N_{m,M}$ of sampled individuals. I write its characteristics as $W_m\coloneqq(Z_m,X_m)$ and its observed share as
\begin{equation}\label{e:observedshare}
\widehat s_m\coloneqq\frac1{N_{m,M}}\sum_{i=1}^{N_{m,M}}Y_{im},
\end{equation}
where $Y_{im}$ is the indicator for choosing alternative 1. Market characteristics $W_m$ are independent and identically distributed across markets. Conditional on these characteristics, choices are independent across individuals and markets, with
\[
Y_{im}\mid W_m\sim\operatorname{Bernoulli}(s_m),
\qquad\text{where}\qquad s_m\coloneqq\PP(Y_{im}=1\mid W_m).
\]
Suppose the observed choices are generated by the model at an unknown parameter value $\theta_0\coloneqq(\alpha_0,\delta_0,\lambda_0,G_0)$. The population choice probability is
\begin{equation}\label{e:populationprob}
\pi(z,x)\coloneqq\PP(Y=1\mid Z=z,X=x)=b(z,x;\theta_0).
\end{equation}
Thus $s_m=\pi(Z_m,X_m)$.

\Needspace{8\baselineskip}
\subsection{Identification}\label{s:id}\label{s:idproblem}
In this subsection, I establish conditions under which population choice probabilities identify $\theta_0$ in the RI model with two alternatives. Subsection~\ref{s:multiple} extends the identification result to multiple alternatives. I write $\mathcal Z_x\coloneqq\supp(Z\mid X=x)$ for the conditional support of $Z$. Identification requires a mapping from these probabilities to the parameters:
\begin{definition}[Identification]\label{d:identification}
The parameter $\theta_0$ is \emph{identified} if there exists a mapping
$\mathcal T$, known to the analyst, that recovers it from the
population choice probabilities:
\begin{equation}\label{e:identificationmap}
\theta_0=\mathcal T\!\left(
\left\{\left\{\pi(z,x)\right\}_{z\in\mathcal Z_x}\right\}_{x\in\mathcal X}
\right).
\end{equation}
\end{definition}

\Needspace{8\baselineskip}
I begin with normalization and regularity assumptions for the RI model.
\begin{condition}{ID1}{}\label{a:ID1}
\begin{enumerate}
\item The taste shocks are independent of the shifters:
\begin{equation}\label{e:exclusion}
\Omega_{im}\indep(Z_m,X_m).
\end{equation}
\item $\E\Omega=0$ and $\Var(\Omega)=1$.
\item $\delta_0\ne0$.
\item $\lambda_0:\mathcal X\to[0,\infty)$ is continuous.
\item For every $x\in\mathcal X$ with $\lambda_0(x)=0$, the function $z\mapsto b(z,x;\theta_0)$ is continuous on $\mathcal Z_x$.
\end{enumerate}
\end{condition}

Assumption~ID1(a) means that changing either shifter leaves the taste-shock distribution $G_0$ unchanged. Assumption~ID1(b) fixes payoff location and scale, and ID1(c) requires the utility shifter to affect payoffs. Assumptions~ID1(d)--(e), together with ID3(b), make $(z,x)\mapsto b(z,x;\theta_0)$ jointly continuous on the observed support. The distribution of observed choices and shifters determines conditional choice probabilities only up to changes on sets of covariate values with probability zero. Joint continuity makes their values uniquely determined at every support point, as required by Definition~\ref{d:identification}.

\Needspace{8\baselineskip}
I next turn to the variation in information costs used for identification.
\begin{condition}{ID2}{Arbitrarily low information costs}\label{a:ID2}
\begin{equation}\label{e:lowcost}
\inf_{x\in\mathcal X}\lambda_0(x)=0.
\end{equation}
\end{condition}

Assumption~ID2 means that some values of the cost shifter make learning nearly effortless.\footnote{Both \citet[equation (15)]{brownjeon} and \citet[equation (16)]{joo} use exponential information-cost specifications. For a specification $\lambda_0(x)=\exp(x'\beta_0)$ with coefficient vector $\beta_0$, ID2 holds if $\supp(X'\beta_0)=\R$, since the index can then become arbitrarily negative.} With finitely many cost categories, the assumption requires one category to have zero cost. With a continuous shifter, costs can approach zero without reaching it.

\Needspace{9\baselineskip}
For the utility shifter, I state conditions on its distribution and observed support conditional on the information-cost shifter.
\begin{condition}{ID3}{Support of the utility shifter}\label{a:ID3}
\begin{enumerate}
\item For each $x$, $Z\mid X=x$ has a continuous distribution
with interval support $\mathcal Z_x$.
\item Define
\[
\mathcal X_0\coloneqq\{x\in\mathcal X:\lambda_0(x)=0\}.
\]
For every $x\in\mathcal X_0$,
\[
\PP\!\left\{
-\Omega\in\alpha_0+\delta_0\operatorname{int}(\mathcal Z_x)
\right\}=1.
\]
For every $x\notin\mathcal X_0$, there exists
\[
z_x^*\in\operatorname{int}(\mathcal Z_x),
\qquad\text{and}\qquad 0<\pi(z_x^*,x)<1.
\]
\end{enumerate}
\end{condition}
Assumption~ID3(b) requires an observed utility-shifter value at which both alternatives have positive choice probability when information is costly. For example, there must be an observed wage at which some workers take the task and others do not. When information is free, observed variation in known payoffs must cover the thresholds $-\Omega$ for choosing alternative 1. In the wage example, the observed wage range must cover workers' reservation wages. If this range has a lowest or highest value, no positive fraction of workers can have a reservation wage equal to that value.

\Needspace{8\baselineskip}
The identification argument requires the choice-probability function $z\mapsto b(z,x;\theta_0)$ to be identified on $\R$ for each $x$. At positive information cost, analyticity allows probabilities on an observed interval to determine values outside that interval. Together with ID1--ID2, the support conditions in ID3 give the following result.
\begin{theorem}[Identification]\label{t:identification}
Under Assumptions~ID1--ID3, the true parameter
\[
\theta_0=(\alpha_0,\delta_0,\lambda_0,G_0)
\]
is identified.
\end{theorem}

\Needspace{4\baselineskip}
I briefly outline the proof; Section~\ref{a:binaryproof} gives the complete argument. Throughout this explanation, assume without loss of generality that $\delta_0>0$. I identify the parameters by interpreting $z\mapsto b(z,x;\theta_0)$ as the CDF of a latent threshold for each fixed $x$. The starting point is the connection with \citet{lewbel}: consider the full-information binary choice rule
\[
Y=\ind\{\alpha_0+\delta_0 Z\geq-\Omega\},
\qquad\text{where }\Omega\indep Z\quad\text{and}\quad\delta_0>0.
\]
Define the threshold
\[
S\coloneqq-\frac{\alpha_0+\Omega}{\delta_0},
\qquad\text{so}\qquad \alpha_0+\delta_0 S=-\Omega.
\]
Rearranging the payoff comparison gives $Y=\ind\{S\leq Z\}$. Since $S$ depends only on $\Omega$, it is independent of $Z$, and hence
\[
\PP(Y=1\mid Z=z)=\PP(S\leq z).
\]
Thus the choice probability as a function of $z$ is the CDF of $S$. When $Z$ is a wage, $S$ is a reservation wage: alternative 1 is chosen when the wage reaches this threshold. The mean-zero and unit-variance normalizations for $\Omega$ in ID1(b) give moments that recover the payoff coefficients:
\[
\E S=-\frac{\alpha_0}{\delta_0},
\qquad\text{and}\qquad \Var(S)=\frac1{\delta_0^2}.
\]

I construct a similar threshold in the RI model from the conditional choice probability $p_{\theta_0}(\omega;z,x)$ in \eqref{e:conditionalRI}. Write $\pi(z,x)\coloneqq b(z,x;\theta_0)$ for the choice probability at every $z\in\R$, including values outside the observed support. Let $\ell(p)\coloneqq\log\{p/(1-p)\}$ and let $\varepsilon$ be standard logistic, independent of $\Omega$, with CDF $\sigma(t)\coloneqq(1+e^{-t})^{-1}$. At positive cost and an interior probability, rewriting \eqref{e:conditionalRI} gives
\begin{align*}
p_{\theta_0}(\omega;z,x)
&=\sigma\!\left(\ell(\pi(z,x))+
\frac{\alpha_0+\delta_0z+\omega}{\lambda_0(x)}\right)\\
&=\PP\!\left\{\varepsilon\leq\ell(\pi(z,x))+
\frac{\alpha_0+\delta_0z+\omega}{\lambda_0(x)}\right\}.
\end{align*}
Averaging over $\Omega\sim G_0$, independently of $\varepsilon$, therefore gives
\begin{equation}\label{e:RIthreshold}
\pi(z,x)=\PP\{h_x(z)\geq-\Omega+\lambda_0(x)\varepsilon\},
\end{equation}
where
\begin{equation}\label{e:h}
h_x(z)\coloneqq\alpha_0+\delta_0z+\lambda_0(x)\ell(\pi(z,x)).
\end{equation}
For $x$ with $\lambda_0(x)>0$, define
\[
I_x\coloneqq\{z\in\R:0<\pi(z,x)<1\}.
\]
The RI model implies that $I_x$ is a nonempty open interval and that $h_x:I_x\to\R$ is strictly increasing and onto. Its inverse therefore converts the threshold in \eqref{e:RIthreshold} into units of the utility shifter.

\begin{definition}[Switch point]\label{d:switchpoint}
For each $x\in\mathcal X$, define the \emph{switch point} $S_x$ by
\[
S_x\coloneqq
\begin{cases}
h_x^{-1}\!\left(-\Omega+\lambda_0(x)\varepsilon\right),
&\lambda_0(x)>0,\\[4pt]
-\dfrac{\alpha_0+\Omega}{\delta_0},&\lambda_0(x)=0,
\end{cases}
\]
where $\Omega\sim G_0$ and $\varepsilon$ is standard logistic and independent of $\Omega$.
\end{definition}

At positive cost, the switch point $S_x$ has CDF $\pi(\cdot,x)$ because $h_x$ is strictly increasing: for $z\in I_x$,
\[
\PP(S_x\leq z)
=\PP\{-\Omega+\lambda_0(x)\varepsilon\leq h_x(z)\}
=\pi(z,x).
\]
The same CDF equality holds for every real value of the utility shifter, at both positive and zero cost:\footnote{At positive cost, $S_x\in I_x$ almost surely. To the left of $I_x$, both $\PP(S_x\leq z)$ and $\pi(z,x)$ equal zero; to its right, both equal one. Continuity gives the same equalities at any finite endpoints. At zero cost, the equality follows from $S_x=-(\alpha_0+\Omega)/\delta_0$, using the right-continuous choice-probability function described in Subsection~\ref{s:modelprobabilities}.}
\begin{equation}\label{e:SxCDF}
\PP(S_x\leq z)=\pi(z,x),\qquad z\in\R.
\end{equation}

The switch-point distribution provides moments that identify the payoff coefficients and information costs. These moments depend on the CDF $z\mapsto\pi(z,x)$ on $\R$, which ID3 determines from the observed probabilities by the support and analyticity argument above. At positive information cost, applying $h_x$ to the switch point gives
\begin{equation}\label{e:switchidentity}
-\Omega+\lambda_0(x)\varepsilon
=\alpha_0+\delta_0S_x+\lambda_0(x)\ell(\pi(S_x,x)).
\end{equation}
I write $\mu(x)\coloneqq\E S_x$ and $V(x)\coloneqq\Var(S_x)$. I also define the following covariance $C(x)$, which is positive for a nondegenerate switch-point distribution:
\begin{equation}\label{e:observablemoments}
C(x)\coloneqq\Cov\bigl(S_x,\ell(\pi(S_x,x))\bigr).
\end{equation}

At positive cost, $\pi(S_x,x)$ is uniform on $(0,1)$, so $\ell(\pi(S_x,x))$ has the same mean and variance as the standard logistic variable $\varepsilon$. Taking means and variances therefore cancels the two logistic variance terms and gives
\begin{equation}\label{e:momentidentity}
\mu(x)=-\frac{\alpha_0}{\delta_0},\qquad\text{and}\qquad
\delta_0^2V(x)+2\delta_0\lambda_0(x)C(x)=1.
\end{equation}
At zero cost, $S_x=-(\alpha_0+\Omega)/\delta_0$ gives the same restrictions. Equation \eqref{e:momentidentity} explains what information costs do to switch-point dispersion: at positive cost, the positive covariance term makes the variance smaller than the full-information value $1/\delta_0^2$. That value is attained exactly when information is free, since $\delta_0$ and $C(x)$ are positive.

The supremum of the switch-point variances $V(x)$ over $x\in\mathcal X$ identifies $\delta_0$. As information cost approaches zero, switch-point variance approaches its full-information value; Cauchy--Schwarz bounds the difference from that value by a constant times $\lambda_0(x)$. Under ID2, these variances can therefore be arbitrarily close to the full-information value. The common mean then identifies the intercept:
\begin{align}
\delta_0&=\left(\sup_{x\in\mathcal X}V(x)\right)^{-1/2},\label{e:deltarecovery}\\
\text{and}\qquad\alpha_0&=-\delta_0\mu(x).\label{e:alpharecovery}
\end{align}
Once $(\alpha_0,\delta_0)$ are known, solving the same moment identity gives $\lambda_0(x)$ at each $x$:
\begin{equation}\label{e:lambdarecovery}
\lambda_0(x)=\frac{1-\delta_0^2V(x)}{2\delta_0C(x)}.
\end{equation}
\begin{figure}[htbp]
\centering
\includegraphics[width=.97\textwidth]{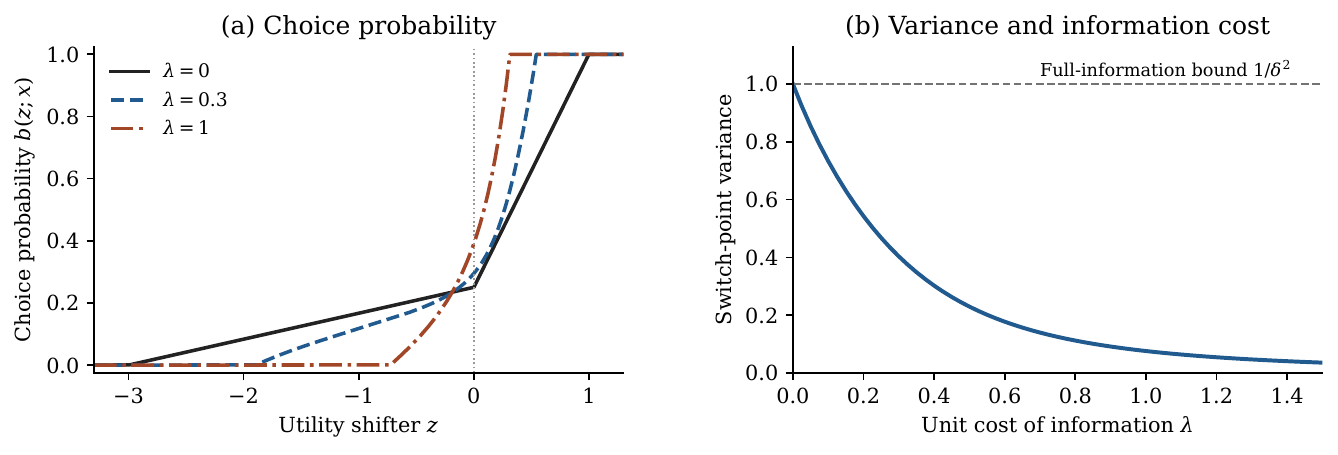}
\caption{Choice probabilities and switch-point variances. The illustration uses $\alpha=0$, $\delta=1$, and an asymmetric taste density equal to $3/4$ on $[-1,0]$ and $1/12$ on $[0,3]$. This distribution has mean zero and variance one. The switch points have the same mean, but their variances fall as information costs rise. I compute the choice probabilities by deterministic integration of the RI model.}\label{f:switchpoints}
\end{figure}

\Needspace{6\baselineskip}
Deconvolution identifies $G_0$ if information costs are positive at some $x$. The identified coefficients $(\alpha_0,\delta_0)$ and cost function $\lambda_0$ make $h_x$ known, so the identified distribution of $S_x$ determines that of $h_x(S_x)$. At such an $x$, \eqref{e:switchidentity} expresses $h_x(S_x)$ as the sum of $-\Omega$ and an independent logistic variable with known scale $\lambda_0(x)$. Writing $\varphi_V(t)\coloneqq\E e^{itV}$, independence gives
\begin{equation}\label{e:deconvolution}
\varphi_{-\Omega}(t)=
\frac{\varphi_{h_x(S_x)}(t)}{\varphi_\varepsilon(\lambda_0(x)t)}.
\end{equation}
The denominator is known and has no real zeros, so the ratio identifies the characteristic function, and hence the distribution, of $-\Omega$. Changing its sign gives the distribution of $\Omega$, namely $G_0$. If costs are zero everywhere, $G_0$ is directly the distribution of $-\alpha_0-\delta_0S_x$. These calculations recover $\theta_0$ from the choice probabilities. The estimator in Section~\ref{s:estimation} uses the RI optimality equation directly, without first estimating switch-point moments or extending the observed choice-probability functions beyond the observed support.

\subsubsection{Extension: Multiple Alternatives}\label{s:multiple}
With multiple alternatives, identification must recover both the marginal taste-shock distributions and their dependence. I consider alternatives $0,1,\ldots,J$, with payoffs
\begin{equation}\label{e:multpayoffs}
U_0=0,\qquad\text{and}\qquad
U_j=\alpha_j+\delta_jz_j+\Omega_j,
\qquad j=1,\ldots,J.
\end{equation}
The payoff of alternative 0 is normalized to zero, and $z\coloneqq(z_1,\ldots,z_J)$ is the vector of utility shifters. The unit cost of information $\lambda(x)$ is common across the alternatives. The individual learns about $\Omega\coloneqq(\Omega_1,\ldots,\Omega_J)$, whose joint distribution $G$ is independent of the shifters. In this extension, I omit subscript 0 on the true parameter and write $\theta\coloneqq(\alpha,\delta,\lambda,G)$, where $\alpha\coloneqq(\alpha_1,\ldots,\alpha_J)'$ and $\delta\coloneqq(\delta_1,\ldots,\delta_J)'$. Dependence among taste shocks allows a worker who finds one kind of task attractive to find another kind attractive as well. I first recover the probability of choosing each alternative $j$ when only $j$ and alternative 0 are available. I then use the probability of choosing alternative 0 when all alternatives are available. Appendix~\ref{a:multipleidentification} states the formal conditions and theorem; Section~\ref{a:multipleproof} gives the proof.

To fix the common payoff scale, I normalize $\E_G\Omega_j=0$ for every $j$ and $\Var_G(\Omega_1)=1$. Only the variance of the first taste shock is normalized; the other marginal variances will be recovered. Changing the sign of each utility shifter when necessary allows me to assume $\delta_j>0$ without loss of generality. I let $b(z,x;\theta)$ collect the $J+1$ probabilities of choosing each alternative. At positive information cost, this vector maximizes the following objective:
\begin{equation}\label{e:multshare}
b(z,x;\theta)\coloneqq\argmax_{q\in\Delta_J}
\lambda(x)\E_G\log\!\left(q_0+\sum_{j=1}^Jq_j e^{U_j/\lambda(x)}\right),
\end{equation}
where $\Delta_J\coloneqq\{q:q_j\geq0,\ \sum_{j=0}^Jq_j=1\}$. Assumption~A2 rules out an almost-sure linear dependence among $1,e^{\Omega_1/\lambda(x)},\ldots,e^{\Omega_J/\lambda(x)}$ and ensures uniqueness of the choice-probability vector.

At positive cost, when all alternatives except 0 and $j$ have zero choice probability, the objective in \eqref{e:multshare} is the binary RI objective:
\[
\max_{p\in[0,1]}\lambda(x)\E_G
\log\!\left(1-p+p e^{U_j/\lambda(x)}\right).
\]
The taste-shock distribution in this binary problem is the marginal distribution of $\Omega_j$. Write $p_j(z_j;x)$ for the probability of choosing alternative $j$ in this binary problem. Let $S_x^{(j)}$ be the switch point with CDF $p_j(\cdot;x)$. Its moments are
\[
\begin{aligned}
\mu_j(x)&\coloneqq\E S_x^{(j)},\qquad V_j(x)\coloneqq\Var(S_x^{(j)}),\\
C_j(x)&\coloneqq\Cov\!\left(S_x^{(j)},\ell(p_j(S_x^{(j)};x))\right).
\end{aligned}
\]

The functions $p_j(\cdot;x)$ can be recovered from choice probabilities when all alternatives are available. At positive information cost and fixed $x$, the full-consideration region is the set of utility-shifter vectors $z$ at which every alternative has positive choice probability. Analyticity determines the probabilities throughout this connected region from their values on an observed open subset. For each $u\in\R$ with $0<p_j(u;x)<1$, I construct covariate vectors $z_n$ with $z_{n,j}=u$ for every $n$, such that every alternative has positive choice probability along the sequence, while the probabilities of alternatives other than $0$ and $j$ approach zero. The limiting probability of choosing alternative $j$ solves the binary RI maximization problem and therefore equals $p_j(u;x)$. Continuity and monotonicity then determine $p_j(\cdot;x)$ also where it equals zero or one. Section~\ref{a:multipleproof} proves this result using the moment conditions in Appendix~\ref{a:multipleidentification}. At zero cost, full conditional support identifies the full-information choice probabilities for every $z\in\R^J$.

The functions $p_1(\cdot;x)$, $x\in\mathcal X$, identify $\alpha_1$, $\delta_1$, and the common cost function $\lambda$ by the formulas for two alternatives, since $\Var_G(\Omega_1)=1$. For alternatives $j=2,\ldots,J$, the marginal taste variances have not been normalized. I write $\sigma_j^2\coloneqq\Var_G(\Omega_j)$ and use the corresponding moment identity
\begin{equation}\label{e:multmoment}
\delta_j^2V_j(x)+2\delta_j\lambda(x)C_j(x)=\sigma_j^2,
\qquad\text{and}\qquad \mu_j(x)=-\frac{\alpha_j}{\delta_j}.
\end{equation}
The right-hand side is an unknown marginal variance, but the cost function $\lambda$ on the left-hand side has already been recovered from the functions $p_1(\cdot;x)$, $x\in\mathcal X$.

Subtracting \eqref{e:multmoment} at two values of $x$ cancels the unknown variance $\sigma_j^2$, because it is the same in both equations. Suppose $\lambda(x)>0$ at some $x\in\mathcal X$. At two values $x,x'$ with $V_j(x)\ne V_j(x')$, the difference identifies $\delta_j$ because $\delta_j>0$ and $\lambda$ is already known:
\begin{equation}\label{e:multslope}
\delta_j=
\frac{2\{\lambda(x')C_j(x')-\lambda(x)C_j(x)\}}
{V_j(x)-V_j(x')}.
\end{equation}
The mean then gives the intercept,
\[
\alpha_j=-\delta_j\mu_j(x),
\]
and substituting the coefficients into either variance equation gives $\sigma_j^2$. For a nondegenerate marginal taste-shock distribution, $V_j(x)$ approaches its full-information value as information costs approach zero. At positive cost it is smaller, so there are two values $x,x'$ with different switch-point variances.

To identify dependence among the taste shocks, I use the probability of choosing alternative 0 when all alternatives are available. At positive information cost, I use an observed region where every alternative has positive choice probability. Write $b^j(z,x)$ for the population choice probability of alternative $j$, the $j$th component of $b(z,x;\theta)$. Let $\eta_0,\ldots,\eta_J$ be independent standard Gumbel variables, independent of $\Omega$. Conditional on $\Omega$, the logit probability of choosing alternative $j$ equals the probability that $j$ maximizes
\[
U_j+\lambda(x)\log b^j(z,x)+\lambda(x)\eta_j,
\qquad j=0,\ldots,J.
\]
Alternative 0 is selected only when its value exceeds that of every other alternative. Its choice probability therefore involves all taste-shock components jointly. I define $\zeta_j\coloneqq\eta_j-\eta_0$ and the known thresholds
\begin{equation}\label{e:multthreshold}
c_j(z,x)\coloneqq-\alpha_j-\delta_jz_j+
\lambda(x)\log\frac{b^0(z,x)}{b^j(z,x)}.
\end{equation}
The probability of these simultaneous inequalities is
\begin{equation}\label{e:jointCDF}
b^0(z,x)=\PP\{\Omega_j+\lambda(x)\zeta_j\leq c_j(z,x)
\text{ for every }j\}.
\end{equation}
Equation \eqref{e:jointCDF} expresses the probability of choosing alternative 0 as the joint CDF of $H_x\coloneqq\Omega+\lambda(x)\zeta$, evaluated at the thresholds determined by the identified coefficients and choice probabilities.

The distribution of $H_x$ is identified from its CDF on an open set of thresholds. The observed full-consideration region gives an open set of thresholds $(c_1(z,x),\ldots,c_J(z,x))$.\footnote{Section~\ref{a:multipleproof} shows that the map $z\mapsto(c_1(z,x),\ldots,c_J(z,x))$ has a nonsingular Jacobian on this region, so its image is open.} At positive cost, adding the independent vector $\lambda(x)\zeta$ to $\Omega$ produces a random vector with a real-analytic density. The CDF in \eqref{e:jointCDF} determines this density on the open set, and real analyticity determines it on $\R^J$.

Deconvolution then recovers the joint taste-shock distribution $G$ from the distribution of $H_x$. Independence of $\eta_0,\ldots,\eta_J$ and the standard Gumbel characteristic function \citep[Section~3]{janson2010} give the characteristic function of $\zeta$:
\[
\varphi_\zeta(t)=
\Gamma\!\left(1+i\sum_{j=1}^Jt_j\right)
\prod_{j=1}^J\Gamma(1-it_j),\qquad t\in\R^J.
\]
Here $\Gamma$ is the gamma function, and $\varphi_\zeta(t)$ is nonzero for every $t\in\R^J$.\footnote{The gamma function has no zeros, and the arguments in this formula have real part one, away from its poles \citep[Section~5.2(i)]{dlmf}.} Independence of $\Omega$ and $\zeta$ therefore gives
\begin{equation}\label{e:jointdeconv}
\varphi_G(t)=\frac{\varphi_{H_x}(t)}{\varphi_\zeta(\lambda(x)t)}.
\end{equation}
This ratio identifies the characteristic function, and hence the joint distribution, of the taste-shock vector.\footnote{The components of $\zeta$ are dependent because each subtracts the same Gumbel variable $\eta_0$. The deconvolution uses their known joint distribution.}

If $\lambda(x)=0$ for every $x\in\mathcal X$, full conditional support identifies the payoff coefficients and joint distribution $G$. The binary formulas identify $\alpha_1$, $\delta_1$, and the zero cost function. For $j=2,\ldots,J$, consider $z_1=t$ and $z_j=rt+u$, with $r>0$, $u\in\R$, and the other shifters fixed. As $t\to\infty$, the ratio $r=\delta_1/\delta_j$ is the only value for which alternatives 1 and $j$ can both have positive limiting choice probabilities for some $u$. The known value of $\delta_1$ fixes the remaining slopes, and the binary switch-point means identify the intercepts. The probability of choosing alternative 0 directly identifies the joint taste-shock distribution because
\[
b^0(z,x)=G(-\alpha_1-\delta_1z_1,\ldots,-\alpha_J-\delta_Jz_J),
\qquad z\in\R^J.
\]

\section{Estimation}\label{s:estimation}
In this section, I construct a sieve minimum-distance estimator using market shares and establish its consistency. For ease of exposition, I first present the estimator and consistency argument for two alternatives. Subsection~\ref{s:multestimation} extends the estimation method to multiple alternatives.
\Needspace{8\baselineskip}
\subsection{The Estimator}
I construct a minimum-distance estimator by measuring how closely the observed market shares satisfy the RI equation \eqref{e:fixedpoint}. I write $w\coloneqq(z,x)$ and $s(w)\coloneqq\pi(z,x)$, and let $\sigma(u)\coloneqq(1+e^{-u})^{-1}$ be the logistic CDF. At positive cost and an input probability $q\in(0,1)$, $\Psi$ averages the conditional choice probabilities computed from $\theta$ and $q$. I also define it at probabilities zero and one and at zero cost:
\begin{equation}\label{e:Psi}
\Psi(\theta;w,q)\coloneqq
\begin{cases}
\displaystyle\int\sigma\!\left(\ell(q)+\frac{\alpha+\delta z+\omega}{\lambda(x)}\right)dG(\omega),
&\lambda(x)>0,\quad0<q<1,\\[8pt]
q,&\lambda(x)>0,\quad q\in\{0,1\},\\[3pt]
\PP_G\{\alpha+\delta z+\Omega>0\},&\lambda(x)=0.
\end{cases}
\end{equation}
The RI residual is the difference between $\Psi(\theta;w,q)$ and the input probability $q$:
\begin{equation}\label{e:residual}
r(\theta;w,q)\coloneqq\Psi(\theta;w,q)-q.
\end{equation}
At the true parameter and population choice probability,
\begin{equation}\label{e:trueresidual}
r(\theta_0;W_m,s_m)=0.
\end{equation}

\Needspace{8\baselineskip}
I use finite-dimensional approximations to estimate $\lambda$ and $G$. Let $\Theta$ be the set of parameters satisfying the model restrictions and $\Theta_M\subseteq\Theta$ a finite-dimensional subset, called a sieve. In the data, I replace the population choice probability $s_m$ by the observed market share $\widehat s_m$ and use the average squared residual:
\begin{equation}\label{e:criterion}
\widehat Q_M(\theta)\coloneqq\frac1M\sum_{m=1}^M
r(\theta;W_m,\widehat s_m)^2.
\end{equation}

A nonnegative basis expansion can approximate the continuous function $\lambda$, which gives the unit cost of information, while probability masses on an increasingly rich grid or a histogram with increasingly many bins can approximate the taste-shock distribution. The probability weights sum to one and satisfy the payoff normalizations. I define the estimator $\widehat\theta_M\in\Theta_M$ as an approximate minimizer of the sample criterion, allowing an optimization error $\epsilon_M$:
\begin{equation}\label{e:approxmin}
\widehat Q_M(\widehat\theta_M)
\leq\inf_{\theta\in\Theta_M}\widehat Q_M(\theta)+\epsilon_M,
\qquad\text{where}\qquad \epsilon_M=o_p(1).
\end{equation}

\Needspace{8\baselineskip}
\subsection{Consistency}
The consistency analysis begins with the following conditions on sample sizes.
\begin{condition}{E1}{}\label{a:E1}
The market sizes satisfy
\[
M\to\infty\qquad\text{and}\qquad N_M\coloneqq \min_{m\leq M}N_{m,M}\to\infty.
\]
\end{condition}

Assumption~E1 requires both the number of markets and the number of observations within each market to increase.

The following assumption collects the restrictions on the parameter space $\Theta$.
\begin{condition}{E2}{Parameter space and regularity}\label{a:E2}
\begin{enumerate}
\item $\Theta$ contains $\theta_0$ and is compact under
\begin{equation}\label{e:metric}
\begin{aligned}
d(\theta,\widetilde\theta)
\coloneqq{}&|\alpha-\widetilde\alpha|+|\delta-\widetilde\delta|\\
&+\|\lambda-\widetilde\lambda\|_\infty+d_L(G,\widetilde G),
\end{aligned}
\end{equation}
where $d_L$ is the L\'evy metric for weak convergence.
\item For every $\theta=(\alpha,\delta,\lambda,G)\in\Theta$,
\[
\int\omega\,dG(\omega)=0,\qquad
\int\omega^2\,dG(\omega)=1,\qquad\text{and}\qquad
\delta\ne0.
\]
The function $\lambda:\mathcal X\to[0,\infty)$, which gives the unit cost of information, is continuous and satisfies
\[
\inf_{x\in\mathcal X}\lambda(x)=0.
\]
\item For every $\theta\in\Theta$ and every $x\in\mathcal X$ with $\lambda(x)=0$,
either $\mathcal Z_x=\R$, or the function $z\mapsto b(z,x;\theta)$ is real analytic on $\R$.
\end{enumerate}
\end{condition}

Under Assumption~E2(a), convergence in the metric $d$ means convergence of $(\alpha,\delta)$, uniform convergence of $\lambda$, and weak convergence of $G$. Assumption~E2(b) imposes the payoff normalization, nonzero-slope, and information-cost restrictions in ID1(b)--(d) and ID2 on every parameter value in $\Theta$. At zero information cost, E2(c) requires either full conditional support of $Z$, or a choice-probability function that is real analytic on $\R$. Together with ID1--ID3, E2(b)--(c) ensure that the population criterion is uniquely minimized at $\theta_0$.

\Needspace{8\baselineskip}
The remaining condition concerns approximation of $\theta_0$ by parameter values in $\Theta_M$.
\begin{condition}{E3}{Sieve approximation}\label{a:E3}
A deterministic sequence $\{\theta_M^*\}_{M=1}^\infty$ satisfies
\[
\theta_M^*\in\Theta_M\subseteq\Theta\qquad\text{and}\qquad
d(\theta_M^*,\theta_0)\to0\quad(M\to\infty).
\]
\end{condition}

Assumption~E3 means that the finite-dimensional approximations in $\Theta_M$ can approach $\theta_0$ as the sample grows.

\Needspace{8\baselineskip}
\begin{theorem}[Consistency]\label{t:consistency}
Let $\widehat\theta_M\in\Theta_M$ satisfy \eqref{e:approxmin}.
Under ID1--ID3 and E1--E3,
\[
d(\widehat\theta_M,\theta_0)\pto0\qquad\text{as }M\to\infty.
\]
\end{theorem}

Briefly, the proof first shows that replacing population choice probabilities with market shares has a vanishing effect on the criterion. It then uses identification and compactness of $\Theta$ to establish convergence to $\theta_0$; Section~\ref{a:consistencyproof} gives the full argument.

\subsubsection{Extension: Multiple Alternatives}\label{s:multestimation}
With multiple alternatives, I minimize the sum of squared residuals for alternatives 1 through $J$. For the $J+1$ alternatives in Section~\ref{s:multiple}, let $\mathbf q\coloneqq(q_1,\ldots,q_J)'$ collect these input probabilities and set $q_0\coloneqq1-\sum_{j=1}^Jq_j$. The probability of choosing alternative 0 is determined by the others, so only $J$ residual components are needed. Write $v_j(\omega;z)\coloneqq\alpha_j+\delta_jz_j+\omega_j$ and $v_0(\omega;z)\coloneqq0$. At positive cost, the averaged conditional choice probability is
\begin{equation}\label{e:multiPsi}
\Psi_j(\theta;w,\mathbf q)\coloneqq
\int\frac{q_j e^{v_j(\omega;z)/\lambda(x)}}
{q_0+\sum_{k=1}^J q_k e^{v_k(\omega;z)/\lambda(x)}}\,dG(\omega),
\qquad j=1,\ldots,J.
\end{equation}
At zero cost, I use the full-information probabilities that each alternative has the highest payoff. When information is free at the true parameter, Assumption~A6(b) rules out payoff ties. I define the column vector
\[
\mathbf r(\theta;w,\mathbf q)
\coloneqq\bigl(\Psi_j(\theta;w,\mathbf q)-q_j\bigr)_{j=1}^J.
\]
Let $\widehat s_{m,j}$ be the observed share choosing alternative $j$ in market $m$, and set $\widehat{\mathbf s}_m\coloneqq(\widehat s_{m,1},\ldots,\widehat s_{m,J})'$. Summing the squared residuals gives
\begin{equation}\label{e:multicriterion}
\widehat Q_M(\theta)\coloneqq\frac1M\sum_{m=1}^M
\norm{\mathbf r(\theta;W_m,\widehat{\mathbf s}_m)}_2^2.
\end{equation}
The sieve approximates the common unit cost of information as a function of the cost shifter and the joint taste-shock distribution. Positive bounded market weights can be included as in \eqref{e:weightedcriterion}.

For fixed $J$, the consistency proof controls each residual component separately and then sums the bounds. The proof also requires continuity and identification conditions for the model with multiple alternatives; Section~\ref{a:consistencyproof} outlines these requirements.

\section{Inference}\label{s:inference}
In this section, I construct confidence regions for the model parameters and confidence intervals for functions of these parameters using test inversion. For ease of exposition, I first present the procedure for two alternatives. Subsection~\ref{s:inferencevalidity} establishes the asymptotic validity of the intervals, and Subsection~\ref{s:multinference} extends the inference procedure to multiple alternatives.

\subsection{Confidence Intervals}
Information costs and sieve probability weights can equal zero, placing estimates on the boundary of the parameter space. A normal approximation derived for interior parameters need not apply in these cases. I instead test $H_0:\theta_0=\theta$ for $\theta\in\Theta_M$ using the RI residuals and collect the parameter values that are not rejected. This gives a confidence region without requiring asymptotic normality of the estimator.

At the true parameter $\theta_0$, $r(\theta_0;W_m,s_m)=0$. The observed share $\widehat s_m$ need not equal the population probability $s_m$, so $r(\theta_0;W_m,\widehat s_m)$ need not be zero. For exposition, I take $N_{m,M}=N_M$ in every market and write
\[
r_m(\theta,q)\coloneqq r(\theta;W_m,q),\qquad\text{and}\qquad
A_m(\theta,q)\coloneqq\partial_qr_m(\theta,q).
\]
\Needspace{8\baselineskip}
At $\theta_0$ and $0<s_m<1$, a Taylor expansion of the residual gives
\begin{equation}\label{e:Taylor}
r_m(\theta_0,\widehat s_m)
=A_m(\theta_0,s_m)(\widehat s_m-s_m)
+O_p((\widehat s_m-s_m)^2).
\end{equation}
Conditional on $W_m$, the first-order term in \eqref{e:Taylor} has variance
\[
\frac{A_m(\theta_0,s_m)^2s_m(1-s_m)}{N_M}.
\]
For a fixed market with $0<s_m<1$ and $A_m(\theta_0,s_m)\ne0$, the binomial central limit theorem therefore gives
\[
\frac{N_M r_m(\theta_0,\widehat s_m)^2}
{A_m(\theta_0,s_m)^2s_m(1-s_m)}
\dto\chi^2_1
\qquad(N_M\to\infty).
\]

I sum the squared residuals divided by their estimated variances:
\begin{equation}\label{e:teststat}
T_M(\theta)\coloneqq\sum_{m\in\widehat{\mathcal A}_M(\theta)}
\frac{N_M r_m(\theta,\widehat s_m)^2}
{A_m(\theta,\widehat s_m)^2\widehat s_m(1-\widehat s_m)},
\end{equation}
where, for fixed $\eta\in(0,1/2)$ and $a_*>0$,
\begin{equation}\label{e:actualretained}
\widehat{\mathcal A}_M(\theta)\coloneqq
\{m:\eta\leq\widehat s_m\leq1-\eta,
\ |A_m(\theta,\widehat s_m)|\geq a_*\}.
\end{equation}
For rejection probability $\gamma\in(0,1)$, I use the critical value $\chi^2_{1-\gamma,|\widehat{\mathcal A}_M(\theta)|}$, the $(1-\gamma)$ quantile of the chi-square distribution with $|\widehat{\mathcal A}_M(\theta)|$ degrees of freedom. Restricting the sum to $\widehat{\mathcal A}_M(\theta)$ excludes markets with observed shares near zero or one or residual derivatives close to zero. Without trimming, small denominators can magnify individual terms in $T_M(\theta)$ and invalidate the chi-square approximation as $M$ and $N_M$ grow together.

I invert this test by collecting the sieve parameters that are not rejected:
\begin{equation}\label{e:CR}
\mathcal C^s_{M,\gamma}\coloneqq
\left\{\theta\in\Theta_M:
T_M(\theta)\leq
\chi^2_{1-\gamma,|\widehat{\mathcal A}_M(\theta)|}\right\}.
\end{equation}
If $\widehat{\mathcal A}_M(\theta)=\varnothing$, then $T_M(\theta)=0$, so $\theta\in\mathcal C^s_{M,\gamma}$. I obtain intervals for parameters and counterfactuals by minimizing and maximizing a scalar function $\phi:\Theta\to\R$ over the confidence region:
\begin{equation}\label{e:projection}
L_M\coloneqq\inf_{\theta\in\mathcal C^s_{M,\gamma}}\phi(\theta),
\qquad\text{and}\qquad
U_M\coloneqq\sup_{\theta\in\mathcal C^s_{M,\gamma}}\phi(\theta).
\end{equation}
The endpoints $L_M$ and $U_M$ may be infinite. For example, $\phi(\theta)$ can equal a payoff coefficient such as $\delta$, the information cost $\lambda(x)$ at a fixed $x$, or a counterfactual change in task-taking probability.

\Needspace{8\baselineskip}
\subsection{Asymptotic Validity of the Confidence Intervals}\label{s:inferencevalidity}
The following assumptions give asymptotic coverage for the intervals in \eqref{e:projection}. I use the sampling model in Section~\ref{s:observations}, with a common within-market sample size $N_M$ and the fixed thresholds $(\eta,a_*)$ in \eqref{e:actualretained}.

For a market with $0<s_m<1$ and $\lambda_0(X_m)>0$, the derivative of the RI residual at the true parameter satisfies
\begin{equation}\label{e:truthderivative}
A_m(\theta_0,s_m)=
-\frac{\Var\{p_{\theta_0}(\Omega;W_m)\}}{s_m(1-s_m)}.
\end{equation}
At zero cost, the residual subtracts its input probability from a full-information probability, so the derivative is $-1$.

\Needspace{9\baselineskip}
For the inference result, I first specify how $M$ and $N_M$ grow together.

\begin{condition}{F1}{Joint growth}\label{a:F1}
Each market has $N_M$ choices, with
\[
M,N_M\to\infty\qquad\text{and}\qquad M/\sqrt{N_M}\to0.
\]
\end{condition}

F1 controls errors in the Taylor approximation and estimated variances as the number of markets increases. Together with F3, it makes trimming based on estimated shares agree with trimming based on population probabilities with probability approaching one.

Write $s(w)\coloneqq \pi(z,x)$. For $w$ with $0<s(w)<1$, let
\[
A_0(w)\coloneqq \partial_qr(\theta_0;w,s(w)).
\]
Here $A_0(w)$ is the derivative of the residual with respect to the input probability $q$, evaluated at $\theta_0$ and $q=s(w)$. I next state a condition on the proportion of markets retained by the test.

\begin{condition}{F2}{Positive probability of satisfying the trimming conditions}\label{a:F2}
\[
\PP\{\eta<s(W)<1-\eta,\ |A_0(W)|>a_*\}>0.
\]
\end{condition}

Assumption~F2 ensures that a positive proportion of the population of markets satisfies both trimming conditions strictly at $\theta_0$.

Let $\theta_M^*=(\alpha_M^*,\delta_M^*,\lambda_M^*,G_M^*)\in\Theta_M$ be a deterministic approximation to $\theta_0$, with $d(\theta_M^*,\theta_0)\to0$, as in E3. For $0<s(w)<1$, define
\[
A_M(w)\coloneqq \partial_qr(\theta_M^*;w,s(w))
\]
and the set of covariate values retained when the test uses population probabilities:
\begin{equation}\label{e:populationretained}
\mathcal R_M\coloneqq 
\{w:\eta\leq s(w)\leq1-\eta,\ |A_M(w)|\geq a_*\}.
\end{equation}

I next consider how share-estimation errors can change which residuals are used to calculate the test statistic. For $0<s(w)<1$, I define the distance to the nearest trimming cutoff:
\begin{equation}\label{e:cutoffdistance}
\Delta_M(w)\coloneqq \min\left\{
|s(w)-\eta|,\ |s(w)-(1-\eta)|,
\bigl||A_M(w)|-a_*\bigr|
\right\}.
\end{equation}
Set $\Delta_M(w)\coloneqq \eta$ when $s(w)\in\{0,1\}$. When $s(w)=0$ or $s(w)=1$, the observed share equals $s(w)$ almost surely, so sampling error cannot change whether the market is retained.

\Needspace{9\baselineskip}
The following assumption gives a probability bound for $\Delta_M(W)$.
\begin{condition}{F3}{Little probability near the trimming cutoffs}\label{a:F3}
There exist $C<\infty$, $t_0>0$ and an integer $M_0$ such that
\[
\PP\{\Delta_M(W)\leq t\}\leq Ct
\quad\text{for every }M\geq M_0\text{ and }0<t\leq t_0.
\]
\end{condition}

F3 limits the probability that a small error in an estimated market share changes whether that market's residual is used to calculate the test statistic.

Replacing $\theta_0$ by its sieve approximation $\theta_M^*$ can produce a nonzero residual even when population choice probabilities are used. I therefore impose the following assumption.

\begin{condition}{F4}{Negligible residual bias}\label{a:F4}
Define
\begin{equation}\label{e:residualbias}
\rho_M\coloneqq \sup_{w\in\mathcal R_M}|r(\theta_M^*;w,s(w))|.
\end{equation}
Require $M^{1/4}\sqrt{N_M}\,\rho_M\to0$.
\end{condition}

F4 is an undersmoothing condition. It makes the sieve approximation bias negligible for the normal approximation to the centered and scaled test statistic.

Recall that $\phi:\Theta\to\R$ is the scalar function of interest and $[L_M,U_M]$ is the interval defined in \eqref{e:projection}. The test is inverted over the sieve $\Theta_M$, while the reported interval is intended to cover the true value $\phi(\theta_0)$. I therefore impose the following assumption.

\begin{condition}{F5}{Negligible approximation error in the reported function}\label{a:F5}
Let $\phi_M^*\coloneqq \phi(\theta_M^*)$ and
$E_M\coloneqq \{\theta_M^*\in\mathcal C^s_{M,\gamma}\}$.
For some deterministic $\varepsilon_M\downarrow0$,
\[
|\phi_M^*-\phi(\theta_0)|=o(\varepsilon_M),
\]
\[
\text{and}\qquad
\PP\!\left\{[\phi_M^*-\varepsilon_M,\phi_M^*+\varepsilon_M]
\subseteq[L_M,U_M]\mid E_M\right\}\to1.
\]
\end{condition}

Here $E_M$ is the event that the test does not reject $\theta_M^*$. F1--F4 establish coverage for the sieve approximation $\phi(\theta_M^*)$. F5 ensures that, when $\theta_M^*$ is accepted, the difference between $\phi(\theta_M^*)$ and $\phi(\theta_0)$ is small enough for the same interval to contain $\phi(\theta_0)$ as well.

\Needspace{8\baselineskip}
\begin{theorem}[Coverage for functions of the parameters]\label{t:coverage}
Under E3, ID1(c), ID3(a), and F1--F5,
\[
\liminf_{M\to\infty}\PP\{\phi(\theta_0)\in[L_M,U_M]\}\geq1-\gamma.
\]
\end{theorem}

Section~\ref{a:inferenceproof} gives the proof. Theorem~\ref{t:coverage} requires no asymptotic normality of the sieve estimator, so parameter constraints may bind.

\subsubsection{Extension: Multiple Alternatives}\label{s:multinference}
With multiple alternatives, share-estimation errors are correlated, so I use a covariance matrix to standardize the residuals. Choosing one alternative excludes choosing another. I use the $J$ shares and residuals for alternatives 1 through $J$ from Section~\ref{s:multestimation}, and let $\mathbf s_m$ denote the vector of their population choice probabilities. For conditionally independent multinomial choices, the covariance of $\widehat{\mathbf s}_m$ conditional on $W_m$ is
\[
\frac{\Sigma_m}{N_M},\qquad\text{where}\qquad
\Sigma_m\coloneqq\operatorname{diag}(\mathbf s_m)-\mathbf s_m\mathbf s_m'.
\]
Let $B_m(\theta,\mathbf q)\coloneqq\partial_{\mathbf q'}\mathbf r(\theta;W_m,\mathbf q)$ be the Jacobian of the residual vector with respect to $\mathbf q$. At the true parameter, the leading residual covariance is
\[
\frac{1}{N_M}B_m(\theta_0,\mathbf s_m)\Sigma_m
B_m(\theta_0,\mathbf s_m)'.
\]
An estimate of $N_M$ times this covariance is
\[
\begin{aligned}
\widehat V_m(\theta)
&\coloneqq B_m(\theta,\widehat{\mathbf s}_m)\widehat\Sigma_m
B_m(\theta,\widehat{\mathbf s}_m)',\\
\text{where}\qquad\widehat\Sigma_m
&\coloneqq\operatorname{diag}(\widehat{\mathbf s}_m)
-\widehat{\mathbf s}_m\widehat{\mathbf s}_m'.
\end{aligned}
\]

The trimming rule now excludes markets with small choice probabilities or a nearly singular residual covariance matrix. The shares of every alternative, including alternative 0, must be bounded away from zero, and the smallest eigenvalue of $\widehat V_m(\theta)$ must be bounded away from zero. For retained markets, the squared standardized residual in the binary statistic is replaced by
\[
N_M\,\mathbf r(\theta;W_m,\widehat{\mathbf s}_m)'
\widehat V_m(\theta)^{-1}
\mathbf r(\theta;W_m,\widehat{\mathbf s}_m).
\]
Extending the proof requires the Jacobian $B_m(\theta,\mathbf q)$ in place of the scalar derivative $A_m(\theta,q)$ and the covariance matrix above to standardize the residual vector. The counterparts of F1--F4 must control sampling and sieve-approximation errors, retain a positive fraction of markets, and limit how often errors in estimated shares change whether a market's residual is used to calculate the statistic. These conditions permit the Taylor-expansion and trimming comparisons used in the binary proof. The test then uses a chi-square critical value with $J$ degrees of freedom per retained market. Test inversion gives a confidence region, and minimizing and maximizing $\phi$ over this region gives its interval. As in F5, when the test does not reject the sieve approximation $\theta_M^*$, the interval must contain $\phi(\theta_0)$ as well as $\phi(\theta_M^*)$, with probability approaching one.

\section[Empirical Application: Assessing Tasks in an Online Labour Market]{Empirical Application: Assessing Tasks\\in an Online Labour Market}\label{s:application}
\subsection{Data and Specification}
I use workers' decisions on Amazon Mechanical Turk, an online market for short assignments, to examine wage responses under different information costs. A worker sees a posting's reward and description but may still be uncertain how demanding the assignment would be for her. I relate the estimated unit cost of information to how concretely the description specifies the work.

The posting data collected by \citet{dube} record rewards, requester identifiers, titles, descriptions, and the number of assignments still available at different times. I retain postings with a positive reward, an identified requester, at least two recorded counts of available assignments within the first 30 minutes, and at least ten assignments available at the first observation. To compare similar work at different rewards, I then select groups from the same requester with matching task descriptions that contain at least four postings, at least three distinct rewards, and a highest reward at least twice the lowest. These restrictions yield 174 postings. Requiring a recorded count of available assignments at the common three-hour endpoint leaves 137 postings in 35 groups, observed between June 1 and August 20, 2017. The repeated tasks provide within-group reward variation, and the common endpoint gives a comparable observation window for task taking.

Workers choose whether to take a task after viewing its posting. To estimate the model parameters, I need to estimate the probability that a worker takes the task. I use each posting's task-taking share, or market share, to estimate its task-taking probability. I add up each decrease in the number of available assignments during the three-hour window, ignoring increases. I divide this total by the calibrated number of workers who viewed the posting.\footnote{I assume a common rate of ten worker arrivals per initial assignment over 90 minutes. For the three-hour window, the calibrated worker count is therefore twenty times the posting's initial number of assignments. \citet{chilton2010} document that workers sort postings by the number of available assignments and concentrate on the first few pages of results. Larger batches can therefore attract more visits by appearing earlier in these results.}

I classify each posting's description by whether it specifies what workers must do and how much work is required. \emph{Task specificity} requires a concrete action and its target, such as finding an agent's email address. \emph{Workload specificity} requires a stated duration, a numerical amount of work, or an effort limit, such as inspecting at most three search results. A broad label such as ``survey'' need not specify a concrete task, and ``quick'' alone does not specify a numerical workload. Titles and descriptions are read together, with identical text assigned the same code. Combining the two criteria determines the description category $x_m$ in Table~\ref{t:categories}. The two middle categories need not have a predetermined cost ranking: one specifies the activity, while the other specifies the workload.

\begin{table}[htbp]
\centering
\small
\caption{Task description categories}\label{t:categories}
\begin{tabular}{@{}clrp{6.1cm}@{}}
\toprule
$x$ & Category & Postings & Illustrative description\\
\midrule
0 & Both specific & 29 & Find the agent's email address; inspect up to three search results.\\
1 & Task only & 83 & Find the agent's email address.\\
2 & Workload only & 9 & Complete a two-minute study.\\
3 & Neither specific & 16 & Complete a survey.\\
\bottomrule
\end{tabular}
\par\smallskip
\begin{minipage}{\textwidth}\footnotesize
Notes: The examples show how I classify descriptions according to whether they specify what workers must do and how much work is required. Task specificity requires a concrete action and target; workload specificity requires a numerical duration, quantity, or effort limit. The sample contains 44 distinct posting texts. Category values are labels: they impose neither equal distances nor a ranking of the two middle categories.
\end{minipage}
\end{table}

\Needspace{7\baselineskip}
I estimate how the posted reward affects the probability that a worker takes a task. Worker $i$ decides whether to take the task in posting $m$, with the payoff from not taking it normalized to zero. The payoff from taking it and the unit cost of information are
\begin{equation}\label{e:empiricalutility}
v_{im}=a_{c(m)}+\delta z_m+\Omega_{im},
\qquad\text{and}\qquad \lambda_m\coloneqq\lambda(x_m).
\end{equation}
The index $c(m)$ groups postings from the same requester with matching task descriptions. The intercept $a_{c(m)}$ allows average payoffs to differ across these groups, while $\Omega_{im}$ represents uncertain worker-specific suitability. I estimate the reward coefficient from reward variation within groups, defining the reward index as
\[
z_m\coloneqq\log r_m-\overline{\log r}_{c(m)},
\]
where $r_m$ is the posted reward and the group average is computed from the estimation sample. Within each task group, I treat the posted reward as independent of $\Omega_{im}$.

Descriptions that specify both the activity and the workload can make it easier for workers to assess personal difficulty. I estimate a separate information cost for each description category, without imposing a ranking or setting any category's cost to zero. I approximate the taste-shock distribution by a 41-bin histogram with mean zero and variance one. I jointly estimate the bin probabilities, the 35 task-group intercepts, the reward coefficient, and four information costs by minimum distance.

\subsection{Estimated Information Costs}
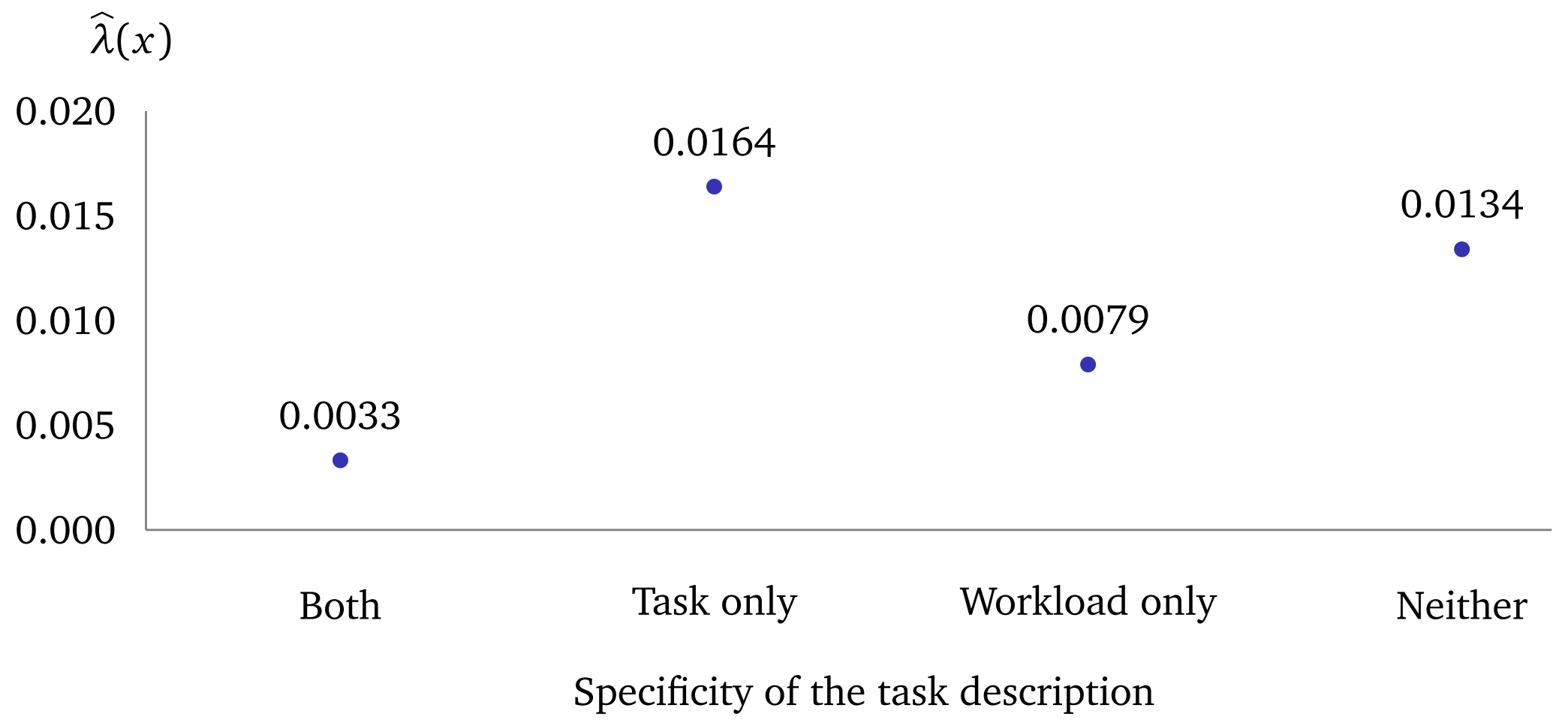
\begin{figure}[!htb]
\centering
\resizebox{.88\textwidth}{!}{\setlength{\unitlength}{1mm}
\begin{picture}(108,47)
\put(9,12){\color{gray}\line(1,0){94}}
\put(9,12){\color{gray}\line(0,1){28}}
\put(8,45){\makebox(0,0){\scriptsize $\widehat\lambda(x)$}}
\put(7,12.000){\makebox(0,0)[r]{\scriptsize 0.000}}
\put(7,19.000){\makebox(0,0)[r]{\scriptsize 0.005}}
\put(7,26.000){\makebox(0,0)[r]{\scriptsize 0.010}}
\put(7,33.000){\makebox(0,0)[r]{\scriptsize 0.015}}
\put(7,40.000){\makebox(0,0)[r]{\scriptsize 0.020}}
\put(22,16.65323){\color{songblue}\circle*{1.2}}
\put(22,19.65323){\makebox(0,0){\scriptsize 0.0033}}
\put(22,7){\makebox(0,0){\scriptsize Both}}
\put(47,34.95207){\color{songblue}\circle*{1.2}}
\put(47,37.95207){\makebox(0,0){\scriptsize 0.0164}}
\put(47,7){\makebox(0,0){\scriptsize Task only}}
\put(72,23.05865){\color{songblue}\circle*{1.2}}
\put(72,26.05865){\makebox(0,0){\scriptsize 0.0079}}
\put(72,7){\makebox(0,0){\scriptsize Workload only}}
\put(97,30.76081){\color{songblue}\circle*{1.2}}
\put(97,33.76081){\makebox(0,0){\scriptsize 0.0134}}
\put(97,7){\makebox(0,0){\scriptsize Neither}}
\put(57,1){\makebox(0,0){\scriptsize Specificity of the task description}}
\end{picture}}
\caption{Estimated unit costs of information by task-description category.}\label{f:empiricalcosts}
\end{figure}

The estimated unit cost of information is lowest when both the task and workload are specific (Figure~\ref{f:empiricalcosts}). It is approximately 0.0033 when both are specific, 0.0164 when only the task is specific, 0.0079 when only the workload is specific, and 0.0134 when neither criterion is met.

Table~\ref{t:costCI} reports nominal 95\% intervals computed by simulation-based test inversion for the posting counts. The last column reports the corresponding intervals for a larger simulated sample generated at the estimated parameters.

\begin{table}[!htb]
\centering
\small
\caption{Information costs and uncertainty}\label{t:costCI}
\begin{tabular}{@{}lccc@{}}
\toprule
Quantity & Estimate & Actual data & Simulated sample\\
\midrule
Both specific & 0.0033 & $[0,8]$ & $[0,0.0555]$\\
Task only & 0.0164 & $[0,0.5953]$ & $[0.0123,0.3017]$\\
Workload only & 0.0079 & $[0,2.1085]$ & $[0.0031,0.1107]$\\
Neither specific & 0.0134 & $[0,8]$ & $[0.0043,0.2540]$\\
Wage coefficient $\delta$ & 0.0017 & $[10^{-6},1.2694]$ & $[0.0015,0.0421]$\\
\bottomrule
\end{tabular}
\par\smallskip
\begin{minipage}{\textwidth}\footnotesize
Notes: The first four rows report unit costs of information; the last reports the wage coefficient. Brackets report nominal 95\% confidence intervals. ``Actual data'' uses the 137 observed postings. The simulated-sample intervals use a larger simulated sample, treating the estimated model parameters as true.
\end{minipage}
\end{table}

\begin{samepage}
\subsection{Wage Responses under Different Information Costs}
I compare wage responses under estimated, halved, and zero information costs, holding the payoff coefficients and taste-shock distribution fixed. For all 137 postings, I double the posted reward and compute choice probabilities at both the original and doubled reward under each cost specification. I let workers choose their information acquisition separately at each reward. I report the percentage-point change in the average task-taking probability and its \mbox{log-arc elasticity}.\footnote{Let $\bar b_1^{\kappa}$ and $\bar b_2^{\kappa}$ be the equally weighted average task-taking probabilities at the original and doubled reward, with information costs $\kappa\widehat\lambda$ for $\kappa\in\{1,1/2,0\}$. The percentage-point change is $100(\bar b_2^{\kappa}-\bar b_1^{\kappa})$, and the log-arc elasticity is
\begin{equation}\label{e:arcelasticity}
\mathcal E^{\kappa}\coloneqq
\frac{\log(\bar b_2^{\kappa}/\bar b_1^{\kappa})}{\log2}.
\end{equation}}

\end{samepage}

\begin{table}[htbp]
\centering
\small
\caption{Wage doubling for all 137 postings}\label{t:wageregimes}
\begin{tabular}{@{}lrrrr@{}}
\toprule
Information regime & Before (\%) & After (\%) & Change (pp) & Elasticity\\
\midrule
Estimated costs & 2.62 & 3.20 & 0.577 & 0.287\\
Half costs & 8.90 & 9.32 & 0.426 & 0.068\\
Full information & 12.87 & 13.21 & 0.335 & 0.037\\
\bottomrule
\end{tabular}
\par\smallskip
\begin{minipage}{\textwidth}\footnotesize
Notes: Before and After are average task-taking probabilities at the original and doubled rewards. Change is measured in percentage points (pp). The same postings and estimated payoffs enter every regime. Rewards double for all postings; the within-group mean log rewards calculated from the original sample remain fixed. Probabilities are averaged with equal posting weights. Elasticity is defined by \eqref{e:arcelasticity}.
\end{minipage}
\end{table}

\begin{samepage}
With the estimated payoff distribution, lower information costs reduce wage elasticity and raise the task-taking probability at the original reward. In Table~\ref{t:wageregimes}, the elasticity falls from 0.287 at estimated costs to 0.068 at half costs and 0.037 under full information. Doubling rewards raises the average task-taking probability by about 22.0\%, 4.8\%, and 2.6\% in these three cases. At the original reward, the average task-taking probability rises from 2.62\% to 8.90\% and 12.87\% as costs fall.

\end{samepage}

\begin{samepage}
Learning at a lower unit cost can help workers recognize that a task suits them and take it at the original reward, leaving fewer of these workers to change their decision when the wage increases. Under full information and the estimated payoff distribution, relatively few additional workers find the task worthwhile when the reward doubles. The absolute increase in the average task-taking probability falls from 0.577 to 0.335 percentage points, and the higher starting probability further reduces the percentage response. Learning can also reveal that an assignment is unsuitable and lead a worker to decline a task she would otherwise take. The net effects on task-taking probabilities and elasticities therefore depend on the payoff distribution and rewards.

\end{samepage}

Table~\ref{t:counterfactualCI} reports nominal 95\% confidence intervals computed by the same numerical test-inversion procedure. The full-information effect compares task-taking probabilities at zero and estimated costs at the original reward; the elasticity difference is the estimated-cost elasticity minus the full-information elasticity. The last column reports intervals for the larger simulated sample.

\begin{table}[htbp]
\centering
\small
\caption{Counterfactual estimates and uncertainty}\label{t:counterfactualCI}
\setlength{\tabcolsep}{5pt}
\begin{tabular}{@{}lrcc@{}}
\toprule
Quantity & Estimate & Actual data & Simulated sample\\
\midrule
Wage effect (pp) & 0.577 & $[0.000,18.675]$ & $[0.527,1.299]$\\
Full-information effect (pp) & 10.253 & $[-4.390,14.945]$ & $[3.970,14.532]$\\
Estimated-cost elasticity & 0.287 & $[0.000,2.353]$ & $[0.072,0.305]$\\
Half-cost elasticity & 0.068 & $[0.000,2.196]$ & $[0.025,0.196]$\\
Full-information elasticity & 0.037 & $[0.000,1.931]$ & $[0.010,0.131]$\\
Elasticity difference & 0.250 & $[-0.504,1.627]$ & $[0.000,0.277]$\\
\bottomrule
\end{tabular}
\par\smallskip
\begin{minipage}{\textwidth}\footnotesize
Notes: Brackets report nominal 95\% confidence intervals. ``Actual data'' uses the 137 observed postings. The simulated-sample intervals use a larger simulated sample, treating the estimated model parameters as true. ``pp'' denotes percentage points. The wage effect is the change in average task-taking probability when rewards double at estimated information costs. The full-information effect is the change in average task-taking probability from estimated costs to zero costs at the original rewards. Elasticities measure the log change in average task-taking probability divided by $\log 2$ when rewards double. The elasticity difference is the estimated-cost elasticity minus the full-information elasticity.
\end{minipage}

\end{table}

\Needspace{8\baselineskip}
\subsection{Wage Responses under a Cardell Taste-Shock Distribution}
I compare the wage responses predicted by the flexible specification with those predicted under a Cardell taste-shock distribution, a family used by \citet{brownjeon}. I normalize the Cardell distribution to mean zero and variance one and estimate its shape jointly with the other parameters. The postings, utility controls, and cost categories are the same as in the flexible specification.

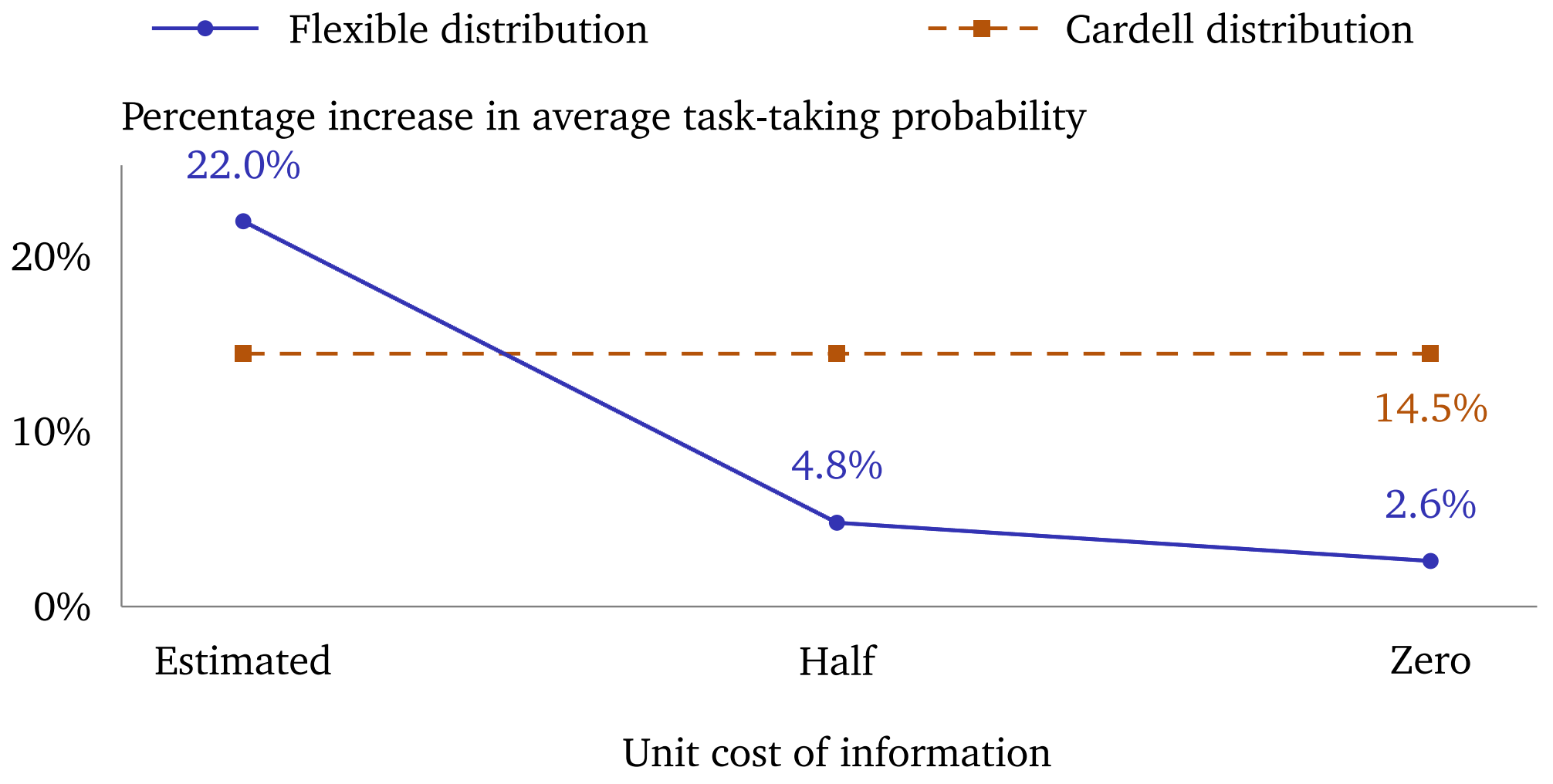
\begin{figure}[htbp]
\centering
\resizebox{.88\textwidth}{!}{
\begingroup
\definecolor{cardellorange}{RGB}{180,83,9}
\setlength{\unitlength}{1mm}
\begin{picture}(108,54)
{\color{songblue}\linethickness{.7pt}\put(12,50){\line(1,0){7}}\put(15.5,50){\circle*{1.2}}}
\put(21,50){\makebox(0,0)[l]{\scriptsize Flexible distribution}}
{\color{cardellorange}\linethickness{.7pt}\multiput(63,50)(2,0){4}{\line(1,0){1.2}}\put(66.5,50){\makebox(0,0){\rule{1.1mm}{1.1mm}}}}
\put(72,50){\makebox(0,0)[l]{\scriptsize Cardell distribution}}
\put(10,44){\makebox(0,0)[l]{\scriptsize Percentage increase in average task-taking probability}}
{\color{gray}\linethickness{.35pt}\put(10,12){\line(1,0){93}}\put(10,12){\line(0,1){29}}}
\put(8,12){\makebox(0,0)[r]{\scriptsize 0\%}}
\put(8,23.5){\makebox(0,0)[r]{\scriptsize 10\%}}
\put(8,35){\makebox(0,0)[r]{\scriptsize 20\%}}
{\color{cardellorange}\linethickness{.7pt}
\multiput(18,28.619847)(2.4,0){33}{\line(1,0){1.4}}
\put(18,28.619847){\makebox(0,0){\rule{1.1mm}{1.1mm}}}
\put(57,28.619847){\makebox(0,0){\rule{1.1mm}{1.1mm}}}
\put(96,28.619847){\makebox(0,0){\rule{1.1mm}{1.1mm}}}}
\put(96,26){\makebox(0,0)[t]{\scriptsize\color{cardellorange}14.5\%}}
{\color{songblue}\linethickness{.7pt}
\qbezier(18,37.320094)(37.5,27.414417)(57,17.508740)
\qbezier(57,17.508740)(76.5,16.251732)(96,14.994724)
\put(18,37.320094){\circle*{1.2}}
\put(57,17.508740){\circle*{1.2}}
\put(96,14.994724){\circle*{1.2}}}
\put(18,40.1){\makebox(0,0)[b]{\scriptsize\color{songblue}22.0\%}}
\put(57,20.3){\makebox(0,0)[b]{\scriptsize\color{songblue}4.8\%}}
\put(96,17.8){\makebox(0,0)[b]{\scriptsize\color{songblue}2.6\%}}
\put(18,8.5){\makebox(0,0){\scriptsize Estimated}}
\put(57,8.5){\makebox(0,0){\scriptsize Half}}
\put(96,8.5){\makebox(0,0){\scriptsize Zero}}
\put(57,2.5){\makebox(0,0){\scriptsize Unit cost of information}}
\end{picture}
\endgroup}
\caption{Percentage response to wage doubling under flexible and Cardell taste-shock distributions. Each model is estimated on the same sample. The Cardell estimates assign zero cost to all four categories, so its three information regimes coincide.}\label{f:Cardell}
\end{figure}

\begin{samepage}
The Cardell estimates assign zero information cost to every category. They predict a 14.5\% increase in the average task-taking probability when rewards double, with log-arc elasticity approximately 0.195. Since the estimated costs are already zero, halving or removing them leaves that prediction unchanged. The flexible specification produces distinct responses under estimated, halved, and zero information costs in Figure~\ref{f:Cardell}.

\end{samepage}

The flexible specification also attains a smaller minimum-distance criterion than the Cardell specification. The values are $7.711\times10^{-5}$ and $2.550\times10^{-4}$, respectively, a reduction of approximately 70\%. This difference is not a formal test of the Cardell family.

\section{Conclusion}\label{s:conclusion}
I establish semiparametric identification of the taste-shock distribution and the unit cost of information as a function of observed characteristics in an RI discrete choice model. For binary choice, I interpret the probability of choosing alternative 1, as a function of the utility shifter, as the CDF of a switch point. The switch point's mean, its variance when information is free, and the covariance in \eqref{e:observablemoments} identify the payoff coefficients and information costs. Deconvolution then recovers the taste-shock distribution. The argument uses optimal learning under Shannon information costs without imposing parametric forms on the taste-shock distribution or on how the unit cost of information varies with observed characteristics.

I estimate the model from market shares using the RI optimality equation. Test inversion gives confidence regions that allow zero information costs and zero probability weights in the taste-shock distribution sieve. The resulting intervals for functions of the parameters have coverage under the sampling, trimming, and approximation conditions stated in the paper.

In the Mechanical Turk application, the estimated unit cost of information varies across description categories. With a flexible taste-shock distribution, lower unit costs of information raise the average task-taking probability at the original reward while reducing wage responsiveness. The Cardell specification assigns zero costs and gives different counterfactual predictions. These estimates illustrate why the taste-shock distribution matters for empirical RI analysis.

\bibliographystyle{plainnat}
\bibliography{references}

\begin{thebibliography}{38}
\providecommand{\natexlab}[1]{#1}
\providecommand{\url}[1]{\texttt{#1}}
\expandafter\ifx\csname urlstyle\endcsname\relax
  \providecommand{\doi}[1]{doi: #1}\else
  \providecommand{\doi}{doi: \begingroup \urlstyle{rm}\Url}\fi

\bibitem[Afrouzi(2024)]{afrouzi}
Hassan Afrouzi.
\newblock Strategic inattention, inflation dynamics, and the nonneutrality of
  money.
\newblock \emph{Journal of Political Economy}, 132\penalty0 (10):\penalty0
  3378--3420, 2024.
\newblock \doi{10.1086/730201}.

\bibitem[Angeletos and Sastry(2025)]{angeletossastry}
George-Marios Angeletos and Karthik~A. Sastry.
\newblock Inattentive economies.
\newblock \emph{Journal of Political Economy}, 133\penalty0 (7):\penalty0
  2265--2319, 2025.
\newblock \doi{10.1086/734870}.

\bibitem[Billingsley(1995)]{billingsley}
Patrick Billingsley.
\newblock \emph{Probability and Measure}.
\newblock John Wiley \& Sons, New York, third edition, 1995.

\bibitem[Brown and Jeon(2024)]{brownjeon}
Zach~Y. Brown and Jihye Jeon.
\newblock Endogenous information and simplifying insurance choice.
\newblock \emph{Econometrica}, 92\penalty0 (3):\penalty0 881--911, 2024.
\newblock \doi{10.3982/ECTA18555}.

\bibitem[Caplin and Dean(2015)]{caplindean}
Andrew Caplin and Mark Dean.
\newblock Revealed preference, rational inattention, and costly information
  acquisition.
\newblock \emph{American Economic Review}, 105\penalty0 (7):\penalty0
  2183--2203, 2015.
\newblock \doi{10.1257/aer.20140117}.

\bibitem[Caplin et~al.(2019)Caplin, Dean, and Leahy]{cdl}
Andrew Caplin, Mark Dean, and John Leahy.
\newblock Rational inattention, optimal consideration sets, and stochastic
  choice.
\newblock \emph{Review of Economic Studies}, 86\penalty0 (3):\penalty0
  1061--1094, 2019.
\newblock \doi{10.1093/restud/rdy037}.

\bibitem[Caplin et~al.(2022)Caplin, Dean, and Leahy]{cdl2022}
Andrew Caplin, Mark Dean, and John Leahy.
\newblock Rationally inattentive behavior: Characterizing and generalizing
  {Shannon} entropy.
\newblock \emph{Journal of Political Economy}, 130\penalty0 (6):\penalty0
  1676--1715, 2022.
\newblock \doi{10.1086/719276}.

\bibitem[Cardell(1997)]{cardell}
N.~Scott Cardell.
\newblock Variance components structures for the extreme-value and logistic
  distributions with application to models of heterogeneity.
\newblock \emph{Econometric Theory}, 13\penalty0 (2):\penalty0 185--213, 1997.
\newblock \doi{10.1017/S0266466600005727}.

\bibitem[Cattaneo et~al.(2020)Cattaneo, Ma, Masatlioglu, and
  Suleymanov]{cattaneo}
Matias~D. Cattaneo, Xinwei Ma, Yusufcan Masatlioglu, and Elchin Suleymanov.
\newblock A random attention model.
\newblock \emph{Journal of Political Economy}, 128\penalty0 (7):\penalty0
  2796--2836, 2020.
\newblock \doi{10.1086/706861}.

\bibitem[Chen(2007)]{chen}
Xiaohong Chen.
\newblock Large sample sieve estimation of semi-nonparametric models.
\newblock In James~J. Heckman and Edward~E. Leamer, editors, \emph{Handbook of
  Econometrics}, volume~6B, chapter~76, pages 5549--5632. Elsevier, 2007.
\newblock \doi{10.1016/S1573-4412(07)06076-X}.

\bibitem[Chilton et~al.(2010)Chilton, Horton, Miller, and Azenkot]{chilton2010}
Lydia~B. Chilton, John~J. Horton, Robert~C. Miller, and Shiri Azenkot.
\newblock Task search in a human computation market.
\newblock In \emph{Proceedings of the ACM SIGKDD Workshop on Human
  Computation}, pages 1--9. Association for Computing Machinery, 2010.
\newblock \doi{10.1145/1837885.1837889}.

\bibitem[Dean and Neligh(2023)]{deanneligh}
Mark Dean and Nathaniel Neligh.
\newblock Experimental tests of rational inattention.
\newblock \emph{Journal of Political Economy}, 131\penalty0 (12):\penalty0
  3415--3461, 2023.
\newblock \doi{10.1086/725174}.

\bibitem[Dube et~al.(2020)Dube, Jacobs, Naidu, and Suri]{dube}
Arindrajit Dube, Jeff Jacobs, Suresh Naidu, and Siddharth Suri.
\newblock Monopsony in online labor markets.
\newblock \emph{American Economic Review: Insights}, 2\penalty0 (1):\penalty0
  33--46, 2020.
\newblock \doi{10.1257/aeri.20180150}.

\bibitem[Engh(2026)]{engh2026}
Christopher~W. Engh.
\newblock A comment on rational inattention with arbitrary choice sets, 2026.
\newblock URL \url{https://arxiv.org/abs/2603.15548v1}.
\newblock arXiv:2603.15548v1.

\bibitem[Fosgerau et~al.(2020)Fosgerau, Melo, de~Palma, and Shum]{fosgerau}
Mogens Fosgerau, Emerson Melo, Andr\'{e} de~Palma, and Matthew Shum.
\newblock Discrete choice and rational inattention: A general equivalence
  result.
\newblock \emph{International Economic Review}, 61\penalty0 (4):\penalty0
  1569--1589, 2020.
\newblock \doi{10.1111/iere.12469}.

\bibitem[Fox et~al.(2012)Fox, Kim, Ryan, and Bajari]{fox2012}
Jeremy~T. Fox, Kyooil Kim, Stephen~P. Ryan, and Patrick Bajari.
\newblock The random coefficients logit model is identified.
\newblock \emph{Journal of Econometrics}, 166\penalty0 (2):\penalty0 204--212,
  2012.
\newblock \doi{10.1016/j.jeconom.2011.09.002}.

\bibitem[Gualdani and Sinha(2024)]{gualdani}
Cristina Gualdani and Shruti Sinha.
\newblock Identification in discrete choice models with imperfect information.
\newblock \emph{Journal of Econometrics}, 244\penalty0 (1):\penalty0 105854,
  2024.
\newblock \doi{10.1016/j.jeconom.2024.105854}.

\bibitem[Hoeffding(1963)]{hoeffding}
Wassily Hoeffding.
\newblock Probability inequalities for sums of bounded random variables.
\newblock \emph{Journal of the American Statistical Association}, 58\penalty0
  (301):\penalty0 13--30, 1963.
\newblock \doi{10.1080/01621459.1963.10500830}.

\bibitem[Iaria and Wang(2025)]{iariawang2025}
Alessandro Iaria and Ao~Wang.
\newblock Real analytic discrete choice models of demand: Theory and
  implications.
\newblock \emph{Econometric Theory}, 41:\penalty0 1080--1128, 2025.
\newblock \doi{10.1017/S0266466624000148}.

\bibitem[Janson(2010)]{janson2010}
Svante Janson.
\newblock Moments of {Gamma} type and the {Brownian} supremum process area.
\newblock \emph{Probability Surveys}, 7:\penalty0 1--52, 2010.
\newblock \doi{10.1214/10-PS160}.
\newblock URL \url{https://arxiv.org/abs/1002.4135}.

\bibitem[Janson(2013)]{janson2013}
Svante Janson.
\newblock Further examples with moments of {Gamma} type, 2013.
\newblock URL \url{https://arxiv.org/abs/1204.5637}.
\newblock arXiv:1204.5637v2, revised February 6, 2013.

\bibitem[Joo(2023)]{joo}
Joonhwi Joo.
\newblock Rational inattention as an empirical framework for discrete choice
  and consumer-welfare evaluation.
\newblock \emph{Journal of Marketing Research}, 60\penalty0 (2):\penalty0
  278--298, 2023.
\newblock \doi{10.1177/00222437221110173}.

\bibitem[Krantz and Parks(2002)]{krantzparks}
Steven~G. Krantz and Harold~R. Parks.
\newblock \emph{A Primer of Real Analytic Functions}.
\newblock Birkh\"auser, Boston, second edition, 2002.
\newblock \doi{10.1007/978-0-8176-8134-0}.

\bibitem[Lewbel(1997)]{lewbel}
Arthur Lewbel.
\newblock Semiparametric estimation of location and other discrete choice
  moments.
\newblock \emph{Econometric Theory}, 13\penalty0 (1):\penalty0 32--51, 1997.
\newblock \doi{10.1017/S0266466600005636}.

\bibitem[Liao(2024)]{liao}
Moyu Liao.
\newblock Identification of a rational inattention discrete choice model.
\newblock \emph{Journal of Econometrics}, 240\penalty0 (1):\penalty0 105670,
  2024.
\newblock \doi{10.1016/j.jeconom.2024.105670}.

\bibitem[Ma\'{c}kowiak and Wiederholt(2026)]{mw2026}
Bartosz Ma\'{c}kowiak and Mirko Wiederholt.
\newblock Rational inattention during a randomized controlled trial.
\newblock \emph{American Economic Review: Insights}, 8\penalty0 (1):\penalty0
  19--35, 2026.
\newblock \doi{10.1257/aeri.20240509}.

\bibitem[Manzini and Mariotti(2014)]{manzini}
Paola Manzini and Marco Mariotti.
\newblock Stochastic choice and consideration sets.
\newblock \emph{Econometrica}, 82\penalty0 (3):\penalty0 1153--1176, 2014.
\newblock \doi{10.3982/ECTA10575}.

\bibitem[Masatlioglu et~al.(2012)Masatlioglu, Nakajima, and Ozbay]{masatlioglu}
Yusufcan Masatlioglu, Daisuke Nakajima, and Erkut~Y. Ozbay.
\newblock Revealed attention.
\newblock \emph{American Economic Review}, 102\penalty0 (5):\penalty0
  2183--2205, 2012.
\newblock \doi{10.1257/aer.102.5.2183}.

\bibitem[Mat\v{e}jka and McKay(2015)]{mm}
Filip Mat\v{e}jka and Alisdair McKay.
\newblock Rational inattention to discrete choices: A new foundation for the
  multinomial logit model.
\newblock \emph{American Economic Review}, 105\penalty0 (1):\penalty0 272--298,
  2015.
\newblock \doi{10.1257/aer.20130047}.

\bibitem[McFadden(1973)]{mcfadden}
Daniel McFadden.
\newblock Conditional logit analysis of qualitative choice behavior.
\newblock In Paul Zarembka, editor, \emph{Frontiers in Econometrics}, pages
  105--142. Academic Press, New York, 1973.

\bibitem[Munkres(2000)]{munkres}
James~R. Munkres.
\newblock \emph{Topology}.
\newblock Prentice Hall, Upper Saddle River, NJ, second edition, 2000.

\bibitem[Newey and McFadden(1994)]{newey}
Whitney~K. Newey and Daniel McFadden.
\newblock Large sample estimation and hypothesis testing.
\newblock In Robert~F. Engle and Daniel~L. McFadden, editors, \emph{Handbook of
  Econometrics}, volume~4, chapter~36, pages 2111--2245. Elsevier, 1994.
\newblock \doi{10.1016/S1573-4412(05)80005-4}.

\bibitem[{NIST}(2026)]{dlmf}
{NIST}.
\newblock {Digital Library of Mathematical Functions}, 2026.
\newblock URL \url{https://dlmf.nist.gov/5}.
\newblock Version 1.2.8, released September 15, 2026; Chapter 5.

\bibitem[Sims(2003)]{sims}
Christopher~A. Sims.
\newblock Implications of rational inattention.
\newblock \emph{Journal of Monetary Economics}, 50\penalty0 (3):\penalty0
  665--690, 2003.
\newblock \doi{10.1016/S0304-3932(03)00029-1}.

\bibitem[Song and Stern(2025)]{songstern}
Wenting Song and Samuel Stern.
\newblock Firm inattention and the efficacy of monetary policy: A text-based
  approach.
\newblock \emph{Review of Economic Studies}, 92\penalty0 (5):\penalty0
  3438--3469, 2025.
\newblock \doi{10.1093/restud/rdae102}.

\bibitem[Train(2002)]{train}
Kenneth Train.
\newblock Discrete choice methods with simulation, 2002.
\newblock URL \url{https://eml.berkeley.edu/books/train1201.pdf}.
\newblock Manuscript dated March 8, 2002.

\bibitem[Wang(2023)]{wang2023}
Ao~Wang.
\newblock Sieve {BLP}: A semi-nonparametric model of demand for differentiated
  products.
\newblock \emph{Journal of Econometrics}, 235:\penalty0 325--351, 2023.
\newblock \doi{10.1016/j.jeconom.2022.04.002}.

\bibitem[Weber et~al.(2025)Weber, Candia, Afrouzi, Ropele, Lluberas, Frache,
  Meyer, Kumar, Gorodnichenko, Georgarakos, Coibion, Kenny, and Ponce]{weber}
Michael Weber, Bernardo Candia, Hassan Afrouzi, Tiziano Ropele, Rodrigo
  Lluberas, Serafin Frache, Brent Meyer, Saten Kumar, Yuriy Gorodnichenko,
  Dimitris Georgarakos, Olivier Coibion, Geoff Kenny, and Jorge Ponce.
\newblock Tell me something {I} don't already know: Learning in low- and
  high-inflation settings.
\newblock \emph{Econometrica}, 93\penalty0 (1):\penalty0 229--264, 2025.
\newblock \doi{10.3982/ECTA22764}.

\end{thebibliography}
\clearpage
\appendix
\setcounter{theorem}{3}
\section{Identification for Choices among Multiple Alternatives}\label{a:multipleidentification}

This appendix states the model, assumptions, and identification theorem
for the extension in Section~\ref{s:multiple}. The proof is in
Section~\ref{a:multipleproof}.

The alternatives are $\{0,1,\ldots,J\}$, with $J\geq1$, and
\[
\begin{gathered}
v_j(\Omega;z)\coloneqq a_j(z)+\Omega_j,
\\
\text{where}\qquad a_j(z)\coloneqq \alpha_j+\delta_jz_j\quad(j=1,\ldots,J),
\qquad\text{and}\qquad a_0(z)=\Omega_0=0.
\end{gathered}
\]
Write $\alpha\coloneqq (\alpha_1,\ldots,\alpha_J)'\in\R^J$, $\delta\coloneqq (\delta_1,\ldots,\delta_J)'\in\R^J$, and $a(z)\coloneqq (a_1(z),\ldots,a_J(z))'\in\R^J$. The joint taste-shock distribution $G$ has finite second moments,
$\E_G\Omega=0$, and $\Var_G(\Omega_1)=1$.
I assume $\delta_j\ne0$ for every $j$.
I orient each shifter so that $\delta_j>0$.
The unit cost of information is a continuous function $\lambda:\mathcal X\to[0,\infty)$ of the cost shifter $x$.
Write $\mathcal Z_x\coloneqq \supp(Z\mid X=x)$ and
$\mathcal X_+\coloneqq \{x\in\mathcal X:\lambda(x)>0\}$.

At positive cost, $b(z;x)\coloneqq (b^0(z;x),\ldots,b^J(z;x))$ solves
\begin{equation}\label{e:appendix-multshare}
\begin{gathered}
\max_{q\in\Delta_J}\Phi(q;a(z),x),\\
\text{where}\qquad\Phi(q;a,x)\coloneqq \lambda(x)\E_G\log
\left\{q_0+\sum_{j=1}^Jq_j e^{(a_j+\Omega_j)/\lambda(x)}\right\},
\end{gathered}
\end{equation}
where $\Delta_J\coloneqq \{q\in[0,1]^{J+1}:\sum_{j=0}^Jq_j=1\}$.
At zero cost, individuals select the alternative with the highest payoff.
The consideration set is
\[
\mathcal A(z;x)\coloneqq \{j\in\{0,\ldots,J\}:b^j(z;x)>0\}.
\]
The population choice probabilities are collected in
\[
\mathcal B_{\mathrm{obs}}
\coloneqq \left\{\left\{(b^0(z;x),\ldots,b^J(z;x))\right\}_{z\in\mathcal Z_x}\right\}_{x\in\mathcal X}.
\]

For the extension to multiple alternatives, I first state the assumptions on taste shocks and individuals' prior beliefs.
\begin{condition}{A1}{Independence from shifters and a common prior}
$\Omega\indep(Z,X)$, $\Omega\sim G$, and individuals use $G$ as their
common prior.
\end{condition}

Individuals know the joint distribution $G$, but initially do not know their own taste-shock vector $\Omega$. Changing either shifter leaves this distribution unchanged.

\Needspace{8\baselineskip}
I next state a condition on the taste-shock distribution $G$.
\begin{condition}{A2}{Exponential non-collinearity}
For every $x\in\mathcal X_+$, if a vector
$u=(u_0,\ldots,u_J)\in\R^{J+1}$ satisfies
\[
u_0+\sum_{j=1}^J u_j e^{\Omega_j/\lambda(x)}=0
\quad G\text{-almost surely},
\]
then $u=0$.
\end{condition}

Assumption~A2 makes the optimal probability vector $b(z;x)$ unique at positive information cost. For example, a joint taste density that is positive on a nonempty open set satisfies A2, including when the taste shocks are correlated.

Identification of the payoff coefficients $(\alpha,\delta)$ also uses the following condition on observed shifter values.
\begin{condition}{A3}{Positive choice probabilities in the observed support}
\[
\text{For every }x\in\mathcal X_+,\quad
\exists z_x^\circ\in\operatorname{int}(\mathcal Z_x):
\quad b^j(z_x^\circ;x)>0\quad(j=0,\ldots,J).
\]
\end{condition}

For fixed $x\in\mathcal X_+$, continuity implies that every alternative also has positive choice probability at shifter vectors sufficiently close to $z_x^\circ$. Some alternatives may have zero probability at other observed shifter values.

The condition on variation in information costs carries over from the binary model.
\begin{condition}{A4}{Arbitrarily low information costs}
\[
\inf_{x\in\mathcal X}\lambda(x)=0.
\]
\end{condition}

I also use the following moment conditions on $G$.
\begin{condition}{A5}{Exponential moments}
For every $t>0$,
\[
\E_G e^{t\Omega_k}<\infty\qquad\text{and}\qquad
\E_G e^{t(\Omega_k-\Omega_j)}<\infty
\quad(j\ne k,\ j,k\in\{1,\ldots,J\}).
\]
\end{condition}

Any bounded taste-shock distribution satisfies these moment conditions.

\Needspace{8\baselineskip}
At zero information cost, I state conditions on the observed support $\mathcal Z_x$ and the pairwise taste differences $\Omega_j-\Omega_k$.
\begin{condition}{A6}{Choice at zero information cost}
\begin{enumerate}
\item For every $x\in\mathcal X$ with $\lambda(x)=0$, $\mathcal Z_x=\R^J$.
\item If the unit cost of information satisfies $\lambda(x)=0$ at some
cost-shifter value $x\in\mathcal X$, every pairwise taste difference has
a continuous distribution:
\[
\PP_G(\Omega_j-\Omega_k=t)=0
\quad\text{for every }t\in\R\text{ and }0\leq j\ne k\leq J.
\]
\end{enumerate}
\end{condition}

At zero information cost, Assumption~A6(a) gives full joint support $\mathcal Z_x=\R^J$, and A6(b) rules out payoff ties with positive probability at each fixed $z$. The choice probabilities therefore do not depend on how ties are broken.

\Needspace{8\baselineskip}
The identification argument differs according to whether information has a positive cost at any $x$. If some $x$ has $\lambda(x)>0$, I recover the payoff coefficients using the choice-probability functions when only alternatives 0 and $j$ are available, together with variation in information costs. I then recover $G$ from the probability of choosing alternative 0 when all alternatives are available. If $\lambda(x)=0$ for every $x$, full conditional support identifies the payoff coefficients and $G$ from full-information choices. The following theorem covers both cases; Section~\ref{a:multipleproof} gives its proof.

\begin{theorem}[Identification with multiple alternatives]\label{t:multipleidentification}
Suppose A1--A6 hold, $\delta_j\ne0$ for $j=1,\ldots,J$,
$\E_G\|\Omega\|^2<\infty$, $\E_G\Omega=0$, and $\Var_G(\Omega_1)=1$.
The observed family $\mathcal B_{\mathrm{obs}}$ identifies
\[
\theta\coloneqq (\alpha,\delta,\lambda,G).
\]
\end{theorem}

\setcounter{theorem}{4}
\clearpage
\section{Proofs}\label{a:proofs}
\numberwithin{equation}{subsection}
\numberwithin{lemma}{subsection}
\subsection{Auxiliary Results}\label{a:auxiliary}
I first state and prove lemmas for the binary RI model used in the proofs
of Theorems~\ref{t:identification}--\ref{t:coverage}. The first two establish
properties of choice-probability functions and switch-point moments for identification.
The next four concern residual continuity, replacing population choice probabilities
with sample shares, binomial sampling moments, and a conditional central
limit theorem. I restate the binary choice problem and residual so that
these lemmas can be read independently of the main text.

I write $w\coloneqq (z,x)$,
$\theta\coloneqq (\alpha,\delta,\lambda,G)$, and
\[
v(\omega;z)\coloneqq \alpha+\delta z+\omega,\qquad
\sigma(u)\coloneqq (1+e^{-u})^{-1},\qquad
\ell(q)\coloneqq \log\{q/(1-q)\}\quad(0<q<1).
\]
For a fixed value $x$ of the cost shifter with $\lambda(x)>0$, the population
choice probability $b(z,x;\theta)$ maximizes
\begin{equation}\label{e:aux-shareproblem}
q\longmapsto\lambda(x)\int
\log\{1-q+q e^{v(\omega;z)/\lambda(x)}\}\,dG(\omega),
\qquad q\in[0,1].
\end{equation}
At zero cost, I use the full-information comparison and its
right-continuous CDF version when defining switch points. When discussing
market shares, $s(w)\coloneqq b(z,x;\theta_0)$ denotes the true population choice probability.
The residual used below is $r(\theta;w,q)\coloneqq \Psi(\theta;w,q)-q$, where
\[
\Psi(\theta;w,q)\coloneqq 
\begin{cases}
\displaystyle\int\sigma\!\left(\ell(q)+\frac{v(\omega;z)}{\lambda(x)}\right)dG(\omega),
&\lambda(x)>0,\ 0<q<1,\\[6pt]
q,&\lambda(x)>0,\ q\in\{0,1\},\\[2pt]
\PP_G\{v(\Omega;z)>0\},&\lambda(x)=0.
\end{cases}
\]
The uses of dominated convergence and interchange of integrals below
follow \citet[Theorems~16.4 and 18.3]{billingsley}.

\Needspace{8\baselineskip}
\begin{lemma}[Binary RI choice probabilities]\label{l:completecurve}
Fix $x$, write $\lambda_x\coloneqq \lambda(x)>0$, and suppose $G$ is
nondegenerate with $\E_G|\Omega|<\infty$ and $\delta>0$.
The maximizer $b(z,x;\theta)$ in \eqref{e:aux-shareproblem} is unique and
continuous in $z$. Define its set of utility-shifter values giving
interior probabilities by
\[
\mathcal I_{\theta,x}\coloneqq \{z\in\R:0<b(z,x;\theta)<1\}.
\]
The set $\mathcal I_{\theta,x}$ is a nonempty open interval.
The function $z\mapsto b(z,x;\theta)$ is real analytic and strictly increasing
on $\mathcal I_{\theta,x}$, with range $(0,1)$. It equals
zero to the left of this interval and one to the right. Moreover,
\[
\begin{aligned}
h_{\theta,x}:\mathcal I_{\theta,x}&\longrightarrow\R,\\
z&\longmapsto\alpha+\delta z+\lambda_x\ell(b(z,x;\theta))
\end{aligned}
\]
is an increasing bijection. If two choice-probability functions at positive information costs
$z\mapsto b(z,x;\theta)$ and $z\mapsto b(z,x;\widetilde\theta)$ agree on a
nonempty open interval where both take values in $(0,1)$, then
$b(z,x;\theta)=b(z,x;\widetilde\theta)$ for every $z\in\R$.
\end{lemma}

\begin{proof}
For this proof only, I suppress the fixed $x$ and write $\lambda\coloneqq \lambda_x$, $b(z)\coloneqq b(z,x;\theta)$, and $h(z)\coloneqq h_{\theta,x}(z)$.

I first describe the solution with $0<b(z)<1$ as the inverse of an
increasing function. This will also determine where $b(z)$ equals zero
or one, and why observations on an interval with $0<b(z)<1$ determine
$b(z)$ for every $z\in\R$.

Write $\bar a\coloneqq \alpha+\delta z$ and let $L\coloneqq -\Omega+\lambda\varepsilon$, where $\varepsilon$ is standard logistic and independent of $\Omega$. With $\sigma(u)\coloneqq (1+e^{-u})^{-1}$,
\[
F_L(h)\coloneqq \PP(L\leq h)
=\int\sigma\!\left(\frac{h+\omega}{\lambda}\right)dG(\omega).
\]
The integrand is strictly increasing in $h$, lies in $(0,1)$, and tends to zero and one at the two ends of $\R$. Bounded convergence gives the same limits for $F_L$. Differentiation under the integral gives
\[
F_L'(h)=\frac1\lambda\int
\sigma\!\left(\frac{h+\omega}{\lambda}\right)
\left\{1-\sigma\!\left(\frac{h+\omega}{\lambda}\right)\right\}dG(\omega)>0.
\]
The derivative is bounded by $1/(4\lambda)$. Thus $F_L$ is continuous and strictly increasing from zero to one, and its inverse $F_L^{-1}(q)$ is defined for every $q\in(0,1)$.

For real analyticity, apply Lemma~\ref{l:gumbelconvolution} with
$J=1$ and mixing variable $-\Omega$. The difference of two independent
standard Gumbel variables is standard logistic \citep[Section~3.1]{train},
so that lemma applies
directly to $L\coloneqq -\Omega+\lambda\varepsilon$. In particular, $F_L$ is real
analytic. This smoothing property does not require $G$ to have a density.

I next relate each interior probability to the known payoff index that makes it optimal.

For $q\in(0,1)$ and $\omega\in\R$, define
\[
p_q(\omega)\coloneqq \sigma\!\left(\frac{F_L^{-1}(q)+\omega}{\lambda}\right),
\]
and define the payoff-index map
\[
a:(0,1)\longrightarrow\R,\qquad
q\longmapsto a(q)\coloneqq F_L^{-1}(q)-\lambda\ell(q).
\]
By construction $\E_Gp_q(\Omega)=q$. Hence
\begin{align*}
\E_G[p_q(\Omega)(1-p_q(\Omega))]
&=\E_Gp_q(\Omega)-\E_G[p_q(\Omega)^2]\\
&=q-q^2-\Var_G(p_q(\Omega))
=q(1-q)-\Var_G(p_q(\Omega)).
\end{align*}
Since $\omega\mapsto p_q(\omega)$ is strictly increasing and $G$ is nondegenerate, $\Var_G(p_q(\Omega))>0$. Consequently
\[
0<F_L'(F_L^{-1}(q))
=\frac{q(1-q)-\Var_G(p_q(\Omega))}{\lambda}
<\frac{q(1-q)}{\lambda}.
\]
Differentiating $a(q)$ now yields
\begin{equation}\label{e:aprimpositive}
a'(q)=\frac1{F_L'(F_L^{-1}(q))}
-\frac{\lambda}{q(1-q)}>0.
\end{equation}
The real-analytic inverse-function theorem
\citep[Theorem~1.5.3]{krantzparks} applies because $F_L'(h)>0$ for every $h\in\R$.
Together with closure under composition, it makes $a(q)$ analytic.
Equation \eqref{e:aprimpositive} makes it strictly increasing. Its image $I_a\coloneqq a((0,1))$ is a nonempty open interval, which may be bounded or unbounded on either side.

The function $a(q)$ also determines when the maximizing probability equals
zero or one.

Let the choice-probability objective, in payoff units, be
\[
J_{\bar a}(q)\coloneqq \lambda\int\log\{1-q+q e^{(\bar a+\omega)/\lambda}\}\,dG(\omega),
\qquad 0\leq q\leq1.
\]
The bound
\[
\left|\lambda\log\{1-q+q e^{(\bar a+\omega)/\lambda}\}\right|
\leq|\bar a|+|\omega|
\]
is integrable because $\E_G|\Omega|<\infty$.
Lemma~\ref{l:shareobjectivecontinuity}, with $J=1$, therefore gives
finiteness and continuity of $q\mapsto J_{\bar a}(q)$ on $[0,1]$.
Nondegeneracy of $G$ also gives A2 in this binary case: a relation
$u_0+u_1e^{\Omega/\lambda}=0$ with $u_1\ne0$ would force $\Omega$
to be constant, while $u_1=0$ requires $u_0=0$.
Lemma~\ref{l:shareobjectiveunique} then gives strict concavity and a
unique optimal choice probability.

For $q\in(0,1)$, choose $\eta>0$ so that
$q\in[\eta,1-\eta]$. With $R(\omega)\coloneqq e^{(\bar a+\omega)/\lambda}>0$,
\[
\left|\lambda\frac{R(\omega)-1}{1-q+qR(\omega)}\right|
\leq\frac{\lambda}{\eta}.
\]
This bound justifies differentiation under the integral on any such
interval of input probabilities. The logit identity then gives
\begin{align*}
J_{\bar a}'(q)
&=\lambda\int\frac{e^{(\bar a+\omega)/\lambda}-1}
{1-q+q e^{(\bar a+\omega)/\lambda}}\,dG(\omega)\\
&=\frac{\lambda}{q(1-q)}
\{F_L(\bar a+\lambda\ell(q))-q\}.
\end{align*}
Because $F_L(a(q)+\lambda\ell(q))=q$ and $F_L$ is strictly increasing,
\[
\operatorname{sign}J_{\bar a}'(q)=\operatorname{sign}\{\bar a-a(q)\}.
\]
If $\bar a\in I_a$, there is exactly one $q$ with $\bar a=a(q)$.
The derivative is positive below that probability and negative above it, so
the unique optimum is $q=a^{-1}(\bar a)$. If $\bar a$ is at or below
a finite left endpoint of $I_a$, $J_{\bar a}'(q)<0$ for every $q\in(0,1)$
and the optimum is zero. If $\bar a$ is at or above a finite right
endpoint, the optimum is one.

The probabilities $a^{-1}(\bar a)$ approach zero and one at finite endpoints
of $I_a$. To prove convergence to zero at a finite left endpoint $L\coloneqq \inf I_a$, take any sequence
$\bar a_n\in I_a$ with $\bar a_n\to L$ and put
$q_n\coloneqq a^{-1}(\bar a_n)$. If $q_n$ did not tend to zero, some
subsequence would satisfy $q_n\geq\varepsilon$ for a fixed
$\varepsilon\in(0,1)$. Monotonicity of $a$ would give
$\bar a_n=a(q_n)\geq a(\varepsilon)>L$ along that subsequence,
contradicting $\bar a_n\to L$. Thus $q_n\to0$.
At a finite right endpoint $R\coloneqq \sup I_a$, the same reasoning uses
$q_n\leq1-\varepsilon$ and
$\bar a_n\leq a(1-\varepsilon)<R$ to show $q_n\to1$.

Thus the interior utility-shifter interval is
\[
I\coloneqq \{z:\alpha+\delta z\in I_a\},
\]
and on this interval
\[
b(z)=a^{-1}(\alpha+\delta z)\quad(z\in I).
\]
The inverse-function theorem and \eqref{e:aprimpositive} make $b$
analytic and strictly increasing on $I$. As $q$ ranges over $(0,1)$,
the corresponding $z$ ranges over $I$, so $b(I)=(0,1)$.
The limits just proved show that $b$ remains continuous where it
equals zero or one. Finally, on $I$,
\[
h(z)=F_L^{-1}(b(z)).
\]
Since $F_L^{-1}$ maps $(0,1)$ onto $\R$, $h$ maps $I$ bijectively and increasingly onto $\R$.

It remains to show why equality on an observed open interval where
both probabilities lie in $(0,1)$ implies equality for every real $z$.

Consider two such functions $z\mapsto b_1(z)$ and $z\mapsto b_2(z)$ with interior intervals $I_1$ and $I_2$, agreeing on a nonempty observed open interval. The identity theorem says that two real-analytic functions agreeing on an open interval agree throughout their connected common domain. By \citet[Corollary 1.2.6]{krantzparks}, they therefore agree on $I_1\cap I_2$.

Write $I_i=(L_i,R_i)$, allowing an endpoint to be infinite.
If $L_1<L_2$, their nonempty intersection implies $L_2<R_1$.
Thus $L_2$ is finite and lies in $I_1$. Choose
$z_n\in I_1\cap I_2$ with $z_n\downarrow L_2$.
The second function satisfies $b_2(z_n)\to0$, whereas continuity of the
first gives $b_1(z_n)\to b_1(L_2)>0$. This contradicts their equality
on the intersection. Interchanging the functions rules out $L_2<L_1$,
so $L_1=L_2$. The corresponding argument at the right endpoints uses
the limit one and gives $R_1=R_2$.
Hence $I_1=I_2$, and their zero and one extensions also agree.
\end{proof}

\Needspace{8\baselineskip}
\begin{lemma}[Switch-point moments]\label{l:switchmoments}
Suppose $\E_G\Omega=0$, $\Var_G(\Omega)=\tau^2\in(0,\infty)$, and $\delta>0$. Fix $x$ with $\lambda_x\coloneqq \lambda(x)\geq0$, and write $\pi(z,x)\coloneqq b(z,x;\theta)$ for every $z\in\R$. If $\lambda_x=0$, suppose this function is continuous. If $\lambda_x>0$, set
\[
I_x\coloneqq \{z:0<\pi(z,x)<1\},\qquad
h_x:I_x\to\R,\quad
h_x(z)\coloneqq \alpha+\delta z+\lambda_x\ell(\pi(z,x)).
\]
For $\Omega\sim G$ and an independent standard logistic $\varepsilon$, define the switch point
\[
S_x\coloneqq 
\begin{cases}
h_x^{-1}(-\Omega+\lambda_x\varepsilon),&\lambda_x>0,\\[3pt]
-(\alpha+\Omega)/\delta,&\lambda_x=0.
\end{cases}
\]
Its CDF is $\pi(\cdot,x)$. The moments
\[
\mu(x)\coloneqq \E S_x,\qquad V(x)\coloneqq \Var(S_x),\qquad
C(x)\coloneqq \Cov(S_x,\ell(\pi(S_x,x)))
\]
are finite, $V(x)>0$, $C(x)>0$, and
\begin{equation}\label{e:general-switch-moments}
\mu(x)=-\frac{\alpha}{\delta},\qquad
\delta^2V(x)+2\delta\lambda_xC(x)=\tau^2.
\end{equation}
Moreover,
\begin{equation}\label{e:variance-gap-proof}
0\leq\tau^2-\delta^2V(x)
\leq\frac{2\pi\tau}{\sqrt3}\lambda_x,
\end{equation}
and $V(x)=\tau^2/\delta^2$ if and only if $\lambda_x=0$.
\end{lemma}

\begin{proof}
I first verify that the switch point has the stated CDF. At positive cost and $z\in I_x$, the Mat\v ejka--McKay conditional choice formula and its average give
\begin{align*}
\pi(z,x)
&=\int
\frac{\pi(z,x)e^{(\alpha+\delta z+\omega)/\lambda_x}}
{1-\pi(z,x)+\pi(z,x)e^{(\alpha+\delta z+\omega)/\lambda_x}}\,dG(\omega)\\
&=\int\sigma\!\left(\frac{h_x(z)+\omega}{\lambda_x}\right)dG(\omega)\\
&=\PP\{h_x(z)\geq-\Omega+\lambda_x\varepsilon\}.
\end{align*}
Lemma~\ref{l:completecurve} makes $h_x$ an increasing bijection from $I_x$ onto $\R$. Thus
\[
\PP(S_x\leq z)=\pi(z,x)\qquad(z\in I_x).
\]
Since $S_x$ lies in $I_x$ almost surely, the equality also holds to its left and right, where both probabilities are zero and one, respectively. Continuity gives equality at finite endpoints. At zero cost, $S_x=-(\alpha+\Omega)/\delta$ gives
\[
\PP(S_x\leq z)=\PP_G(\alpha+\delta z+\Omega\geq0)=\pi(z,x),
\]
where continuity rules out a positive probability of a payoff tie at any $z$. This also shows that $S_x$ is nondegenerate: at positive cost its CDF increases from zero to one on $I_x$, and at zero cost its variance is $\tau^2/\delta^2>0$.

The continuous CDF implies that $\pi(S_x,x)$ is uniform on $(0,1)$. To verify this directly, fix $u\in(0,1)$ and choose $z_u$ with $\pi(z_u,x)=u$. Any interval on which the CDF equals $u$ has probability zero. Consequently
\[
\PP\{\pi(S_x,x)\leq u\}
=\PP(S_x\leq z_u)=u.
\]
In particular, the variable
\[
\eta_x\coloneqq \ell(\pi(S_x,x))
\]
is well defined almost surely and is standard logistic. Both $\varepsilon$ and $\eta_x$ have mean zero and variance $\pi^2/3$. These moments follow, for example, from the representation of a standard logistic variable as the difference of two independent standard Gumbel variables, whose variance is $\pi^2/6$ \citep[Section~3.1]{train}. The variables $\varepsilon$ and $\eta_x$ need not be independent.

Let $L_x\coloneqq -\Omega+\lambda_x\varepsilon$. Applying $h_x$ to the definition of $S_x$ at positive cost gives
\begin{equation}\label{e:switchpoint-identity-proof}
L_x=\alpha+\delta S_x+\lambda_x\eta_x.
\end{equation}
At zero cost, the same identity follows directly from $S_x=-(\alpha+\Omega)/\delta$. Independence of $\Omega$ and $\varepsilon$ gives
\[
\E L_x=0,\qquad
\E L_x^2=\Var(L_x)=\tau^2+\lambda_x^2\frac{\pi^2}{3}<\infty.
\]
Rearranging \eqref{e:switchpoint-identity-proof} bounds the switch-point second moment:
\[
\E S_x^2
\leq\frac{3}{\delta^2}
\left\{\E L_x^2+\alpha^2+\lambda_x^2\E\eta_x^2\right\}<\infty.
\]
Thus $\mu(x)$, $V(x)$, and $C(x)$ are finite. Taking means in \eqref{e:switchpoint-identity-proof} gives $0=\alpha+\delta\mu(x)$. Taking variances gives
\[
\tau^2+\lambda_x^2\frac{\pi^2}{3}
=\delta^2V(x)+2\delta\lambda_x C(x)
+\lambda_x^2\frac{\pi^2}{3}.
\]
The common logistic variance terms cancel, proving \eqref{e:general-switch-moments}. The covariance term remains because $S_x$ and $\eta_x$ generally depend on each other.

I next establish $C(x)>0$. Let $S_x'$ be an independent copy of $S_x$. Expansion and independence give
\[
2C(x)=\E\!\left[(S_x-S_x')
\{\ell(\pi(S_x,x))-\ell(\pi(S_x',x))\}\right].
\]
The product is nonnegative because the CDF and log odds are increasing. The two CDF values are independent continuous uniform variables, so they differ almost surely. Monotonicity then makes the product strictly positive almost surely. Its expectation is finite by the second-moment bound, and hence $C(x)>0$.

It follows from \eqref{e:general-switch-moments} that
\[
\tau^2-\delta^2V(x)=2\delta\lambda_xC(x)\geq0,
\]
with equality exactly when $\lambda_x=0$. Cauchy--Schwarz and this inequality give
\[
C(x)\leq\frac{\pi}{\sqrt3}\sqrt{V(x)},\qquad
\delta\sqrt{V(x)}\leq\tau.
\]
Therefore
\[
0\leq\tau^2-\delta^2V(x)
=2\delta\lambda_xC(x)
\leq\frac{2\pi}{\sqrt3}\lambda_x\delta\sqrt{V(x)}
\leq\frac{2\pi\tau}{\sqrt3}\lambda_x,
\]
which proves \eqref{e:variance-gap-proof}.
\end{proof}

\Needspace{8\baselineskip}
\begin{lemma}[Residual derivatives and continuity]\label{l:residualregularity}
For each $\eta\in(0,1/2)$ there is a finite constant $K_\eta$, independent of $\theta$ and $w$, such that
\begin{equation}\label{e:residual-derivative-bounds}
|r(\theta;w,q)|\leq1,\qquad
|\partial_qr(\theta;w,q)|+|\partial_q^2r(\theta;w,q)|\leq K_\eta
\end{equation}
for $q\in[\eta,1-\eta]$. In particular, $q\mapsto\partial_qr(\theta;w,q)$ is uniformly Lipschitz on this interval.
Fix $w\coloneqq (z,x)$ and $q\in(0,1)$, and suppose $d(\theta_n,\theta)\to0$, where $\theta\coloneqq (\alpha,\delta,\lambda,G)$.
If $\lambda(x)>0$, or if $\lambda(x)=0$ and $G\{-\alpha-\delta z\}=0$, then
\[
\begin{aligned}
r(\theta_n;w,q)&\longrightarrow r(\theta;w,q),\\
\partial_qr(\theta_n;w,q)&\longrightarrow\partial_qr(\theta;w,q).
\end{aligned}
\]
For each fixed $\theta\coloneqq (\alpha,\delta,\lambda,G)$ with $\delta\ne0$,
ID3(a) implies
\[
\PP\{\lambda(X)=0,\ G\{-\alpha-\delta Z\}>0\}=0.
\]
\end{lemma}

\begin{proof}
Recall $r(\theta;w,q)\coloneqq \Psi(\theta;w,q)-q$. Because $\Psi(\theta;w,q)$ and $q$ lie in $[0,1]$, $|r(\theta;w,q)|\leq1$. I next bound derivatives with respect to the input probability and then examine continuity in the parameter, including a zero-cost limit.

When $\lambda(x)>0$, define
\[
p_q(\omega)\coloneqq \sigma\!\left(\ell(q)+\frac{\alpha+\delta z+\omega}{\lambda(x)}\right).
\]
Using $\sigma'(u)=\sigma(u)(1-\sigma(u))$ and $\ell'(q)=1/\{q(1-q)\}$ gives
\[
\partial_qp_q(\omega)=\frac{p_q(\omega)(1-p_q(\omega))}{q(1-q)}.
\]
Differentiating once more and collecting terms gives
\[
\begin{aligned}
\partial_q^2p_q(\omega)
&=\frac{p_q(\omega)(1-p_q(\omega))\{(1-2p_q(\omega))-(1-2q)\}}{q^2(1-q)^2}\\
&=\frac{2p_q(\omega)(1-p_q(\omega))(q-p_q(\omega))}{q^2(1-q)^2}.
\end{aligned}
\]
Let $d_\eta\coloneqq \eta(1-\eta)>0$. For $q\in[\eta,1-\eta]$ and $\omega\in\R$,
\[
|\partial_qp_q(\omega)|\leq\frac1{4d_\eta},
\qquad
|\partial_q^2p_q(\omega)|\leq\frac1{2d_\eta^2}.
\]
These bounds do not involve the payoff, the cost, or $G$. They justify differentiation under the integral. Since
\[
\begin{aligned}
\partial_qr(\theta;w,q)&=\int\partial_qp_q(\omega)\,dG(\omega)-1,\\
\partial_q^2r(\theta;w,q)&=\int\partial_q^2p_q(\omega)\,dG(\omega),
\end{aligned}
\]
the constant $K_\eta\coloneqq 1+1/(4d_\eta)+1/(2d_\eta^2)$ suffices in \eqref{e:residual-derivative-bounds}. When $\lambda(x)=0$, $\Psi(\theta;w,q)$ is independent of $q$, so $\partial_qr(\theta;w,q)=-1$ and $\partial_q^2r(\theta;w,q)=0$. The mean-value theorem also gives
\[
|\partial_qr(\theta;w,t)-\partial_qr(\theta;w,q)|
\leq K_\eta|t-q|
\]
for two shares in the trimmed interval.

When $\lambda_0(x)>0$, set $\theta=\theta_0$ and $q=s(w)\in(0,1)$. The population choice probability $q$ satisfies $\E p_q(\Omega)=q$. Thus
\begin{align*}
\partial_qr(\theta_0;w,q)
&=\frac{\E[p_q(\Omega)(1-p_q(\Omega))]}{q(1-q)}-1\\
&=\frac{q(1-q)-\Var(p_q(\Omega))}{q(1-q)}-1
=-\frac{\Var(p_q(\Omega))}{q(1-q)}.
\end{align*}
This verifies \eqref{e:truthderivative}.

I next show that the residual $r(\theta_n;w,q)$ and its derivative $\partial_qr(\theta_n;w,q)$ converge to their values at $\theta$, under the conditions in the lemma.

For the sequence in the statement, the representation theorem in \citet[Theorem~25.6]{billingsley}
allows weak convergence $G_n\Rightarrow G$ to be represented by random variables $\Omega_n\sim G_n$ and $\Omega\sim G$ on a common probability space with $\Omega_n\to\Omega$ almost surely. This representation lets me compare the integrands directly. Metric convergence also gives
\[
\alpha_n\to\alpha,\qquad \delta_n\to\delta,\qquad
\lambda_n(x)\to\lambda(x).
\]
Write $v_n\coloneqq \alpha_n+\delta_n z+\Omega_n$ and $v\coloneqq \alpha+\delta z+\Omega$. The integrand defining $\Psi(\theta_n;w,q)$ is
\[
p_n(q)\coloneqq 
\begin{cases}
\sigma\!\left(\ell(q)+v_n/\lambda_n(x)\right),&\lambda_n(x)>0,\\
\ind\{v_n>0\},&\lambda_n(x)=0.
\end{cases}
\]
If $\lambda(x)>0$, then $\lambda_n(x)>0$ eventually. Ordinary continuity gives, almost surely,
\[
\begin{aligned}
p_n(q)&\longrightarrow p(q)\coloneqq \sigma\!\left(\ell(q)+v/\lambda(x)\right),\\
\partial_qp_n(q)&=\frac{p_n(q)(1-p_n(q))}{q(1-q)}
\longrightarrow\frac{p(q)(1-p(q))}{q(1-q)}.
\end{aligned}
\]
The integrands satisfy $0\leq p_n(q)\leq1$ and $|\partial_qp_n(q)|\leq1/\{4q(1-q)\}$. Because $q$ is fixed in $(0,1)$, bounded convergence applies to both expectations. Subtracting $q$ from the first expectation and $1$ from the second proves the two limits in the statement.

If $\lambda(x)=0$, then, outside the payoff-tie event $v=0$, $v_n\to v$ with the same sign eventually. Along terms with $\lambda_n(x)>0$,
\[
\begin{aligned}
v>0&\ \Longrightarrow\ 
\ell(q)+v_n/\lambda_n(x)\to+\infty,\quad p_n(q)\to1,\\
v<0&\ \Longrightarrow\ 
\ell(q)+v_n/\lambda_n(x)\to-\infty,\quad p_n(q)\to0.
\end{aligned}
\]
In either case
\[
\partial_qp_n(q)=\frac{p_n(q)(1-p_n(q))}{q(1-q)}\to0.
\]
Along terms with $\lambda_n(x)=0$, $p_n(q)=\ind\{v_n>0\}$ and $\partial_qp_n(q)=0$. Therefore $p_n(q)\to\ind\{v>0\}$ and $\partial_qp_n(q)\to0$ along the full sequence. The condition $G\{-\alpha-\delta z\}=0$ makes these limits almost sure. Applying bounded convergence with the same bounds gives
\[
\begin{aligned}
r(\theta_n;w,q)&\longrightarrow\PP_G(v>0)-q=r(\theta;w,q),\\
\partial_qr(\theta_n;w,q)&\longrightarrow-1=\partial_qr(\theta;w,q).
\end{aligned}
\]

To apply this continuity argument across markets for a fixed
$\theta\coloneqq (\alpha,\delta,\lambda,G)$, I must show that
$G\{-\alpha-\delta Z\}>0$ has probability zero.

Let $\mathcal D_G\coloneqq \{\omega:G\{\omega\}>0\}$ be the set of point masses of a fixed $G$. This set is countable. With $\delta\ne0$, a payoff tie requires
\[
Z\in\mathcal T_\theta
\coloneqq \{(-\alpha-\omega)/\delta:\omega\in\mathcal D_G\}.
\]
The set $\mathcal T_\theta$ is also countable. ID3(a) gives
$\PP(Z\in\mathcal T_\theta\mid X=x)=0$ at each $x$, and integration over $X$ gives probability zero across markets. This assertion is for each fixed candidate; the argument does not require a single tie-free event for every candidate simultaneously.
\end{proof}

For the share-sampling results, markets have characteristics $W_m$,
population choice probabilities $s_m\coloneqq s(W_m)$, and deterministic sizes $N_{m,M}$.
Conditional on the characteristics, the count $N_{m,M}\widehat s_m$ is
binomial with success probability $s_m$. The two criteria are
\[
\widehat Q_M(\theta)\coloneqq M^{-1}\sum_m r(\theta;W_m,\widehat s_m)^2,
\qquad Q_M(\theta)\coloneqq M^{-1}\sum_m r(\theta;W_m,s_m)^2.
\]
To bound the change in a squared residual when its input probability changes from $s$ to $t$, define
\begin{equation}\label{e:H}
H(t,s)\coloneqq 
\begin{cases}
\min\{1,\tfrac12|\ell(t)-\ell(s)|+2|t-s|\},&t,s\in(0,1),\\
0,&t=s\in\{0,1\},\\
1,&\text{otherwise}.
\end{cases}
\end{equation}

\Needspace{8\baselineskip}
\begin{lemma}[Share substitution]\label{l:sharesubstitution}
Under the sampling conditions and E1, the function $H:[0,1]^2\to[0,1]$ defined above satisfies
\begin{equation}\label{e:sharesub}
\sup_{\theta\in\Theta}|\widehat Q_M(\theta)-Q_M(\theta)|
\leq\frac1M\sum_{m=1}^M H(\widehat s_m,s_m)\pto0.
\end{equation}
\end{lemma}

\begin{proof}
I first bound $|r(\theta;w,t)^2-r(\theta;w,s)^2|$ by a function of
$t$ and $s$ that does not depend on $\theta$. I then show that the
average bound $M^{-1}\sum_m H(\widehat s_m,s_m)$ tends to zero.

For two input probabilities $t,s$, write $r_t\coloneqq r(\theta;w,t)$ and $r_s\coloneqq r(\theta;w,s)$. Since $|r_t|,|r_s|\leq1$,
\begin{align*}
|r_t^2-r_s^2|
&=|r_t-r_s||r_t+r_s|\\
&\leq2|r_t-r_s|
\leq2|\Psi(\theta;w,t)-\Psi(\theta;w,s)|+2|t-s|.
\end{align*}
When $\lambda(x)>0$ and $t,s\in(0,1)$,
$\sup_u|\sigma'(u)|=1/4$ implies
\[
|\Psi(\theta;w,t)-\Psi(\theta;w,s)|
\leq\tfrac14|\ell(t)-\ell(s)|.
\]
Thus $|r_t^2-r_s^2|\leq\tfrac12|\ell(t)-\ell(s)|+2|t-s|$.
When $\lambda(x)=0$, $\Psi(\theta;w,q)$ does not depend on $q$,
so $|r_t^2-r_s^2|\leq2|t-s|$.
Since $r_t^2,r_s^2\in[0,1]$, $|r_t^2-r_s^2|\leq1$ in all cases.
If $t=s\in\{0,1\}$, then $r_t^2=r_s^2$.
If either input equals zero or one and $t\ne s$, I use the bound one.
These inequalities give exactly $|r_t^2-r_s^2|\leq H(t,s)$.

Applying this market by market,
\[
|\widehat Q_M(\theta)-Q_M(\theta)|
\leq\frac1M\sum_m H(\widehat s_m,s_m).
\]
The right-hand side contains no candidate parameter, so taking the supremum over $\theta$ leaves the same bound.

\Needspace{8\baselineskip}

It remains to show
$\E[M^{-1}\sum_mH(\widehat s_m,s_m)]\to0$.

For fixed market characteristics $w$, let $n$ denote its within-market sample size and define
\[
h_n(w)\coloneqq \E[H(\widehat s,s(w))\mid W=w],
\qquad\text{where}\qquad
n\widehat s\mid W=w\sim\operatorname{Binomial}(n,s(w)).
\]
When $0<s(w)<1$, binomial share consistency gives
$\widehat s\to_p s(w)$ as $n\to\infty$. Since $H(t,s(w))\to0$ as
$t\to s(w)$ and $0\leq H\leq1$, for every $\varepsilon>0$,
\[
h_n(w)\leq\varepsilon+
\PP\{H(\widehat s,s(w))>\varepsilon\mid W=w\},
\qquad\text{so}\qquad \limsup_{n\to\infty}h_n(w)\leq\varepsilon.
\]
Letting $\varepsilon\downarrow0$ gives $h_n(w)\to0$. When $s(w)\in\{0,1\}$, the sampled share equals $s(w)$ exactly, so $h_n(w)=0$.

The minimum sample size $N_M\coloneqq \min_m N_{m,M}$ tends to infinity by E1. Even though the sizes can differ across markets,
\begin{align*}
\E\!\left[\frac1M\sum_mH(\widehat s_m,s_m)\right]
&=\int\frac1M\sum_m h_{N_{m,M}}(w)\,dF_W(w)\\
&\leq\int\sup_{n\geq N_M}h_n(w)\,dF_W(w)\longrightarrow0.
\end{align*}
The inequality uses $N_{m,M}\geq N_M$ for every market. For each fixed $w$, $h_n(w)\to0$ means that all sufficiently late terms are small, so $\sup_{n\geq N_M}h_n(w)\to0$. This supremum is bounded by one; bounded convergence justifies the last integral limit. Finally, Markov's inequality gives
\[
\PP\!\left\{\frac1M\sum_mH(\widehat s_m,s_m)>\varepsilon\right\}
\leq\frac1\varepsilon
\E\!\left[\frac1M\sum_mH(\widehat s_m,s_m)\right]\longrightarrow0.\qedhere
\]
\end{proof}

\Needspace{8\baselineskip}
\begin{lemma}[Binomial sampling moments]\label{l:binomialmoments}
Fix $\eta\in(0,1/2)$ and $N\in\mathbb N$. Let $N\widehat s\sim\operatorname{Binomial}(N,s)$ with $s\in[\eta,1-\eta]$, and set
$\xi\coloneqq \sqrt N(\widehat s-s)/\sqrt{s(1-s)}$. Then
\begin{align*}
\E\xi&=0,&\E\xi^2&=1,\\
\E\xi^3&=\frac{1-2s}{\sqrt{Ns(1-s)}},&
\E\xi^4&=3+\frac{1-6s(1-s)}{Ns(1-s)}.
\end{align*}
For each fixed $p\geq2$, $\E|\xi|^p$ is uniformly bounded and
$\E|\widehat s-s|^p\leq C_pN^{-p/2}$, uniformly over this share interval. The same statements hold conditionally on market characteristics.
\end{lemma}

\begin{proof}
Write $N\widehat s=\sum_{i=1}^N Y_i$, with independent Bernoulli variables of success probability $s$, and let $B_i\coloneqq Y_i-s$. Put $D\coloneqq s(1-s)$. The two possible values of $B_i$ give
\[
\E B_i=0,\quad \E B_i^2=D,\quad
\E B_i^3=D(1-2s),\quad
\E B_i^4=D(1-3D).
\]
Also $\xi=(ND)^{-1/2}\sum_i B_i$. Independence and centering immediately give $\E\xi=0$ and $\E\xi^2=1$.

In the expansion of the third power, any term with an index appearing only once has expectation zero. Only terms with all three indices equal remain:
\[
\E\left(\sum_i B_i\right)^3=N\E B_1^3.
\]
In the fourth power, the nonzero terms have either all four indices equal or two pairs of equal indices. There are $3N(N-1)$ terms of the latter type, so
\[
\E\left(\sum_iB_i\right)^4
=N\E B_1^4+3N(N-1)(\E B_1^2)^2.
\]
Dividing by $(ND)^{3/2}$ and $(ND)^2$, respectively, gives
\[
\E\xi^3=\frac{1-2s}{\sqrt{ND}},\qquad
\E\xi^4=\frac{1-3D}{ND}+3\frac{N-1}{N}
=3+\frac{1-6D}{ND}.
\]

For higher absolute moments, \citet[Theorem~2]{hoeffding}, applied to
both tails of the Bernoulli average, and $D\geq\eta(1-\eta)$ give
\[
\PP(|\xi|>t)
=\PP\!\left(|\widehat s-s|>t\sqrt{D/N}\right)
\leq2e^{-2Dt^2}
\leq2e^{-2\eta(1-\eta)t^2}.
\]
Using the tail-integral formula,
\[
\E|\xi|^p
=\int_0^\infty pt^{p-1}\PP(|\xi|>t)\,dt
\leq2p\int_0^\infty t^{p-1}e^{-2\eta(1-\eta)t^2}\,dt<\infty.
\]
The bound depends only on $p$ and $\eta$. Since $\widehat s-s=\sqrt{D/N}\xi$, it also gives the stated share-error bound. Conditional on the market characteristics, $s$ is fixed and the observations remain independent Bernoulli variables, so every calculation applies conditionally as well.
\end{proof}

\Needspace{8\baselineskip}
\begin{lemma}[Conditional central limit theorem]\label{l:conditionalclt}
Let $\mathcal F_M\coloneqq \sigma(W_1,\ldots,W_M)$. For each $M$, suppose there are $k_M$ conditionally independent variables
$X_{M,1},\ldots,X_{M,k_M}$, centered given $\mathcal F_M$. The number of summands $k_M$ is finite and $\mathcal F_M$-measurable. Set
\[
v_M^2\coloneqq \sum_{j=1}^{k_M}\E[X_{M,j}^2\mid\mathcal F_M].
\]
All conditional moments used below are finite. Suppose $\PP(v_M>0)\to1$ and, for some $p>2$,
\[
\Lambda_M\coloneqq 
\frac{\sum_{j=1}^{k_M}\E[|X_{M,j}|^p\mid\mathcal F_M]}{v_M^p}\pto0,
\]
with the ratio defined as zero when $v_M=0$. Define
\[
S_M\coloneqq 
\begin{cases}
v_M^{-1}\sum_{j=1}^{k_M}X_{M,j},&v_M>0,\\
0,&v_M=0
\end{cases}.
\]
Then
\[
S_M\dto N(0,1).
\]
\end{lemma}

\begin{proof}
Once the market characteristics are fixed, the number of summands and their conditional moments are fixed; the sampled choices, and hence the summands themselves, remain random. The ordinary Lyapunov theorem \citep[Theorem~27.3]{billingsley}
can then be applied to these conditional distributions. I spell out why convergence of the random Lyapunov ratio in probability is sufficient.

For a real number $t$, define the conditional CDF
\[
F_M(t)\coloneqq \PP(S_M\leq t\mid\mathcal F_M).
\]
I will first show $F_M(t)\to_p\Phi(t)$, where $\Phi$ is the standard normal CDF. The ordinary Lyapunov theorem concerns deterministic arrays of distributions. Here $\Lambda_M$, $k_M$, and $v_M$ depend on market characteristics, so I first obtain almost-sure convergence along a subsequence.

Take any deterministic subsequence of sample sizes $(M_\ell)$. Since
\[
\Lambda_M+\ind\{v_M=0\}\pto0,
\]
there is a further deterministic subsequence $(M_{\ell_r})$ along which
\[
\Lambda_{M_{\ell_r}}+\ind\{v_{M_{\ell_r}}=0\}\longrightarrow0
\quad\text{almost surely}.
\]
This follows from the subsequence characterization of convergence in
probability \citep[Theorem~20.5]{billingsley}. The indicator implies $v_{M_{\ell_r}}>0$ eventually, almost surely.

Choose the joint conditional distributions of the summands $X_{M,1},\ldots,X_{M,k_M}$ given $\mathcal F_M$. In this application they are supplied directly by the finite product of binomial sampling distributions after $W_1,\ldots,W_M$ have been fixed. On a probability-one set, conditional independence, centering, and the moment identities hold for every sample size; a single such set can be used because there are countably many sample sizes.

Now fix a realization $\omega$ in this set and in the almost-sure convergence event above. For sample size $M_{\ell_r}$, the conditional distribution has exactly $k_{M_{\ell_r}}(\omega)$ independent centered summands and total variance $v_{M_{\ell_r}}(\omega)^2$. Its ordinary Lyapunov ratio is the number $\Lambda_{M_{\ell_r}}(\omega)$, which tends to zero. Theorem~27.3 of \citet{billingsley} therefore gives
\[
F_{M_{\ell_r}}(t)(\omega)\longrightarrow\Phi(t)
\qquad\text{for every real }t.
\]
Thus every subsequence of $F_M(t)$ has a further subsequence converging almost surely to $\Phi(t)$. This proves
\[
F_M(t)\pto\Phi(t).
\]
No independence between samples of different sizes is needed: the Lyapunov theorem uses independence among the summands within each sample.

The conditional convergence just obtained also implies unconditional convergence.

Because $0\leq F_M(t)\leq1$, for every $\varepsilon>0$,
\[
\E|F_M(t)-\Phi(t)|
\leq\varepsilon+
\PP\{|F_M(t)-\Phi(t)|>\varepsilon\}.
\]
The probability tends to zero. Letting $\varepsilon\downarrow0$ gives
$\E|F_M(t)-\Phi(t)|\to0$. Finally, the tower property of conditional expectation
\citep[Theorem~34.4]{billingsley} yields
\[
\PP(S_M\leq t)=\E F_M(t),
\qquad\text{so}\qquad
|\PP(S_M\leq t)-\Phi(t)|
\leq\E|F_M(t)-\Phi(t)|\longrightarrow0.
\]
The standard normal CDF is continuous at every real $t$, so this is precisely $S_M\dto N(0,1)$.
\end{proof}

\Needspace{8\baselineskip}
I now state and prove lemmas for the RI model with multiple alternatives,
used in the identification proof of Theorem~\ref{t:multipleidentification}.
Throughout these lemmas, I use the setup and notation from
Appendix~\ref{a:multipleidentification}.
The first gives the joint distribution of the Gumbel differences used
in that proof. Its special case $J=1$ also gives the logistic distribution
used in the binary model. The subsequent lemmas establish continuity and
uniqueness of the optimal choice probability vector and show how an interior probability
vector determines the known payoff indices.

\begin{lemma}[The Gumbel-difference distribution and its convolution]\label{l:gumbelconvolution}
Let $\eta_0,\ldots,\eta_J$ be independent standard Gumbel variables,
with CDF $\PP(\eta_j\leq y)=\exp(-e^{-y})$, and put
$\zeta\coloneqq (\zeta_1,\ldots,\zeta_J)'$, where $\zeta_j\coloneqq \eta_j-\eta_0$.
Its CDF $F_\zeta$, density $f_\zeta$, and characteristic function
$\varphi_\zeta$ satisfy, for $y,t\in\R^J$,
\begin{align}
F_\zeta(y)&=\frac1{1+\sum_{j=1}^J e^{-y_j}},\label{e:aux-gumbel-cdf}\\
f_\zeta(y)&=\frac{J!\,e^{-\sum_jy_j}}
{(1+\sum_j e^{-y_j})^{J+1}},\label{e:aux-gumbel-density}\\
\varphi_\zeta(t)&=\Gamma\!\left(1+i\sum_{j=1}^Jt_j\right)
\prod_{j=1}^J\Gamma(1-it_j)\ne0\quad(t\in\R^J).
\label{e:aux-gumbel-cf}
\end{align}
If $\Omega\in\R^J$ is any random vector independent of
$\eta_0,\ldots,\eta_J$ and $\lambda>0$, then the CDF $F_H$ and density
$f_H$ of $H\coloneqq \Omega+\lambda\zeta$ are real analytic on $\R^J$.
\end{lemma}
\begin{proof}
The inequality $\zeta\leq y$ means $\eta_j-\eta_0\leq y_j$ for every
$j=1,\ldots,J$. Thus $\{\zeta\leq y\}$ is the event that alternative zero is
chosen when the known payoffs are $0,-y_1,\ldots,-y_J$ and the payoff
shocks are $\eta_0,\ldots,\eta_J$. The multinomial logit formula
\citep[Eq. (3.6)]{train} therefore gives \eqref{e:aux-gumbel-cdf}.
Differentiating once in each coordinate gives
\eqref{e:aux-gumbel-density}.

A standard Gumbel variable has characteristic function
$\Gamma(1-it)$ \citep[Section~3]{janson2010}. Independence gives
\[
\varphi_\zeta(t)
=\E e^{-i(\sum_jt_j)\eta_0}\prod_j\E e^{it_j\eta_j}
=\Gamma\!\left(1+i\sum_jt_j\right)\prod_j\Gamma(1-it_j).
\]
The gamma function $\Gamma$ has no zeros, and all its arguments here have real
part one, away from its poles \citep[Section~5.2(i)]{dlmf}.
This proves \eqref{e:aux-gumbel-cf}, including nonvanishing.

For the analyticity claim, let $G$ denote the distribution of $\Omega$.
Conditioning on $\Omega$ gives
\[
F_H(h)=\int\frac{1}{1+\sum_{j=1}^J
\exp\{(\omega_j-h_j)/\lambda\}}\,dG(\omega).
\]
This is a mixed logit probability for the reference alternative, with
known payoff indices $-h_j/\lambda$ and random intercepts
$\Omega_j/\lambda$. By \citet[Theorem~1(a)]{iariawang2025}, the
non-reference choice probabilities are real analytic for any mixing
distribution. Since $F_H$ is one minus their sum, it is real analytic
on $\R^J$. Its density
$f_H=\partial_{h_1}\cdots\partial_{h_J}F_H$ is real analytic as well,
because derivatives preserve real analyticity. No density or moment
restriction on $G$ is needed.
\end{proof}

\Needspace{8\baselineskip}
\begin{lemma}[Continuity of the choice-probability objective and its unique maximizer]\label{l:shareobjectivecontinuity}
Fix $J\geq1$ and $x$ with $\lambda(x)>0$. Let $\Omega\in\R^J$ have
distribution $G$ satisfying
\[
\E_G\|\Omega\|_1=\sum_{j=1}^J\E_G|\Omega_j|<\infty.
\]
For $a\in\R^J$ and
$\Delta_J\coloneqq \{q\in[0,1]^{J+1}:\sum_{j=0}^Jq_j=1\}$, define
\[
\Phi(q;a,x)\coloneqq \lambda(x)\E_G\log\left\{
q_0+\sum_{j=1}^Jq_j e^{(a_j+\Omega_j)/\lambda(x)}\right\}.
\]
Then $(q,a)\mapsto\Phi(q;a,x)$ is finite and continuous on
$\Delta_J\times\R^J$, including the boundary of $\Delta_J$.
For every $a$, a maximizer over $\Delta_J$ exists. If that maximizer is
unique for every $a$, the map
\[
q^*(\cdot;x):\R^J\longrightarrow\Delta_J,\qquad
 a\longmapsto q^*(a;x)\coloneqq \argmax_{q\in\Delta_J}\Phi(q;a,x)
\]
is continuous.
\end{lemma}
\begin{proof}
Write $\lambda\coloneqq \lambda(x)$ and define the integrand
\[
g(q,a,\omega)\coloneqq \lambda\log\left\{
q_0+\sum_{j=1}^Jq_j e^{(a_j+\omega_j)/\lambda}\right\}.
\]
Because the nonnegative weights $q_0,\ldots,q_J$ sum to one,
\[
\min\{0,a_1+\omega_1,\ldots,a_J+\omega_J\}
\leq g(q,a,\omega)
\leq\max\{0,a_1+\omega_1,\ldots,a_J+\omega_J\}.
\]
In particular,
\[
|g(q,a,\omega)|\leq\|a\|_\infty+\|\omega\|_1.
\]
Since $\E_G\|\Omega\|_1<\infty$, the bound is integrable for fixed
$a$, so $\Phi(q;a,x)$ is finite for every $(q,a)$.

To prove joint continuity, take arbitrary sequences
$q_n\to q$ in $\Delta_J$ and $a_n\to a$ in $\R^J$.
The sequence $a_n$ is bounded, so there is $R<\infty$ with
$\|a_n\|_\infty\leq R$ for all $n$.
For each fixed $\omega$, the sum inside the logarithm is strictly
positive, even if some coordinates of $q$ are zero. Thus
\[
g(q_n,a_n,\omega)\longrightarrow g(q,a,\omega),\qquad
|g(q_n,a_n,\omega)|\leq R+\|\omega\|_1.
\]
The bound is independent of $n$ and integrable under $G$.
Dominated convergence gives
$\Phi(q_n;a_n,x)\to\Phi(q;a,x)$, proving joint continuity.
For fixed $a$, continuity on the compact simplex gives existence of a
maximizer by the extreme value theorem
\citep[Theorem~27.4]{munkres}.

Now suppose the maximizer is unique for every $a$. Fix any
$a_n\to a$ and write $q_n^*\coloneqq q^*(a_n;x)$.
I show $q_n^*\to q^*(a;x)$.
Take any subsequence $\{q_{n_k}^*\}$. Compactness of $\Delta_J$ gives a
further subsequence $q_{n_{k_\ell}}^*\to\bar q\in\Delta_J$.
For every fixed $q\in\Delta_J$, optimality at $a_{n_{k_\ell}}$ gives
\[
\Phi(q_{n_{k_\ell}}^*;a_{n_{k_\ell}},x)
\geq\Phi(q;a_{n_{k_\ell}},x).
\]
Joint continuity permits passage to the limit on both sides:
\[
\Phi(\bar q;a,x)\geq\Phi(q;a,x)
\qquad\text{for every }q\in\Delta_J.
\]
Hence $\bar q$ maximizes the objective at $a$. Uniqueness gives
$\bar q=q^*(a;x)$.
If the original sequence $q_n^*$ did not converge to $q^*(a;x)$,
some subsequence would stay at least a fixed positive distance from
$q^*(a;x)$. The compactness argument just given would produce a
further subsequence converging to $q^*(a;x)$, a contradiction.
Therefore $q^*(a_n;x)\to q^*(a;x)$.
\end{proof}

\Needspace{8\baselineskip}
\begin{lemma}[Uniqueness of optimal choice probabilities]\label{l:shareobjectiveunique}
For the distribution $G$, cost $\lambda(x)>0$, simplex $\Delta_J$, and
objective $\Phi(q;a,x)$ in Lemma~\ref{l:shareobjectivecontinuity}, suppose
A2 holds at $\lambda(x)$: for $u\in\R^{J+1}$,
\[
u_0+\sum_{j=1}^J u_j e^{\Omega_j/\lambda(x)}=0
\quad G\text{-almost surely}
\quad\Longrightarrow\quad u=0.
\]
For every $a\in\R^J$, the function $q\mapsto\Phi(q;a,x)$ is strictly
concave on $\Delta_J$. Its maximizer $q^*(a;x)$ is unique, and the map
$a\mapsto q^*(a;x)$ is continuous.
\end{lemma}
\begin{proof}
Fix $a\in\R^J$ and write $\lambda\coloneqq \lambda(x)$. For $q\in\Delta_J$,
put
\[
D_q(\Omega)\coloneqq q_0+\sum_{j=1}^Jq_j e^{a_j/\lambda}e^{\Omega_j/\lambda}>0.
\]
Take $q,\widetilde q\in\Delta_J$ with $q\ne\widetilde q$.
If $D_q(\Omega)=D_{\widetilde q}(\Omega)$ almost surely, then
\[
(q_0-\widetilde q_0)+\sum_{j=1}^J
(q_j-\widetilde q_j)e^{a_j/\lambda}e^{\Omega_j/\lambda}=0
\quad G\text{-almost surely}.
\]
Apply A2 with
$u_0\coloneqq q_0-\widetilde q_0$ and
$u_j\coloneqq (q_j-\widetilde q_j)e^{a_j/\lambda}$ for $j=1,\ldots,J$.
It gives $u_j=0$ for every $j$. Since $e^{a_j/\lambda}>0$, this would
imply $q=\widetilde q$, a contradiction. Therefore
\[
\PP_G\{D_q(\Omega)\ne D_{\widetilde q}(\Omega)\}>0.
\]
For any $t\in(0,1)$, strict concavity of the logarithm gives
\[
\log\{tD_q(\Omega)+(1-t)D_{\widetilde q}(\Omega)\}
\geq t\log D_q(\Omega)+(1-t)\log D_{\widetilde q}(\Omega),
\]
with a strict inequality on the event just displayed.
All three logarithms are integrable by
Lemma~\ref{l:shareobjectivecontinuity}. Taking expectations and
multiplying by $\lambda>0$ therefore gives
\[
\Phi(tq+(1-t)\widetilde q;a,x)
>t\Phi(q;a,x)+(1-t)\Phi(\widetilde q;a,x).
\]
This is strict concavity. Since a maximizer exists, it is unique.
The continuity conclusion now follows from
Lemma~\ref{l:shareobjectivecontinuity}.
\end{proof}

\Needspace{8\baselineskip}
\begin{lemma}[Interior shares determine a unique payoff vector]\label{l:sharepayoffbijection}
Use the cost $\lambda(x)>0$, distribution $G$, and objective
$\Phi(q;a,x)$ in Lemma~\ref{l:shareobjectivecontinuity}, and suppose A2
holds at $\lambda(x)$. Write
\[
\begin{aligned}
\mathcal D_J&\coloneqq \{p\in[0,1]^J:\textstyle\sum_{j=1}^Jp_j\leq1\},\\
\mathcal D_J^\circ&\coloneqq \{p\in\mathcal D_J:p_j>0\ (j=1,\ldots,J),
\ \textstyle\sum_{j=1}^Jp_j<1\},\\
p_0&\coloneqq 1-\textstyle\sum_{j=1}^Jp_j,\qquad q(p)\coloneqq (p_0,p_1,\ldots,p_J)',
\end{aligned}
\]
so that $\mathcal D_J$ uses the $J$ free coordinates of $\Delta_J$.
Let $\bar b(a;x)$ be the unique maximizer of $\Phi(\cdot;a,x)$, and set
\[
\mathcal W_x\coloneqq \{a\in\R^J:\bar b^j(a;x)>0\ (j=0,\ldots,J)\}.
\]
For every $p\in\mathcal D_J^\circ$, the function
\[
H(a;p,x)\coloneqq \Phi(q(p);a,x)-p'a\qquad(a\in\R^J)
\]
is strictly convex and tends to $+\infty$ as $\|a\|_\infty\to\infty$.
It has a unique minimizer $a^*(p;x)$. The set $\mathcal W_x$ is nonempty
and open, and the maps
\[
\begin{aligned}
B_x:\mathcal W_x&\longrightarrow\mathcal D_J^\circ,
&a&\longmapsto B_x(a)\coloneqq (\bar b^1(a;x),\ldots,\bar b^J(a;x))',\\
a^*(\cdot;x):\mathcal D_J^\circ&\longrightarrow\mathcal W_x,
&p&\longmapsto a^*(p;x)
\end{aligned}
\]
are mutually inverse bijections.
\end{lemma}
\begin{proof}
Lemmas~\ref{l:shareobjectivecontinuity} and
\ref{l:shareobjectiveunique} give a unique choice-probability vector $\bar b(a;x)$
and continuity of $a\mapsto\bar b(a;x)$ on $\R^J$. Consequently
$\mathcal W_x$ is open: if every $\bar b^j(a;x)>0$, continuity keeps
every $\bar b^j(\widetilde a;x)$ positive for $\widetilde a$ sufficiently
close to $a$.
I show that each $p\in\mathcal D_J^\circ$ is attained at exactly one
payoff vector $a$ by minimizing $a\mapsto H(a;p,x)$.

Fix $p\in\mathcal D_J^\circ$, write $\lambda\coloneqq \lambda(x)$, and define
\[
\begin{aligned}
D(a,\omega;p)&\coloneqq p_0+\sum_{j=1}^Jp_j e^{(a_j+\omega_j)/\lambda},\\
\pi_0(a,\omega;p)&\coloneqq p_0/D(a,\omega;p),\\
\pi_j(a,\omega;p)&\coloneqq p_j e^{(a_j+\omega_j)/\lambda}/D(a,\omega;p)
\quad(j=1,\ldots,J).
\end{aligned}
\]
All $\pi_j(a,\omega;p)$ are positive and sum to one. Differentiating
with respect to $a$ gives
\[
\begin{aligned}
\nabla_a H(a;p,x)&=\E_G\pi_{1:J}(a,\Omega;p)-p,\\
\nabla_a^2 H(a;p,x)&=\frac1\lambda\E_G\{
\operatorname{diag}(\pi_{1:J})-\pi_{1:J}\pi_{1:J}'\},
\end{aligned}
\]
where $\pi_{1:J}\coloneqq (\pi_1,\ldots,\pi_J)'$, evaluated at
$(a,\Omega;p)$ in the second line. These differentiations under the
expectation are valid because $0<\pi_j<1$ and
\[
\partial_{a_k}\pi_j
=\lambda^{-1}\pi_j(\ind\{j=k\}-\pi_k)
\quad(j,k=1,\ldots,J)
\]
is bounded independently of $\Omega$.

For $u\in\R^J$, put $u_0\coloneqq 0$. The Hessian's quadratic form is
\[
u'\nabla_a^2 H(a;p,x)u
=\frac1\lambda\E_G\left\{
\sum_{j=0}^J\pi_j u_j^2-\left(\sum_{j=0}^J\pi_j u_j\right)^2
\right\}.
\]
For each $\Omega$, the expression in braces is the variance of the
numbers $u_0,\ldots,u_J$ under the positive probabilities
$\pi_0,\ldots,\pi_J$. It can be zero only if all these numbers are
equal. Since $u_0=0$, it is strictly positive for every $u\ne0$.
Thus $\nabla_a^2H(a;p,x)$ is positive definite and
$H(\cdot;p,x)$ is strictly convex.

To establish existence of its minimizer, put
$p_{\min}\coloneqq \min_{0\leq j\leq J}p_j>0$ and $a_0=0$. Choosing the
largest of $a_0,\ldots,a_J$ in the denominator gives
\[
\lambda\log D(a,\Omega;p)
\geq\lambda\log p_{\min}+\max_{0\leq j\leq J}a_j
-\|\Omega\|_\infty.
\]
\Needspace{8\baselineskip}
The first-moment assumption implies $\E_G\|\Omega\|_\infty<\infty$.
Writing $a_{\max}\coloneqq \max_{0\leq j\leq J}a_j$ and
$a_{\min}\coloneqq \min_{0\leq j\leq J}a_j$, I obtain
\[
\begin{aligned}
H(a;p,x)
&\geq a_{\max}-\sum_{j=0}^Jp_j a_j
  +\lambda\log p_{\min}-\E_G\|\Omega\|_\infty.
\end{aligned}
\]
The first difference on the right satisfies
\[
\begin{aligned}
a_{\max}-\sum_{j=0}^Jp_j a_j
&=\sum_{j=0}^Jp_j(a_{\max}-a_j)\\
&\geq p_{\min}(a_{\max}-a_{\min})
\geq p_{\min}\|a\|_\infty.
\end{aligned}
\]
The last inequality uses $a_0=0$. Therefore $H(a;p,x)\to+\infty$
as $\|a\|_\infty\to\infty$. In particular, outside a sufficiently
large closed ball its value exceeds $H(0;p,x)$. Continuity and the
extreme value theorem on that ball give a minimizer; strict convexity
makes it unique. Denote it by $a^*(p;x)$.

I next show that the maximizing shares at $a^*(p;x)$ are exactly
$q(p)$. The first-order condition for minimizing $H$ gives
$\E_G\pi_j(a^*(p;x),\Omega;p)=p_j$ for $j=1,\ldots,J$.
Summing these equations and taking complements gives the same
equality for $j=0$. Since all $p_j$ are positive, division gives
\[
\E_G\frac{e^{(a_j^*(p;x)+\Omega_j)/\lambda}}
{D(a^*(p;x),\Omega;p)}=1\quad(j=1,\ldots,J),\qquad
\E_G\frac1{D(a^*(p;x),\Omega;p)}=1.
\]
The derivative of the choice-probability objective in its free coordinates is
\[
\partial_{p_j}\Phi(q(p);a,x)
=\lambda\E_G\frac{e^{(a_j+\Omega_j)/\lambda}-1}{D(a,\Omega;p)}.
\]
This derivative is valid for $p\in\mathcal D_J^\circ$:
each ratio $e^{(a_j+\Omega_j)/\lambda}/D$
is bounded by $1/p_j$, and $1/D\leq1/p_0$, with locally uniform bounds.
The preceding equalities make this derivative zero for every $j$
at $(a^*(p;x),p)$. Strict concavity in the shares then gives
\[
\bar b(a^*(p;x);x)=q(p),\qquad B_x(a^*(p;x))=p.
\]
Thus $a^*(p;x)\in\mathcal W_x$, and $\mathcal W_x$ is nonempty.

Conversely, take $a\in\mathcal W_x$ and put $p\coloneqq B_x(a)$.
Since every component of $q(p)$ is positive, the equations
$\partial_{p_j}\Phi(q(p);a,x)=0$ give a common number $k$:
\[
\E_G\frac{e^{(a_j+\Omega_j)/\lambda}}{D(a,\Omega;p)}
=\E_G\frac1{D(a,\Omega;p)}\eqqcolon k\quad(j=1,\ldots,J).
\]
\Needspace{6\baselineskip}
Multiplying by $p_j$, including the equation for alternative 0 multiplied by $p_0$, and
summing shows $k=\E_G[D/D]=1$.
Hence $\E_G\pi_j(a,\Omega;p)=p_j$ for every $j$ and
$\nabla_aH(a;p,x)=0$. The strictly convex function $H(\cdot;p,x)$
has only one stationary point, so $a=a^*(p;x)$.
This proves both inverse identities and the claimed bijection.
\end{proof}

\Needspace{8\baselineskip}
\begin{lemma}[Analytic choice-probability maps and the value function]\label{l:sharevaluegradients}
Under the conditions and notation of Lemma~\ref{l:sharepayoffbijection},
the mutually inverse maps $B_x$ and $a^*(\cdot;x)$ are real analytic.
Define the optimized value and its convex conjugate by
\[
V_x(a)\coloneqq \max_{q\in\Delta_J}\Phi(q;a,x),\qquad
V_x^*(p)\coloneqq \sup_{a\in\R^J}\{p'a-V_x(a)\}.
\]
The function $V_x$ is finite and convex on $\R^J$. For
$p\in\mathcal D_J^\circ$ and $a\in\mathcal W_x$,
\[
\begin{aligned}
V_x^*(p)&=-H(a^*(p;x);p,x),\\
\nabla_aV_x(a)&=B_x(a),\qquad
\nabla_pV_x^*(p)=a^*(p;x).
\end{aligned}
\]
The matrix $D_aB_x(a)$ is symmetric and positive definite for every
$a\in\mathcal W_x$.
\end{lemma}
\begin{proof}
I first establish the differentiability needed for the gradient
identities. Fix $x$ and write $\lambda\coloneqq \lambda(x)$. For
$(a,p)\in\R^J\times\mathcal D_J^\circ$, define
\[
c_j(a,p)\coloneqq -a_j+\lambda\log(p_0/p_j),\qquad
Q_j(c)\coloneqq \E_G\frac{e^{(\Omega_j-c_j)/\lambda}}
{1+\sum_{k=1}^J e^{(\Omega_k-c_k)/\lambda}}
\quad(j=1,\ldots,J),
\]
and $Q_0(c)\coloneqq 1-\sum_{j=1}^JQ_j(c)$. These mixed logit probabilities
are real analytic in $c$ by \citet[Theorem~1(a)]{iariawang2025}.
Since $(a,p)\mapsto c(a,p)$ is real analytic on the stated domain,
the following two first-order-condition maps are jointly real analytic:
\[
\begin{aligned}
F_j(a,p)&\coloneqq \partial_{p_j}\Phi(q(p);a,x)
=\lambda\{Q_j(c(a,p))/p_j-Q_0(c(a,p))/p_0\},\\
L_j(a,p)&\coloneqq \partial_{a_j}H(a;p,x)=Q_j(c(a,p))-p_j,
\quad j=1,\ldots,J.
\end{aligned}
\]
Write $F\coloneqq (F_1,\ldots,F_J)'$ and $L\coloneqq (L_1,\ldots,L_J)'$.

To apply the implicit-function theorem to $F(a,p)=0$, I verify that
$D_pF(a,p)$ is nonsingular. Put
\[
t_j(a,\Omega)\coloneqq e^{(a_j+\Omega_j)/\lambda}-1,
\qquad t(a,\Omega)\coloneqq (t_1(a,\Omega),\ldots,t_J(a,\Omega))',
\]
and recall the denominator
\[
D(a,\Omega;p)=p_0+\sum_{j=1}^Jp_j e^{(a_j+\Omega_j)/\lambda}.
\]
Differentiating the choice-probability objective gives
\[
D_pF(a,p)=-\lambda\E_G
\frac{t(a,\Omega)t(a,\Omega)'}{D(a,\Omega;p)^2}.
\]
The ratios $t_j/D$ are locally uniformly bounded by $1/p_j+1/p_0$,
which justifies differentiation under the expectation.
For a nonzero $u\in\R^J$, the equality $u't(a,\Omega)=0$ almost surely
would imply
\[
-\sum_{j=1}^J u_j+\sum_{j=1}^J u_j e^{a_j/\lambda}
e^{\Omega_j/\lambda}=0\quad G\text{-almost surely}.
\]
A2 would make every $u_j e^{a_j/\lambda}$ zero, contradicting $u\ne0$.
Thus $u't$ is nonzero with positive probability, and
\[
u'D_pF(a,p)u
=-\lambda\E_G\{(u't)^2/D^2\}<0.
\]
The matrix $D_pF(a,p)$ is therefore negative definite.
The real-analytic implicit-function theorem
\citep[Theorem~2.5.1]{krantzparks}, applied at every
$(a,B_x(a))$ with $a\in\mathcal W_x$,
makes $B_x(a)$ real analytic on $\mathcal W_x$.

For the inverse, $D_aL(a,p)=\nabla_a^2H(a;p,x)$ is positive definite
by the preceding lemma. The same theorem applied to $L(a,p)=0$
makes $a^*(p;x)$ real analytic on $\mathcal D_J^\circ$.
The local solutions agree on overlaps because $B_x(a)$ and
$a^*(p;x)$ are unique at each $a$ and $p$, respectively.

I now relate these inverse maps to $V_x$ and $V_x^*$.
For fixed $q$, the function $a\mapsto\Phi(q;a,x)$ is convex, because
the logarithm of a sum of exponentials with nonnegative weights is
convex. Its maximum over $q$ is convex as well. The bound in
Lemma~\ref{l:shareobjectivecontinuity} gives
\[
|V_x(a)|\leq\|a\|_\infty+\E_G\|\Omega\|_1<\infty.
\]
On $\mathcal W_x$, differentiate
$V_x(a)=\Phi(q(B_x(a));a,x)$.
The derivative through $B_x(a)$ vanishes because the share
first-order conditions are $F(a,B_x(a))=0$. The remaining derivative is
\[
\nabla_aV_x(a)=\E_G\pi_{1:J}(a,\Omega;B_x(a))=B_x(a),
\]
where the final equality was proved in Lemma~\ref{l:sharepayoffbijection}.

Fix $p\in\mathcal D_J^\circ$. For every $a\in\R^J$,
\[
p'a-V_x(a)
\leq p'a-\Phi(q(p);a,x)
=-H(a;p,x)
\leq-H(a^*(p;x);p,x).
\]
\Needspace{8\baselineskip}
Both inequalities are equalities at $a=a^*(p;x)$ because
$\bar b(a^*(p;x);x)=q(p)$. Taking the supremum proves the expression
for $V_x^*(p)$. Equivalently,
\[
V_x^*(p)=p'a^*(p;x)-V_x(a^*(p;x)).
\]
The chain rule gives
\[
\nabla_pV_x^*(p)
=a^*(p;x)+D_pa^*(p;x)'\{p-\nabla_aV_x(a^*(p;x))\}
=a^*(p;x),
\]
since $\nabla_aV_x(a^*(p;x))=B_x(a^*(p;x))=p$.

Finally, $D_aB_x(a)=\nabla_a^2V_x(a)$ is symmetric and positive
semidefinite by convexity of $V_x$. Differentiating the inverse
identity gives
\[
D_pa^*(B_x(a);x)\,D_aB_x(a)=I_J.
\]
Hence $D_aB_x(a)$ is nonsingular. A symmetric positive semidefinite
matrix with no zero eigenvalue is positive definite, proving the last
claim.
\end{proof}

\Needspace{8\baselineskip}
I return to the binary RI model to justify the joint continuity of
$(z,x)\mapsto b(z,x;\theta_0)$ used in the discussion preceding
Theorem~\ref{t:identification}. Assumption~ID1(e) imposes continuity
only in $z$. The next result derives joint continuity
from the RI problem, continuity of $\lambda_0$, and the existing
zero-cost support condition in ID3(b).
\begin{lemma}[Joint continuity of the true choice probability]\label{l:choicejointcontinuity}
Suppose ID1 and ID3 hold. If $\lambda_0(x)=0$ for some $x$, the
distribution $G_0$ has no atoms. At every such $x$,
$z\mapsto b(z,x;\theta_0)$ is continuous on $\R$ and
\[
\inf_{z\in\mathcal Z_x}b(z,x;\theta_0)=0,\qquad
\sup_{z\in\mathcal Z_x}b(z,x;\theta_0)=1.
\]
The map
\[
\operatorname{supp}(Z,X)\longrightarrow[0,1],\qquad
(z,x)\longmapsto b(z,x;\theta_0)
\]
is jointly continuous. In particular, for every sequence
$(z_n,x_n)$ in this support converging to
$(z_\star,x_\star)$ in the support,
\begin{equation}\label{e:jointchoicecontinuity}
b(z_n,x_n;\theta_0)\longrightarrow b(z_\star,x_\star;\theta_0).
\end{equation}
\end{lemma}
\begin{proof}
First fix $x$ with $\lambda_0(x)=0$. The switch point is
$S_x=-(\alpha_0+\Omega)/\delta_0$, and ID3(b) gives
\[
\PP\{S_x\in\operatorname{int}(\mathcal Z_x)\}=1.
\]
\Needspace{6\baselineskip}
An atom of $S_x$ at any $s\in\operatorname{int}(\mathcal Z_x)$
would make the full-information probability
$z\mapsto\PP(\alpha_0+\delta_0z+\Omega>0)$ discontinuous at
$s$, contrary to ID1(e). Thus $S_x$ has no atoms. Since
$\Omega=-\alpha_0-\delta_0S_x$ and $\delta_0\ne0$, $G_0$ has no
atoms either.

Write $F_{S_x}(z)\coloneqq \PP(S_x\leq z)$. The full-information rule gives
\[
b(z,x;\theta_0)=
\begin{cases}
F_{S_x}(z),&\delta_0>0,\\
1-F_{S_x}(z),&\delta_0<0,
\end{cases}
\qquad z\in\R.
\]
Both expressions are continuous on $\R$. Since $S_x$ lies in
$\operatorname{int}(\mathcal Z_x)$ almost surely, $F_{S_x}$ is zero
to the left of $\mathcal Z_x$ and one to its right. Its continuity
and its limits zero and one as $z\to-\infty$ and $z\to+\infty$
therefore give
\[
\inf_{z\in\mathcal Z_x}F_{S_x}(z)=0,\qquad
\sup_{z\in\mathcal Z_x}F_{S_x}(z)=1.
\]
The same infimum and supremum hold for $b(\cdot,x;\theta_0)$
under either sign of $\delta_0$.

I now prove joint continuity. Take a sequence $(z_n,x_n)$ and a
limit $(z_\star,x_\star)$ as in the statement.
The taste-shock distribution is the same $G_0$ along the sequence by ID1(a).
Write
\[
\begin{aligned}
a_n&\coloneqq \alpha_0+\delta_0z_n,&
\lambda_n&\coloneqq \lambda_0(x_n),&
b_n&\coloneqq b(z_n,x_n;\theta_0),\\
a_\star&\coloneqq \alpha_0+\delta_0z_\star,&
\lambda_\star&\coloneqq \lambda_0(x_\star),&
b_\star&\coloneqq b(z_\star,x_\star;\theta_0).
\end{aligned}
\]
Then $a_n\to a_\star$ and $\lambda_n\to\lambda_\star$ by ID1(d).
The normalization in ID1(b) gives
$\E|\Omega|\leq(\E\Omega^2)^{1/2}=1$.

First suppose $\lambda_\star>0$. Then $\lambda_n>0$ eventually.
For $q\in[0,1]$ and $\lambda_0(x)>0$, define the choice-probability objective
\[
J(q;z,x)\coloneqq \lambda_0(x)\int
\log\left\{1-q+q e^{(\alpha_0+\delta_0z+\omega)/\lambda_0(x)}\right\}
\,dG_0(\omega).
\]
If $q_n\to q$, the integrand at $(q_n,z_n,x_n)$ converges pointwise
to the integrand at $(q,z_\star,x_\star)$. It also satisfies
\[
\left|\lambda_n\log\{1-q_n+q_n e^{(a_n+\omega)/\lambda_n}\}\right|
\leq |a_n|+|\omega|.
\]
The sequence $(a_n)$ is bounded, so the right-hand side has an
integrable bound independent of $n$. Dominated convergence gives
$J(q_n;z_n,x_n)\to J(q;z_\star,x_\star)$.

The variance normalization makes $G_0$ nondegenerate, so the binary
maximizer is unique by Lemma~\ref{l:shareobjectiveunique}.
The same continuity-of-maximizers argument proved in
Lemma~\ref{l:shareobjectivecontinuity}, using the compact set $[0,1]$
and the objective convergence just established, gives $b_n\to b_\star$.

\Needspace{7\baselineskip}
Now suppose $\lambda_\star=0$. The full-information rule in \citet[Theorem~1]{mm} gives
\[
b(z,x_\star;\theta_0)
=\PP\{\alpha_0+\delta_0z+\Omega>0\}.
\]
The absence of atoms established at the start of the proof implies
\[
\PP\{a_\star+\Omega=0\}=0.
\]
I first consider terms with $\lambda_n=0$. At these terms,
$b_n=\PP\{a_n+\Omega>0\}$. Since $a_n\to a_\star$ and the
limiting payoff tie has probability zero, bounded convergence gives
$b_n\to\PP\{a_\star+\Omega>0\}=b_\star$.

For terms with $\lambda_n>0$, fix any trial probability $q\in(0,1)$.
Outside the probability-zero limiting payoff tie, the logistic
integrand converges to $\ind\{a_\star+\omega>0\}$.
Since it lies in $[0,1]$, bounded convergence gives
\begin{equation}\label{e:smallcostresiduallimit}
r(\theta_0;(z_n,x_n),q)
=\int\sigma\!\left(\ell(q)+\frac{a_n+\omega}{\lambda_n}\right)dG_0(\omega)-q
\longrightarrow b_\star-q.
\end{equation}
The derivative identity proved in Lemma~\ref{l:completecurve} gives
\[
\partial_q J(q;z_n,x_n)
=\frac{\lambda_n}{q(1-q)}r(\theta_0;(z_n,x_n),q).
\]
For $q<b_\star$, the derivative is positive eventually and concavity
gives $b_n>q$; for $q>b_\star$, it gives $b_n<q$.
Letting these fixed trial probabilities approach $b_\star$ proves $b_n\to b_\star$.
If $b_\star=0$, only the upper bounds $b_n<q$ are needed;
if $b_\star=1$, only the lower bounds $b_n>q$ are needed.
Combining the terms with zero and positive costs proves
\eqref{e:jointchoicecontinuity} and completes the proof.
\end{proof}

Joint continuity also determines the conditional choice probabilities at every point of the observed support. The distribution of the data initially determines these probabilities only up to sets of covariate values with probability zero. If two continuous versions differed at a support point $(z,x)$, continuity would make them differ throughout a neighborhood of that point. Such a neighborhood has positive probability by the definition of support, contradicting almost-everywhere agreement. Thus the choice-probability function in Definition~\ref{d:identification} is uniquely determined on the observed support.

\subsection{Proof of Theorem~1}\label{a:identificationproofs}
\label{a:binaryproof}
\begin{proof}
Without loss of generality, assume $\delta_0>0$. Lemma~\ref{l:choicejointcontinuity} shows that, at every $x$ with $\lambda_0(x)=0$, the function $z\mapsto b(z,x;\theta_0)$ is continuous on $\R$ and its values on $\mathcal Z_x$ can be arbitrarily close to zero and arbitrarily close to one.

Fix $x\in\mathcal X$ and define from the observed probabilities
\[
p_-(x)\coloneqq \inf_{z\in\mathcal Z_x}\pi(z,x),\qquad
p_+(x)\coloneqq \sup_{z\in\mathcal Z_x}\pi(z,x).
\]
If $p_-(x)=0$ and $p_+(x)=1$, monotonicity gives $\pi(z,x)=0$ to the left of $\mathcal Z_x$ and $\pi(z,x)=1$ to its right. This argument applies whether information costs are zero or positive.

If $p_-(x)>0$ or $p_+(x)<1$, then $\lambda_0(x)>0$, because zero cost would imply $p_-(x)=0$ and $p_+(x)=1$. ID3(b) and continuity of $z\mapsto b(z,x;\theta_0)$ then supply a nonempty open interval $J_x\subseteq\operatorname{int}(\mathcal Z_x)$ such that
\[
0<\pi(z,x)<1\qquad(z\in J_x).
\]
Lemma~\ref{l:completecurve} then identifies $z\mapsto b(z,x;\theta_0)$ on $\R$ from its values on $J_x$. Thus, for every $x\in\mathcal X$, the function
\[
\pi(\cdot,x):\R\longrightarrow[0,1],\qquad
z\longmapsto b(z,x;\theta_0)
\]
is identified. I use this notation also for $z\notin\mathcal Z_x$. The function is continuous on $\R$: at positive cost this follows from Lemma~\ref{l:completecurve}, and at zero cost from Lemma~\ref{l:choicejointcontinuity}.

For $x$ with $\lambda_0(x)>0$, define
\[
I_x\coloneqq \{z\in\R:0<\pi(z,x)<1\},\qquad
\begin{aligned}
h_x:I_x&\longrightarrow\R,\\
z&\longmapsto\alpha_0+\delta_0z+\lambda_0(x)\ell(\pi(z,x)).
\end{aligned}
\]
For $z\in I_x$, the conditional choice probability in \eqref{e:conditionalRI} can be written as
\begin{align*}
p_{\theta_0}(\omega;z,x)
&=\frac{\pi(z,x)e^{(\alpha_0+\delta_0z+\omega)/\lambda_0(x)}}
{1-\pi(z,x)+\pi(z,x)e^{(\alpha_0+\delta_0z+\omega)/\lambda_0(x)}}\\
&=\sigma\!\left(\frac{h_x(z)+\omega}{\lambda_0(x)}\right).
\end{align*}
Let $\varepsilon$ be standard logistic and independent of $\Omega\sim G_0$. Since $\sigma$ is its CDF, averaging this equation over $G_0$ gives
\begin{equation}\label{e:proof-logit-threshold}
\begin{aligned}
\pi(z,x)
&=\int\PP\!\left\{\varepsilon\leq
\frac{h_x(z)+\omega}{\lambda_0(x)}\right\}\,dG_0(\omega)\\
&=\PP\{h_x(z)\geq-\Omega+\lambda_0(x)\varepsilon\}.
\end{aligned}
\end{equation}
Lemma~\ref{l:completecurve} makes $h_x:I_x\to\R$ an increasing bijection, so the switch point in Definition~\ref{d:switchpoint} is well defined:
\[
S_x\coloneqq 
\begin{cases}
h_x^{-1}(-\Omega+\lambda_0(x)\varepsilon),&\lambda_0(x)>0,\\[3pt]
-(\alpha_0+\Omega)/\delta_0,&\lambda_0(x)=0.
\end{cases}
\]
For positive cost and $z\in I_x$, \eqref{e:proof-logit-threshold} gives
\[
\PP(S_x\leq z)
=\PP\{-\Omega+\lambda_0(x)\varepsilon\leq h_x(z)\}
=\pi(z,x).
\]
Since $S_x\in I_x$ almost surely, this equality extends to $z\notin I_x$, with equality at any finite endpoints by continuity. At zero cost, continuity of $\pi(\cdot,x)$ and $\delta_0>0$ imply
\[
\PP(\Omega=-\alpha_0-\delta_0z)=0
\qquad(z\in\R).
\]
Consequently, the definition of $S_x$ gives
\[
\PP(S_x\leq z)
=\PP(\alpha_0+\delta_0z+\Omega\geq0)
=\pi(z,x)\qquad(z\in\R)
\]
at zero cost as well. Thus the identified function $\pi(\cdot,x)$ is the CDF of $S_x$ for every $x$.

Lemma~\ref{l:switchmoments} gives the finite, identified moments
\[
\mu(x)\coloneqq \E S_x,\qquad V(x)\coloneqq \Var(S_x),\qquad
C(x)\coloneqq \Cov(S_x,\ell(\pi(S_x,x)))>0.
\]
They are integrals with respect to the identified CDF $\pi(\cdot,x)$. The lemma also gives
\begin{equation}\label{e:proof-identification-moments}
\mu(x)=-\frac{\alpha_0}{\delta_0},\qquad\text{and}\qquad
\delta_0^2V(x)+2\delta_0\lambda_0(x)C(x)=1.
\end{equation}
In particular, $V(x)\leq1/\delta_0^2$ at every $x$. For each positive integer $n$, ID2 supplies $x_n\in\mathcal X$ with $\lambda_0(x_n)<1/n$. The variance bound in Lemma~\ref{l:switchmoments} gives
\[
0\leq1-\delta_0^2V(x_n)
\leq\frac{2\pi}{\sqrt3}\lambda_0(x_n)
<\frac{2\pi}{\sqrt3\,n}\longrightarrow0.
\]
Hence
\[
\sup_{x\in\mathcal X}V(x)=\frac1{\delta_0^2}.
\]
The supremum need not be attained. This equality and the common mean identify
\begin{equation}\label{e:proof-payoff-recovery}
\delta_0=\left(\sup_{x\in\mathcal X}V(x)\right)^{-1/2},
\qquad\text{and}\qquad \alpha_0=-\delta_0\mu(x).
\end{equation}
The sequence $x_n$ proves the equality but is not an input to these formulas. With $\delta_0$ identified, solving \eqref{e:proof-identification-moments} gives
\begin{equation}\label{e:proof-cost-recovery}
\lambda_0(x)=\frac{1-\delta_0^2V(x)}{2\delta_0C(x)},
\qquad x\in\mathcal X.
\end{equation}
The denominator is positive, so this formula also identifies which values of $x$ have zero cost.

I next define a map from the identified choice-probability functions $\pi(\cdot,x)$, $x\in\mathcal X$, to $G_0$. For every $x$ with $\lambda_0(x)>0$, both $h_x$ and the distribution of $S_x$ are now identified. The equality $h_x(S_x)=-\Omega+\lambda_0(x)\varepsilon$ and independence of $\Omega$ and $\varepsilon$ give
\begin{align*}
\int_{I_x}e^{-it h_x(s)}\,d\pi(s,x)
&=\E e^{-it h_x(S_x)}\\
&=\E\!\left[e^{it\Omega}e^{-it\lambda_0(x)\varepsilon}\right]\\
&=\varphi_{G_0}(t)\varphi_\varepsilon(-\lambda_0(x)t),
\end{align*}
where $\varphi_{G_0}(t)\coloneqq \int e^{it\omega}\,dG_0(\omega)$. The integral in the first line is with respect to the switch-point distribution with CDF $s\mapsto\pi(s,x)$. The standard logistic characteristic function $\varphi_\varepsilon$ is nonzero at every real argument. Specifically, \citet[Example B.13, Eq. (B.47)]{janson2013} gives
\[
\varphi_\varepsilon(t)=
\begin{cases}\pi t/\sinh(\pi t),&t\ne0,\\1,&t=0.\end{cases}
\]
For every $x$ with $\lambda_0(x)=0$, the definition of $S_x$ instead gives
\[
\Omega=-\alpha_0-\delta_0S_x,\qquad\text{so}\qquad
\varphi_{G_0}(t)=\int_{\R}e^{-it(\alpha_0+\delta_0s)}\,d\pi(s,x).
\]
In this case continuity also gives the CDF formula
\[
G_0(w)=1-\pi\!\left(\frac{-\alpha_0-w}{\delta_0},x\right)
\qquad(w\in\R).
\]

These equalities apply at every $x$ and do not require selecting one value of the cost shifter. To state the resulting map precisely, define from the identified functions
\begin{equation}\label{e:proof-taste-recovery}
\psi_x(t)\coloneqq 
\begin{cases}
\displaystyle
\frac{\int_{I_x}e^{-it h_x(s)}\,d\pi(s,x)}
{\varphi_\varepsilon(-\lambda_0(x)t)},&\lambda_0(x)>0,\\[10pt]
\displaystyle
\int_{\R}e^{-it(\alpha_0+\delta_0s)}\,d\pi(s,x),&\lambda_0(x)=0.
\end{cases}
\end{equation}
All terms in \eqref{e:proof-taste-recovery} are determined by $\pi(\cdot,x)$ and the parameter formulas \eqref{e:proof-payoff-recovery}--\eqref{e:proof-cost-recovery}. The preceding equalities give $\psi_x=\varphi_{G_0}$ for every $x\in\mathcal X$.
\Needspace{8\baselineskip}
Define\footnote{Let $\Pi$ denote the input probability family in Definition~\ref{d:identification}. Applying \eqref{e:proof-taste-recovery} at each $x\in\mathcal X$ defines the map $\Pi\mapsto\bigcup_x\{\psi_x\}$, where $\psi_x$ denotes the function $t\mapsto\psi_x(t)$. The preceding calculation shows that this union is $\{\varphi_{G_0}\}$. Composing this map with $\{f\}\mapsto f$ defines $\mathcal K$, with output denoted by $\psi$.}
\begin{equation}\label{e:proof-characteristic-map}
\mathcal K\!\left(
\left\{\{\pi(z,x)\}_{z\in\mathcal Z_x}\right\}_{x\in\mathcal X}
\right)\coloneqq \psi.
\end{equation}
Since $\psi=\varphi_{G_0}$, characteristic-function uniqueness identifies $G_0$ \citep[Theorem~26.2]{billingsley}.

Finally, let $m$ denote the common value of $\mu(x)$ and set $v_* \coloneqq  \sup_{x\in\mathcal X}V(x)$. Both are identified, and $0<v_*<\infty$. The calculations above give the following explicit map in Definition~\ref{d:identification}:
\begin{equation}\label{e:proof-explicit-map}
\begin{aligned}
\mathcal T\!\left(
\left\{\{\pi(z,x)\}_{z\in\mathcal Z_x}\right\}_{x\in\mathcal X}
\right)
&\coloneqq \left(
-\frac{m}{\sqrt{v_*}},\quad
\frac1{\sqrt{v_*}},\quad
\left[x\mapsto\frac{v_*-V(x)}{2\sqrt{v_*}\,C(x)}\right],\quad
G_0\right)\\
&=(\alpha_0,\delta_0,\lambda_0,G_0).
\end{aligned}
\end{equation}
Every function on the right is obtained from the input probabilities by the preceding constructions. This proves the theorem.
\end{proof}

\subsection{Proof of Theorem~2}\label{a:consistencyproof}
The estimator $\widehat\theta_M$ approximately minimizes $\theta\mapsto\widehat Q_M(\theta)$,
which uses estimated shares and can be discontinuous when
$\lambda(x)=0$. Fix $w\coloneqq (z,x)$ and the payoff parameters $(\alpha,\delta,G)$, and write $r_t(w,q)$ for the residual when the unit cost at $x$ is $t\geq0$. At an input probability of zero,
\[
r_t(w,0)=
\begin{cases}
0,&t>0,\\[2pt]
\PP_G\{\alpha+\delta z+\Omega>0\},&t=0.
\end{cases}
\]
If this full-information probability is positive, then
\begin{equation}\label{e:zero-cost-loss-jump}
\lim_{t\downarrow0}r_t(w,0)^2=0<r_0(w,0)^2.
\end{equation}
Thus the market's squared residual is zero at positive cost but positive at zero cost. Continuity of the true choice-probability function does not prevent this jump in a candidate's squared residual.

I compare $\widehat Q_M$ with the two criteria formed using
population choice probabilities $s_m\coloneqq s(W_m)$:
\begin{equation}\label{e:auxiliary}
\begin{gathered}
Q_M(\theta)\coloneqq \frac1M\sum_m r(\theta;W_m,s_m)^2,
\qquad\text{and}\qquad
Q_M^\circ(\theta)\coloneqq \frac1M\sum_m f_\theta(W_m),\\
\text{where}\qquad
f_\theta(w)\coloneqq \ind\{0<s(w)<1\}r(\theta;w,s(w))^2,
\end{gathered}
\end{equation}
and the population counterpart of $Q_M^\circ$ is
\[
Q^\circ(\theta)\coloneqq \E f_\theta(W).
\]
The criterion $Q_M^\circ(\theta)$ excludes markets with
$s_m\in\{0,1\}$ and satisfies $0\leq Q_M^\circ(\theta)\leq Q_M(\theta)$
for every $\theta\in\Theta$.
Following \citet[Section~3.1, proof of Theorem~3.1 and Remark 3.1(4)]{chen},
I show that $Q_M^\circ(\widehat\theta_M)\to_p0$ and that
$Q^\circ(\theta)$ is bounded away from zero when $d(\theta,\theta_0)$
is bounded away from zero. The following lemmas establish continuity and uniform
convergence for $Q_M^\circ$, the unique zero of $Q^\circ$, and
$Q_M(\theta_M^*)\to_p0$ for the sieve approximant.

E2(c) applies to the entire compact parameter space $\Theta$, including limits of parameter sequences, rather than only to the finite-dimensional sieves. \footnote{One way to obtain compactness is to use a uniformly bounded, equicontinuous set of cost functions and a weakly closed set of taste-shock distributions with a common moment bound of order greater than two, provided the restrictions in E2(b)--(c) are preserved under limits.} It allows $\lambda(x)>0$ for every $x\in\mathcal X$ at every $\theta\in\Theta$: these costs may approach zero as $x$ varies, as ID2 requires, without having a positive lower bound uniform in $x$.\footnote{If $\mathcal X$ is compact, continuity and strict positivity would imply a positive minimum, so this everywhere-positive special case would be incompatible with ID2.}

The sieve approximations must also satisfy E2(c). Discrete taste-shock distributions are allowed when that condition is satisfied through positive information costs or full conditional support. If a distribution used at zero cost must generate a real-analytic choice-probability function, finite Gaussian mixtures provide one possible approximation.

\begin{lemma}[Continuity and uniform convergence of the auxiliary criterion]\label{l:auxiliaryuniform}
Under the sampling conditions, ID3(a), and E2, $Q^\circ$ is continuous on
$\Theta$ and
\[
\sup_{\theta\in\Theta}|Q_M^\circ(\theta)-Q^\circ(\theta)|\pto0.
\]
\end{lemma}
\begin{proof}
Fix $\theta=(\alpha,\delta,\lambda,G)\in\Theta$ and a sequence
$\vartheta_n\to\theta$ in the metric $d$. On $\{0<s(W)<1\}$,
Lemma~\ref{l:residualregularity} gives continuity of
$\vartheta\mapsto r(\vartheta;W,s(W))$ at $\theta$, except possibly on
\[
\{\lambda(X)=0,\ G\{-\alpha-\delta Z\}>0\}.
\]
This event means that, at zero information cost, the candidate
distribution puts positive probability on equal payoffs for the two
alternatives. Since $\delta\ne0$, ID3(a) and
Lemma~\ref{l:residualregularity} give it probability zero.
On $\{s(W)\in\{0,1\}\}$, $f_\vartheta(W)=0$ for every $\vartheta$,
so continuity holds automatically. Therefore
\[
f_{\vartheta_n}(W)\to f_\theta(W)\quad\text{almost surely},
\qquad 0\leq f_\vartheta(W)\leq1.
\]
Bounded convergence yields $Q^\circ(\vartheta_n)\to Q^\circ(\theta)$. This proves continuity of $Q^\circ$ on the compact parameter space.

For each fixed $\theta$, the preceding argument gives almost-sure
continuity of $\vartheta\mapsto f_\vartheta(W)$ at $\theta$.
The bound $0\leq f_\vartheta(W)\leq1$ gives an integrable envelope,
the market characteristics are
i.i.d., and E2 makes $\Theta$ compact under $d$. Lemma~2.4 of
\citet{newey} therefore gives
\[
\sup_{\theta\in\Theta}|Q_M^\circ(\theta)-Q^\circ(\theta)|\pto0.
\]
The result uses continuity at each fixed parameter with probability one;
it does not require $\vartheta\mapsto r(\vartheta;w,s(w))^2$ to be
continuous at every $\theta\in\Theta$ for every $w$.
As usual, the supremum may be interpreted in outer
probability if needed, as in the qualification following that lemma.

\end{proof}

The next lemma shows that $Q^\circ(\theta)=0$ identifies $\theta_0$.
Its proof must account for the exclusion of markets with
$s(W)\in\{0,1\}$ and for candidates satisfying $\lambda(x)=0$.
\begin{lemma}[The auxiliary criterion has a unique zero]\label{l:auxiliaryunique}
Under ID1--ID3 and E2, $Q^\circ(\theta)=0$ for $\theta\in\Theta$ implies
$\theta=\theta_0$.
\end{lemma}
\begin{proof}
Suppose $Q^\circ(\theta)=0$. Since its integrand is nonnegative,
\[
r(\theta;W,s(W))=0
\quad\text{almost surely on }\{0<s(W)<1\}.
\]
To make the conditional statement precise, define
\[
 g_\theta(x)\coloneqq \E\!\left[
 \ind\{0<s(W)<1\}r(\theta;W,s(W))^2\mid X=x\right].
\]
The tower property gives $\E g_\theta(X)=Q^\circ(\theta)=0$.
Because $g_\theta(X)\geq0$, it follows that $g_\theta(X)=0$ almost surely.
Hence the set
$\mathcal B\coloneqq \{x\in\mathcal X:g_\theta(x)=0\}$ has probability one.
For each $x\in\mathcal B$, the conditional nonnegative integrand is zero
almost surely, giving the stated residual equality for almost every
$z$ with $0<s(z,x)<1$ under $Z\mid X=x$.

At a point with $0<s(z,x)<1$ and $\lambda(x)>0$,
$r(\theta;(z,x),s(z,x))=0$ is the first-order condition for the
strictly concave choice-probability objective. It implies $b(z,x;\theta)=s(z,x)$.
When $\lambda(x)=0$, the same conclusion follows from
\[
r(\theta;(z,x),s(z,x))
=\PP_G\{\alpha+\delta z+\Omega>0\}-s(z,x).
\]
For $x\in\mathcal B$, the equality $b(z,x;\theta)=s(z,x)$ therefore
holds at a dense set of points in every observed open interval with
$0<s(z,x)<1$: every nonempty open subinterval has positive
conditional probability under $Z\mid X=x$.

At zero information cost, global real analyticity of $z\mapsto b(z,x;\theta)$ implies
$0<b(z,x;\theta)<1$ for every $z\in\R$. To see this, first suppose $\delta>0$.
Then $b(z,x;\theta)=\PP_G\{\alpha+\delta z+\Omega>0\}$ is nondecreasing,
with limits zero as $z\to-\infty$ and one as $z\to+\infty$.
If it equals zero at some $z_0\in\R$, monotonicity makes it zero on $(-\infty,z_0)$.
The identity theorem \citep[Corollary 1.2.6]{krantzparks} then makes it zero on $\R$,
contradicting its limit at $+\infty$. If it equals one at $z_0$, the same argument
on $(z_0,\infty)$ contradicts its limit at $-\infty$. For $\delta<0$, the directions
of monotonicity and the two limits reverse, and the same argument applies.
E2(b) excludes $\delta=0$.

Replace $Z$ by $-Z$ if necessary so that $\delta_0>0$.
Under E2, $\mathcal Z_x=\R$ whenever $\lambda_0(x)=0$.
To see this, E2(a)
puts $\theta_0$ in $\Theta$. If $\lambda_0(x)=0$ and
$\mathcal Z_x\ne\R$, E2(c) would make $z\mapsto b(z,x;\theta_0)$ real analytic on $\R$.
The preceding argument would then give $0<b(z,x;\theta_0)<1$ for every $z\in\R$. By ID3(b), $-\Omega$
lies in $\alpha_0+\delta_0\operatorname{int}(\mathcal Z_x)$ almost surely, so
$b(z,x;\theta_0)$ equals zero or one at every $z$ outside
$\mathcal Z_x$. This is a
contradiction. Thus $\lambda_0(x)=0$ implies $\mathcal Z_x=\R$
under E2.

Fix $x\in\mathcal B$. There are observed values $z_1<z_2$ at which
$b(z_i,x;\theta)=s(z_i,x)$ for $i=1,2$, with
$0<s(z_1,x)<s(z_2,x)<1$. When $\lambda_0(x)>0$,
Lemma~\ref{l:completecurve} and ID3(b) supply an observed interval
where $s(\cdot,x)$ is strictly increasing. When $\lambda_0(x)=0$,
$\mathcal Z_x=\R$ by E2, and the continuous, nonconstant
full-information CDF also supplies values $z_1,z_2$ with these properties.
If $\delta<0$, $b(\cdot,x;\theta)$ would be nonincreasing and could
not equal $s$ at both $z_1$ and $z_2$. Thus $\delta>0$ as well.

Now take any open interval contained in
$\operatorname{int}(\mathcal Z_x)$ on which $0<s(z,x)<1$.
For any point $z$ in the interval, choose observed values
$z_n^-\uparrow z$ and $z_n^+\downarrow z$ at which
$b(\cdot,x;\theta)$ and $s(\cdot,x)$ agree.
Since $\delta>0$ and $\delta_0>0$, monotonicity gives
\[
s(z_n^-,x)=b(z_n^-,x;\theta)
\leq b(z,x;\theta)
\leq b(z_n^+,x;\theta)=s(z_n^+,x).
\]
The true choice-probability function $s(\cdot,x)$ is continuous, so both outer
terms tend to $s(z,x)$. Therefore $b(z,x;\theta)=s(z,x)$ at every
point of the interval. This argument also rules out a jump in $z\mapsto b(z,x;\theta)$ at such a point when $\lambda(x)=0$.

I next show $b(z,x;\theta)=s(z,x)$ for every $z\in\R$, at each
$x\in\mathcal B$. Define the interval
\[
I_x\coloneqq \{z\in\R:0<s(z,x)<1\}.
\]
The argument depends on the support $\mathcal Z_x$ and the candidate
cost $\lambda(x)$.

If $\mathcal Z_x=\R$, the preceding argument gives equality on all of
$I_x$. At a finite left boundary of $I_x$, $s(z,x)$ tends to zero;
at a finite right boundary it tends to one. Monotonicity and
$0\leq b(z,x;\theta)\leq1$ then force equality also where
$s(z,x)\in\{0,1\}$. This case includes every $x$ with
$\lambda_0(x)=0$, by E2.

If $\mathcal Z_x\ne\R$, E2 implies
$\lambda_0(x)>0$. ID3(b) then provides a nonempty open interval in
$\mathcal Z_x$ on which $0<s(z,x)<1$.
When $\lambda(x)>0$ also, Lemma~\ref{l:completecurve} gives
$b(z,x;\theta)=b(z,x;\theta_0)$ for every $z\in\R$.

It remains to consider $\mathcal Z_x\ne\R$ and $\lambda(x)=0$.
Because $\lambda(x)=0$ and $\mathcal Z_x\ne\R$, E2(c) makes
$z\mapsto b(z,x;\theta)$ real analytic on $\R$.
The preceding monotonicity argument then gives $0<b(z,x;\theta)<1$ for every $z\in\R$.
The functions $z\mapsto b(z,x;\theta)$ and $z\mapsto s(z,x)$ are analytic on $I_x$, and they agree on an observed open subinterval. The identity theorem \citep[Corollary 1.2.6]{krantzparks} gives agreement throughout $I_x$. If its left endpoint $a$ were finite, continuity would imply
\[
b(a,x;\theta)=\lim_{z\downarrow a}b(z,x;\theta_0)=0,
\]
contradicting $b(a,x;\theta)>0$. A finite right endpoint similarly contradicts $b(z,x;\theta)<1$ for every $z\in\R$. Thus $I_x=\R$, and $b(z,x;\theta)=b(z,x;\theta_0)$ for every $z\in\R$ in this case too.

The choice-probability functions agree, so I write their common value as
\[
F_x(z)\coloneqq b(z,x;\theta)=b(z,x;\theta_0)
\quad(z\in\R,\ x\in\mathcal B).
\]
The function $F_x$ is continuous. When $\lambda_0(x)>0$, continuity follows from Lemma~\ref{l:completecurve}; when $\lambda_0(x)=0$, it follows from ID1(e) and ID3(b). Hence the switch points constructed in Lemma~\ref{l:switchmoments} for $\theta$ and $\theta_0$ have the same continuous CDF $F_x$. In particular, their means, variances, and covariances
\[
\mu(x)\coloneqq \E S_x,\qquad V(x)\coloneqq \Var(S_x),\qquad
C(x)\coloneqq \Cov(S_x,\ell(F_x(S_x)))
\]
are common to both parameter values on $\mathcal B$. The lemma's zero-cost continuity condition therefore holds for both $\theta$ and $\theta_0$ wherever it is needed.

To apply the moment identities on $\mathcal B$, I need costs to approach zero on this set. Let $\lambda^\dagger\in\{\lambda,\lambda_0\}$. For every $\varepsilon>0$, ID2 gives $x_0\in\mathcal X$ with $\lambda^\dagger(x_0)<\varepsilon/2$. Continuity of $\lambda^\dagger$ gives a relative open neighborhood $U$ of $x_0$ on which $\lambda^\dagger(x)<\varepsilon$. Because $x_0\in\supp(X)$,
\[
\PP(X\in U)>0,\qquad
\PP(X\in U\cap\mathcal B)=\PP(X\in U)>0.
\]
Thus $\mathcal B$ contains a point $x$ with $\lambda^\dagger(x)<\varepsilon$. Applying this argument to $\lambda$ and $\lambda_0$ proves
\[
\inf_{x\in\mathcal B}\lambda(x)
=\inf_{x\in\mathcal B}\lambda_0(x)=0.
\]
Lemma~\ref{l:switchmoments}, applied separately at $\theta$ and $\theta_0$, now gives
\[
\sup_{x\in\mathcal B}V(x)=\frac1{\delta^2}=\frac1{\delta_0^2}.
\]
Since $\delta>0$ and $\delta_0>0$, this equality yields
$\delta=\delta_0$. The common mean gives $\alpha=\alpha_0$, and the
covariance identity gives
\[
\lambda(x)=\frac{1-\delta_0^2V(x)}{2\delta_0C(x)}
=\lambda_0(x)\qquad(x\in\mathcal B).
\]
If there is $x\in\mathcal B$ with $\lambda(x)=\lambda_0(x)>0$,
the deconvolution argument in Section~\ref{a:binaryproof} gives
$G=G_0$. If instead $\lambda(x)=\lambda_0(x)=0$ for every
$x\in\mathcal B$, the common switch-point distribution gives $G=G_0$ directly through
$\Omega\overset{d}{=}-\alpha_0-\delta_0S_x$.

Finally, $\lambda$ and $\lambda_0$ agree on the full-measure set $\mathcal B\subseteq\supp(X)$, hence everywhere on that support. If they differed at some point, continuity would give a neighborhood with a nonzero difference and positive support probability, contradicting their almost-everywhere equality. Consequently $\theta=\theta_0$, and $Q^\circ$ has a unique zero.
\end{proof}

To compare $\widehat\theta_M$ with the feasible sieve approximant
$\theta_M^*\in\Theta_M$, I also need $Q_M(\theta_M^*)\to_p0$.
This criterion includes markets with $s_m\in\{0,1\}$.
\begin{lemma}[Vanishing loss at the sieve approximant]\label{l:approximantloss}
Under the sampling conditions, ID1--ID3, and E3,
\[
\E[r(\theta_M^*;W,s(W))^2]\to0,
\qquad Q_M(\theta_M^*)\pto0.
\]
\end{lemma}
\begin{proof}
Write
\[
\theta_M^*=(\alpha_M^*,\delta_M^*,\lambda_M^*,G_M^*).
\]
On $\{0<s(W)<1\}$, E3 and Lemma~\ref{l:residualregularity} give
\[
r(\theta_M^*;W,s(W))\longrightarrow0.
\]
For $w\coloneqq (z,x)$ with $s(w)\in\{0,1\}$, consider $\lambda_0(x)>0$
and $\lambda_0(x)=0$ separately. If $\lambda_0(x)>0$,
$\lambda_M^*(x)>0$ eventually, and $r(\theta_M^*;w,s(w))=0$ by
\eqref{e:Psi}. If $\lambda_0(x)=0$, terms with
$\lambda_M^*(x)>0$ also have $r(\theta_M^*;w,s(w))=0$.
Along terms with $\lambda_M^*(x)=0$, use the indicator-convergence
argument in the proof of Lemma~\ref{l:residualregularity}. It uses
convergence of $(\alpha_M^*,\delta_M^*,G_M^*)$ and
$G_0\{-\alpha_0-\delta_0z\}=0$. This part of the argument compares
the indicators of $\alpha_M^*+\delta_M^*z+\Omega>0$ and
$\alpha_0+\delta_0z+\Omega>0$ and therefore also applies when
$s(w)\in\{0,1\}$. It gives
\[
\Psi(\theta_M^*;w,s(w))
=\PP_{G_M^*}\{\alpha_M^*+\delta_M^*z+\Omega>0\}
\longrightarrow
\PP\{\alpha_0+\delta_0z+\Omega>0\}=s(w).
\]
Subtracting $s(w)$ gives $r(\theta_M^*;w,s(w))\to0$ along these
terms as well. By ID1(c), ID3(a), and
Lemma~\ref{l:residualregularity},
\[
\PP\{\lambda_0(X)=0,\ G_0\{-\alpha_0-\delta_0Z\}>0\}=0.
\]
Thus the limits just proved hold almost surely over $W$.

Every population choice probability satisfies either $0<s(W)<1$ or
$s(W)\in\{0,1\}$. The preceding arguments therefore establish
\[
r(\theta_M^*;W,s(W))\longrightarrow0
\quad\text{almost surely}.
\]
Lemma~\ref{l:residualregularity} also gives
\[
|r(\theta_M^*;W,s(W))|\leq1.
\]
Since $\theta_M^*$ is deterministic and every $W_m$ has the same
distribution as $W$, bounded convergence yields
\[
\E Q_M(\theta_M^*)
=\E[r(\theta_M^*;W,s(W))^2]\longrightarrow0.
\]
Markov's inequality therefore gives $Q_M(\theta_M^*)\to_p0$.
This argument evaluates $Q_M$ only at $\theta_M^*$; it uses no
uniform convergence of $\theta\mapsto Q_M(\theta)$.
\end{proof}

\begin{proof}[Proof of Theorem~\ref{t:consistency}]
The preceding lemmas allow me to show that
$Q_M^\circ(\widehat\theta_M)\to_p0$, then use the unique minimum of
$Q^\circ$ to obtain $\widehat\theta_M\to_p\theta_0$ in the metric $d$.
The inequalities below connect the criterion
$\widehat Q_M$ minimized by the estimator to $Q_M^\circ$.
Let
\[
\zeta_M\coloneqq \sup_{\theta\in\Theta}|\widehat Q_M(\theta)-Q_M(\theta)|.
\]
\Needspace{8\baselineskip}
Lemma~\ref{l:sharesubstitution} gives $\zeta_M=o_p(1)$.
Since $\theta_M^*\in\Theta_M$, minimization up to the allowed error
$\epsilon_M=o_p(1)$ gives
\begin{equation}\label{e:comparisonchain}
\begin{aligned}
0\leq Q_M^\circ(\widehat\theta_M)
&\leq Q_M(\widehat\theta_M)\\
&\leq\widehat Q_M(\widehat\theta_M)+\zeta_M\\
&\leq\widehat Q_M(\theta_M^*)+\zeta_M+\epsilon_M\\
&\leq Q_M(\theta_M^*)+2\zeta_M+\epsilon_M=o_p(1).
\end{aligned}
\end{equation}
The two copies of $\zeta_M$ arise from replacing $\widehat s_m$ by
$s_m$ at $\widehat\theta_M$ and at $\theta_M^*$.
Uniform convergence of $Q_M^\circ$ to $Q^\circ$ then gives
\[
0\leq Q^\circ(\widehat\theta_M)
\leq Q_M^\circ(\widehat\theta_M)
+\sup_{\theta\in\Theta}|Q_M^\circ(\theta)-Q^\circ(\theta)|
\pto0.
\]

For $\varepsilon>0$, let $K_\varepsilon\coloneqq \{\theta\in\Theta:d(\theta,\theta_0)\geq\varepsilon\}$. If this set is empty, the desired probability is zero. Otherwise, the extreme value theorem
\citep[Theorem~27.4]{munkres}, applied to the compact set
$K_\varepsilon$, and the unique zero imply
\[
\kappa_\varepsilon\coloneqq \min_{\theta\in K_\varepsilon}Q^\circ(\theta)>0.
\]
Therefore
\[
\PP\{d(\widehat\theta_M,\theta_0)\geq\varepsilon\}
\leq\PP\{Q^\circ(\widehat\theta_M)\geq\kappa_\varepsilon\}
\longrightarrow0.
\]
This is $d(\widehat\theta_M,\theta_0)\to_p0$.
\end{proof}

\enlargethispage{2\baselineskip}
\Needspace{8\baselineskip}
The same comparison proves consistency with bounded positive weights. The weighted criterion is
\begin{equation}\label{e:weightedcriterion}
\widehat Q_M^w(\theta)\coloneqq \frac1M\sum_{m=1}^M
w(\theta;W_m,\widehat s_m)r(\theta;W_m,\widehat s_m)^2.
\end{equation}
Let $\widehat\theta_M^w$ approximately minimize \eqref{e:weightedcriterion}, with error $\epsilon_M^w=o_p(1)$, and suppose $0<c\leq w(\theta;w,q)\leq C<\infty$ uniformly over $(\theta,w,q)$. The weight bounds give
\[
c\widehat Q_M(\widehat\theta_M^w)
\leq\widehat Q_M^w(\widehat\theta_M^w)
\leq\widehat Q_M^w(\theta_M^*)+\epsilon_M^w
\leq C\widehat Q_M(\theta_M^*)+\epsilon_M^w.
\]
Share substitution and Lemma~\ref{l:approximantloss} imply
\[
\widehat Q_M(\widehat\theta_M^w)
\leq\frac Cc\{Q_M(\theta_M^*)+\zeta_M\}+\frac{\epsilon_M^w}{c}
\pto0.
\]
Hence
\[
Q_M^\circ(\widehat\theta_M^w)
\leq\widehat Q_M(\widehat\theta_M^w)+\zeta_M\to_p0.
\]
Lemma~\ref{l:auxiliaryuniform} then gives
$Q^\circ(\widehat\theta_M^w)\to_p0$.
For $\varepsilon>0$ with $K_\varepsilon\ne\varnothing$,
\[
\PP\{d(\widehat\theta_M^w,\theta_0)\geq\varepsilon\}
\leq\PP\{Q^\circ(\widehat\theta_M^w)\geq\kappa_\varepsilon\}
\longrightarrow0.
\]
If $K_\varepsilon=\varnothing$, the probability on the left is zero.

For the estimator with multiple alternatives in Section~\ref{s:multestimation}, bound the change in each squared residual when population probabilities are replaced by market shares. Summing these $J$ bounds controls the change in the criterion when $J$ is fixed. The auxiliary criterion includes only markets where every alternative has positive population probability. If its summand is almost surely continuous in the parameter, the uniform law of large numbers gives uniform convergence on the compact parameter space. Provided identification gives a unique zero, the preceding approximate-minimization argument proves consistency.

\subsection{Proof of Theorem~3}\label{a:inferenceproof}
I first show that the test accepts the deterministic sieve approximant
$\theta_M^*$ with probability approaching $1-\gamma$. I then use F5 to
show that the reported interval covers $\phi(\theta_0)$ with at least
this limiting probability.
The first part requires three comparisons. Trimming with estimated
probabilities must retain the same markets as trimming with population
probabilities, with probability approaching one. A centered and scaled
statistic that uses population probabilities for trimming and
standardization must have a normal limit. Finally, replacing those
population quantities by their estimates must have a negligible effect.

In the sample with $M$ markets, $N_M$ is the common within-market
sample size in F1. Write $w\coloneqq (z,x)$, $s(w)\coloneqq \pi(z,x)$, and $s_m\coloneqq s(W_m)$ for population
choice probabilities, and $\widehat s_m$ for the observed market shares. Fix the trimming
thresholds $\eta\in(0,1/2)$ and $a_*>0$. The sigma-field
\[
\mathcal F_M\coloneqq \sigma(W_1,\ldots,W_M)
\]
contains the observed market characteristics. Conditional on
$\mathcal F_M$, the population choice probabilities are fixed and the sampled choices
are independent across markets. In particular,
$N_M\widehat s_m\mid\mathcal F_M$ is binomial with success probability
$s_m$.

For $q\in(0,1)$, write $A_m(\theta,q)\coloneqq \partial_qr(\theta;W_m,q)$.
For $0<s(w)<1$, define
\[
A_M(w)\coloneqq \partial_qr(\theta_M^*;w,s(w)),\qquad
\text{and}\qquad
\mathcal R_M\coloneqq \{w:\eta\leq s(w)\leq1-\eta,\ |A_M(w)|\geq a_*\}.
\]
The set $\mathcal R_M$ contains the covariate values retained using population choice probabilities. Their
observed indices and number are
\[
\mathcal A_M\coloneqq \{m\leq M:W_m\in\mathcal R_M\},\qquad
k_M\coloneqq |\mathcal A_M|.
\]
I also write
\[
\widehat{\mathcal A}_M
\coloneqq \widehat{\mathcal A}_M(\theta_M^*),\qquad
\widehat k_M\coloneqq |\widehat{\mathcal A}_M|,
\]
for the retained indices and number obtained from estimated shares
when testing $\theta_M^*$. Thus $\mathcal A_M$ depends on market
characteristics and population choice probabilities, whereas
$\widehat{\mathcal A}_M$ also depends on share-estimation error.
For the proof only, put $d_M\coloneqq \max(k_M,1)$ so that the denominators
$\sqrt{d_M}$ and $\sqrt{2d_M}$ remain positive.
The implemented test continues to use
$\widehat k_M$, and accepts when that count is zero.

For each $m\in\mathcal A_M$, define
\begin{equation}\label{e:inference-proof-variables}
\xi_{m,M}\coloneqq \frac{\sqrt{N_M}(\widehat s_m-s_m)}{\sqrt{s_m(1-s_m)}},
\qquad
\beta_{m,M}\coloneqq \frac{\sqrt{N_M}r(\theta_M^*;W_m,s_m)}
{A_M(W_m)\sqrt{s_m(1-s_m)}}.
\end{equation}
The variable $\xi_{m,M}$ rescales the share error $\widehat s_m-s_m$.
Conditional on $\mathcal F_M$, it has mean zero and variance one.
The quantity $\beta_{m,M}$ rescales $r(\theta_M^*;W_m,s_m)$.
It is fixed conditional on $\mathcal F_M$.

I will compare the implemented statistic with
\[
T_M^\circ\coloneqq \sum_{m\in\mathcal A_M}(\xi_{m,M}+\beta_{m,M})^2.
\]
In $T_M^\circ$, the population choice probabilities $s_m$ determine both
$\mathcal A_M$ and the denominators in $\xi_{m,M}$ and $\beta_{m,M}$.
The sampled shares $\widehat s_m$ still enter through $\xi_{m,M}$.

\begin{lemma}[A positive fraction of markets is retained]\label{l:retainednumber}
Under E3, ID1(c), ID3(a), and F2, there is $p_*>0$ such that
\[
\PP\{k_M\geq p_*M/2\}\to1,
\qquad d_M\to_p\infty,\qquad M/d_M=O_p(1).
\]
\end{lemma}
\begin{proof}
Recall that
\[
A_M(w)\coloneqq \partial_qr(\theta_M^*;w,s(w)),\qquad
A_0(w)\coloneqq \partial_qr(\theta_0;w,s(w)).
\]
E3 and Lemma~\ref{l:residualregularity} imply
$A_M(W)\to A_0(W)$ almost surely whenever $0<s(W)<1$,
except possibly on
\[
\{\lambda_0(X)=0,\ G_0\{-\alpha_0-\delta_0Z\}>0\}.
\]
Lemma~\ref{l:residualregularity} gives this event probability zero:
ID1(c) makes $\delta_0\ne0$, and the continuous conditional distribution
of $Z$ in ID3(a) excludes the countably many values at which $G_0$
puts positive probability on a payoff tie.
Thus $A_M(W)\to A_0(W)$ almost surely on $\{0<s(W)<1\}$.

Define the event that both trimming inequalities hold strictly:
\[
\mathcal B\coloneqq \{\eta<s(W)<1-\eta,\ |A_0(W)|>a_*\}.
\]
Its probability $p_0\coloneqq \PP(\mathcal B)$ is positive by F2.
On the event $\mathcal B$, the population choice probability already satisfies the
population trimming inequalities, and derivative convergence gives
$|A_M(W)|\geq a_*$ for all sufficiently large $M$.
Consequently,
\[
\liminf_{M\to\infty}\ind\{W\in\mathcal R_M\}
\geq\ind\{\mathcal B\}
\quad\text{almost surely}.
\]
Fatou's inequality shows that the probability of satisfying the
population trimming conditions stays bounded away from zero:
\begin{equation}\label{e:retentionbridge}
\begin{aligned}
\liminf_{M\to\infty}\PP\{W\in\mathcal R_M\}
&=\liminf_{M\to\infty}p_M\\
&=\liminf_{M\to\infty}\E\ind\{W\in\mathcal R_M\}\\
&\geq\E\liminf_{M\to\infty}\ind\{W\in\mathcal R_M\}\\
&\geq p_0>0,
\end{aligned}
\end{equation}
where $p_M\coloneqq \PP(W\in\mathcal R_M)$.
Choose $p_*>0$ with $p_M\geq p_*$ for all sufficiently large $M$.

The indicators $\ind\{W_m\in\mathcal R_M\}$ are i.i.d.\ within each
sample because $\theta_M^*$ and $\mathcal R_M$ are deterministic and
$W_1,\ldots,W_M$ are i.i.d. Therefore
\[
\E(k_M/M)=p_M,\qquad
\Var(k_M/M)=\frac{p_M(1-p_M)}{M}\leq\frac1{4M}.
\]
For large $M$, Chebyshev's inequality implies
\[
\PP\{k_M<p_*M/2\}
\leq
\PP\{|k_M/M-p_M|>p_*/2\}
\leq\frac1{p_*^2M}\longrightarrow0.
\]
Hence the number of retained markets diverges with probability
approaching one. In particular,
\[
\PP(k_M=d_M)\to1,\qquad d_M\to_p\infty,\qquad M/d_M=O_p(1).\qedhere
\]
\end{proof}

For $m\in\mathcal A_M$, the population choice probabilities $s_m$ and derivatives
$A_M(W_m)$ satisfy $s_m\in[\eta,1-\eta]$ and $|A_M(W_m)|\geq a_*$.
F4 bounds $r(\theta_M^*;W_m,s_m)$ and hence its standardized value
$\beta_{m,M}$.
\begin{lemma}[The standardized sieve bias is negligible]\label{l:sievebias}
Let
\[
\nu_M\coloneqq \sum_{m\in\mathcal A_M}\beta_{m,M}^2,
\]
with empty sums and maxima set to zero. Under F4,
\[
\max_{m\in\mathcal A_M}|\beta_{m,M}|\to0,
\qquad\text{and}\qquad \nu_M=o(\sqrt M).
\]
Together with Lemma~\ref{l:retainednumber},
$\nu_M/\sqrt{d_M}\to_p0$.
\end{lemma}
\begin{proof}
On $\mathcal A_M$, the trimming rules give
\[
s_m(1-s_m)\geq\eta(1-\eta),\qquad |A_M(W_m)|\geq a_*.
\]
With $\rho_M=\sup_{w\in\mathcal R_M}|r(\theta_M^*;w,s(w))|$ from F4, these bounds give
\[
|\beta_{m,M}|
\leq\frac{\sqrt{N_M}\rho_M}{a_*\sqrt{\eta(1-\eta)}}.
\]
With constants
depending only on the fixed trimming bounds,
\begin{equation}\label{e:inference-proof-bias}
\max_{m\in\mathcal A_M}|\beta_{m,M}|
\leq C\sqrt{N_M}\rho_M\longrightarrow0,\qquad
\nu_M\coloneqq \sum_{m\in\mathcal A_M}\beta_{m,M}^2
\leq CMN_M\rho_M^2=o(\sqrt M).
\end{equation}
To verify both rates, F4 states
$M^{1/4}\sqrt{N_M}\rho_M\to0$, so
\[
\begin{gathered}
\sqrt{N_M}\rho_M
=M^{-1/4}\{M^{1/4}\sqrt{N_M}\rho_M\}\to0,\\
\text{and}\qquad
\frac{\nu_M}{\sqrt M}
\leq C\sqrt M N_M\rho_M^2
=C\{M^{1/4}\sqrt{N_M}\rho_M\}^2\to0.
\end{gathered}
\]
Combining the last bound with Lemma~\ref{l:retainednumber} also yields
$\nu_M/\sqrt{d_M}\to_p0$.
\end{proof}

Before studying the statistic's distribution, I show that estimating
shares does not change which markets enter it asymptotically.
\begin{lemma}[Agreement of population and sample trimming]\label{l:trimmingagreement}
Under F1 and F3,
\[
\PP\{\widehat{\mathcal A}_M\ne\mathcal A_M\}\to0.
\]
\end{lemma}
\begin{proof}
Fix a market $m$ and write $e_{m,M}\coloneqq \widehat s_m-s_m$ for its share error.
The sample and population trimming rules can disagree for two reasons.
First, $\widehat s_m$ and $s_m$ can lie on different sides of either
$\eta$ or $1-\eta$.
Such a crossing requires
\[
\min\{|s_m-\eta|,\ |s_m-(1-\eta)|\}\leq|e_{m,M}|.
\]
Second, if both $\widehat s_m$ and $s_m$ lie in $[\eta,1-\eta]$, the absolute residual
derivative can cross $a_*$. The uniform Lipschitz bound in
Lemma~\ref{l:residualregularity} gives
\[
\bigl||A_m(\theta_M^*,\widehat s_m)|-|A_M(W_m)|\bigr|
\leq K_\eta|e_{m,M}|.
\]
A derivative crossing therefore requires
\[
\bigl||A_M(W_m)|-a_*\bigr|\leq K_\eta|e_{m,M}|.
\]
For $w$ with $0<s(w)<1$, define its distance to the cutoffs by
\[
\Delta_M(w)\coloneqq \min\{|s(w)-\eta|,\ |s(w)-(1-\eta)|,
\bigl||A_M(w)|-a_*\bigr|\}.
\]
When $s(w)\in\{0,1\}$, set $\Delta_M(w)\coloneqq \eta$.
This is the quantity in F3. It is the smallest distance to any trimming
cutoff. For a fixed $K\geq\max(1,K_\eta)$, disagreement between the
sample and population trimming rules implies
\[
\Delta_M(W_m)\leq K|e_{m,M}|.
\]
When $s_m\in\{0,1\}$, the observed share equals $s_m$ almost surely and both trimming rules
discard the market, so the rules cannot disagree.

Applying \citet[Theorem~2]{hoeffding} to the independent Bernoulli
choices conditional on $\mathcal F_M$, and using a union bound for
the two tails, gives
\[
\PP\{|e_{m,M}|\geq u\mid\mathcal F_M\}
\leq2e^{-2N_Mu^2}.
\]
Substituting $u=\Delta_M(W_m)/K$ and then taking expectations shows
\[
\PP(\text{the trimming rules disagree for market }m)
\leq2\E e^{-cN_M\Delta_M(W)^2},
\qquad\text{where}\qquad c\coloneqq 2/K^2.
\]
I next bound this expectation using F3.
For every nonnegative number $u$,
\[
e^{-cN_Mu^2}=\int_u^\infty 2cN_Mt e^{-cN_Mt^2}\,dt.
\]
Taking expectations and interchanging the nonnegative integrals yields
\begin{align*}
\E e^{-cN_M\Delta_M(W)^2}
&=\int_0^\infty 2cN_Mt e^{-cN_Mt^2}
       \PP\{\Delta_M(W)\leq t\}\,dt\\
&\leq2cCN_M\int_0^{t_0}t^2e^{-cN_Mt^2}\,dt
       +e^{-cN_Mt_0^2}\\
&\leq\frac{2cC}{\sqrt{N_M}}
       \int_0^\infty u^2e^{-cu^2}\,du
       +e^{-cN_Mt_0^2}
=O(N_M^{-1/2}).
\end{align*}
The second line uses F3 for $t\leq t_0$ and the bound by one
for larger $t$; the third line uses $u=\sqrt{N_M}t$.
The constants in this bound may depend on the fixed Lipschitz constant $K_\eta$.

A union bound over the $M$ markets and F1 now give
\begin{equation}\label{e:stable-trimming-proof}
\PP\{\widehat{\mathcal A}_M\ne\mathcal A_M\}
\leq CM/\sqrt{N_M}\longrightarrow0.
\end{equation}
Thus, with probability approaching one, replacing the population
choice probabilities with observed shares changes none of the trimming decisions.
\end{proof}

The next result studies $T_M^\circ$, which uses $s_m$ in the
standardization and $\mathcal A_M$ to choose the markets.
Conditioning on $\mathcal F_M$ fixes $\mathcal A_M$ and
leaves independent share errors as its source of randomness.
\Needspace{8\baselineskip}
\begin{lemma}[Normal limit with population standardization]\label{l:populationstatlimit}
Under the sampling conditions, E3, ID1(c), ID3(a), and F1--F4,
\[
\frac{T_M^\circ-d_M}{\sqrt{2d_M}}\dto N(0,1).
\]
\end{lemma}
\begin{proof}
Because $\E(\xi_{m,M}\mid\mathcal F_M)=0$ and
$\E(\xi_{m,M}^2\mid\mathcal F_M)=1$,
\[
\E\{(\xi_{m,M}+\beta_{m,M})^2\mid\mathcal F_M\}
=1+\beta_{m,M}^2.
\]
Define the centered contribution
\[
X_{m,M}
\coloneqq (\xi_{m,M}+\beta_{m,M})^2-(1+\beta_{m,M}^2)
=(\xi_{m,M}^2-1)+2\beta_{m,M}\xi_{m,M}.
\]
Conditional on $\mathcal F_M$, these contributions are independent
and have mean zero. In particular,
\[
T_M^\circ-k_M-\nu_M=\sum_{m\in\mathcal A_M}X_{m,M}.
\]

To calculate their conditional variances, expand the last expression
for $X_{m,M}$:
\begin{align*}
\Var(X_{m,M}\mid\mathcal F_M)
&=\E(\xi_{m,M}^4\mid\mathcal F_M)-1
   +4\beta_{m,M}^2\E(\xi_{m,M}^2\mid\mathcal F_M)\\
&\qquad
   +4\beta_{m,M}\E(\xi_{m,M}^3-\xi_{m,M}\mid\mathcal F_M)\\
&=2+4\beta_{m,M}^2
   +4\beta_{m,M}\E(\xi_{m,M}^3\mid\mathcal F_M)
   +O(N_M^{-1}).
\end{align*}
The final equality uses the binomial formulas in
Lemma~\ref{l:binomialmoments}; their bounds are uniform because
$s_m\in[\eta,1-\eta]$ for every $m\in\mathcal A_M$.
The same lemma gives a uniformly bounded sixth moment of $\xi_{m,M}$.
Using $|a+b|^3\leq4(|a|^3+|b|^3)$, it follows that
\begin{align*}
\E(|X_{m,M}|^3\mid\mathcal F_M)
&\leq4\E(|\xi_{m,M}^2-1|^3\mid\mathcal F_M)
       +32|\beta_{m,M}|^3\E(|\xi_{m,M}|^3\mid\mathcal F_M)\\
&\leq C(1+|\beta_{m,M}|^3).
\end{align*}

Let the total conditional variance be
\[
v_M^2\coloneqq \sum_{m\in\mathcal A_M}\Var(X_{m,M}\mid\mathcal F_M).
\]
The bound in Lemma~\ref{l:sievebias} supplies the deterministic sequence
\[
\begin{gathered}
\bar\beta_M\coloneqq \frac{\sqrt{N_M}\rho_M}{a_*\sqrt{\eta(1-\eta)}}\to0,\\
\text{such that}\qquad
|\beta_{m,M}|\leq\bar\beta_M\quad(m\in\mathcal A_M).
\end{gathered}
\]
Lemma~\ref{l:binomialmoments} also gives
$|\E(\xi_{m,M}^3\mid\mathcal F_M)|\leq C/\sqrt{N_M}$ uniformly over
these markets. Substituting both bounds into the variance expansion yields
\[
\bigl|\Var(X_{m,M}\mid\mathcal F_M)-2\bigr|
\leq C\left(\bar\beta_M^2+\frac{\bar\beta_M}{\sqrt{N_M}}
+\frac1{N_M}\right).
\]
The right-hand side is deterministic, tends to zero, and is the same
for every retained market. Summing over $\mathcal A_M$ gives
\[
|v_M^2-2k_M|
\leq C k_M\left(\bar\beta_M^2+\frac{\bar\beta_M}{\sqrt{N_M}}
+\frac1{N_M}\right).
\]
On the event $k_M>0$, divide by $2k_M$.
Lemma~\ref{l:retainednumber} gives $\PP(k_M=d_M>0)\to1$, so
\[
\frac{v_M^2}{2d_M}\to_p1.
\]
For large $M$, on $\{k_M\geq p_*M/2\}$, $v_M^2\geq k_M$ and
$\sum_{m\in\mathcal A_M}\E(|X_{m,M}|^3\mid\mathcal F_M)\leq Ck_M$.
Therefore
\[
\frac{\sum_{m\in\mathcal A_M}\E(|X_{m,M}|^3\mid\mathcal F_M)}
     {v_M^3}
\leq\frac{Ck_M}{k_M^{3/2}}
=\frac C{\sqrt{k_M}}\longrightarrow0
\]
on that event. Its probability tends to one, so the ratio converges
to zero in probability, and $\PP(v_M>0)\to1$.
As in Lemma~\ref{l:conditionalclt}, the ratio is defined as zero
when $v_M=0$.

To apply that lemma, order the indices in $\mathcal A_M$ as
\[
\mathcal A_M=\{m_{M,1},\ldots,m_{M,k_M}\},
\qquad\text{where }m_{M,1}<\cdots<m_{M,k_M}.
\]
There are $k_M$ summands in this sample. Once the market
characteristics are fixed, the index list, its length, and the
conditional distributions of the contributions
$X_{m_{M,1},M},\ldots,X_{m_{M,k_M},M}$ are fixed.
The contributions themselves remain random because the choices are
sampled. Thus the conditional independence, centering, and
third-moment condition just verified are precisely the hypotheses
of Lemma~\ref{l:conditionalclt}.

\Needspace{6\baselineskip}
Applying Lemma~\ref{l:conditionalclt} to
$X_{m_{M,1},M},\ldots,X_{m_{M,k_M},M}$ gives
\[
\frac{T_M^\circ-k_M-\nu_M}{v_M}\dto N(0,1).
\]
Here the ratio on the left is defined as zero when $v_M=0$.
Since $v_M/\sqrt{2d_M}\to_p1$, Slutsky's theorem
\citep[Section~25, in particular Theorem~25.4 and Example 25.8]{billingsley} yields
\[
\frac{T_M^\circ-k_M-\nu_M}{\sqrt{2d_M}}\dto N(0,1).
\]
By Lemma~\ref{l:retainednumber}, $k_M=d_M$ with probability approaching one.
Lemma~\ref{l:sievebias} gives $\nu_M/\sqrt{d_M}\to_p0$. Hence
\begin{equation}\label{e:comparison-stat-clt}
\frac{T_M^\circ-d_M}{\sqrt{2d_M}}\dto N(0,1).
\end{equation}

\end{proof}

It remains to check the estimated denominators
$A_m(\theta_M^*,\widehat s_m)\sqrt{\widehat s_m(1-\widehat s_m)}$
and residuals $r(\theta_M^*;W_m,\widehat s_m)$ in $T_M(\theta_M^*)$. Their replacement error must be small
relative to $\sqrt{d_M}$, the scale in the preceding normal limit.
\begin{lemma}[Estimated standardization has a negligible effect]\label{l:estimatedstandardization}
Under the conditions of Lemma~\ref{l:populationstatlimit},
\[
\frac{T_M(\theta_M^*)-T_M^\circ}{\sqrt{d_M}}\pto0.
\]
\end{lemma}
\begin{proof}
For $m\in\mathcal A_M\cap\widehat{\mathcal A}_M$, define
the share error $e_{m,M}\coloneqq \widehat s_m-s_m$ and the standardized residual
\[
Z_{m,M}\coloneqq
\frac{\sqrt{N_M}r(\theta_M^*;W_m,\widehat s_m)}
     {A_m(\theta_M^*,\widehat s_m)
      \sqrt{\widehat s_m(1-\widehat s_m)}}.
\]
Its square is the market's contribution to the implemented statistic
\eqref{e:teststat}.
For this calculation I write the population and estimated
denominators as
\[
D_{m,M}\coloneqq A_M(W_m)\sqrt{s_m(1-s_m)},\qquad
\widehat D_{m,M}
\coloneqq A_m(\theta_M^*,\widehat s_m)
 \sqrt{\widehat s_m(1-\widehat s_m)}.
\]
Both have absolute value at least
$a_*\sqrt{\eta(1-\eta)}>0$.
Their signs need not be positive; only this lower bound on their
absolute values is used.

Both shares, and the interval between them, lie in
$[\eta,1-\eta]$. Taylor's theorem and the uniform second-derivative
bound in Lemma~\ref{l:residualregularity} give
\[
\begin{gathered}
r(\theta_M^*;W_m,\widehat s_m)
=r(\theta_M^*;W_m,s_m)+A_M(W_m)e_{m,M}
 +u_{m,M}/\sqrt{N_M},\\
\text{where}\qquad |u_{m,M}|\leq C\sqrt{N_M}e_{m,M}^2.
\end{gathered}
\]
By the definitions of $\beta_{m,M}$ and $\xi_{m,M}$, this is
\[
\sqrt{N_M}r(\theta_M^*;W_m,\widehat s_m)
=D_{m,M}(\beta_{m,M}+\xi_{m,M})+u_{m,M}.
\]
To bound the change in the denominator, define
\[
F_{m,M}:[\eta,1-\eta]\longrightarrow\R,\qquad
q\longmapsto F_{m,M}(q)
\coloneqq A_m(\theta_M^*,q)\sqrt{q(1-q)}.
\]
The product rule gives
\[
F_{m,M}'(q)
=\partial_q^2r(\theta_M^*;W_m,q)\sqrt{q(1-q)}
+A_m(\theta_M^*,q)\frac{1-2q}{2\sqrt{q(1-q)}}.
\]
For $q\in[\eta,1-\eta]$, $q(1-q)\geq\eta(1-\eta)>0$.
Lemma~\ref{l:residualregularity} bounds
$|\partial_qr(\theta_M^*;W_m,q)|$ and
$|\partial_q^2r(\theta_M^*;W_m,q)|$
by $K_\eta$, so $|F_{m,M}'(q)|\leq C_\eta$, uniformly over $m$, $M$,
and $q$ in that interval. The mean-value theorem therefore gives
\[
|\widehat D_{m,M}-D_{m,M}|
=|F_{m,M}(\widehat s_m)-F_{m,M}(s_m)|
\leq C_\eta|e_{m,M}|.
\]

Dividing the Taylor expression by $\widehat D_{m,M}$ and
subtracting $\beta_{m,M}+\xi_{m,M}$ gives the explicit remainder:
\[
R_{m,M}
\coloneqq \frac{(D_{m,M}-\widehat D_{m,M})
       (\beta_{m,M}+\xi_{m,M})+u_{m,M}}
       {\widehat D_{m,M}}.
\]
Put $d_*\coloneqq a_*\sqrt{\eta(1-\eta)}>0$. Since
$|\widehat D_{m,M}|\geq d_*$, the preceding bounds imply
\[
\begin{aligned}
|R_{m,M}|
&\leq d_*^{-1}\left\{
C_\eta|e_{m,M}|(|\beta_{m,M}|+|\xi_{m,M}|)
+C\sqrt{N_M}e_{m,M}^2\right\},\\
|\xi_{m,M}|
&=\frac{\sqrt{N_M}|e_{m,M}|}{\sqrt{s_m(1-s_m)}}
\leq\frac{\sqrt{N_M}|e_{m,M}|}{\sqrt{\eta(1-\eta)}}.
\end{aligned}
\]
Substituting the second inequality into the first shows that the term
$|e_{m,M}|\,|\xi_{m,M}|$ is bounded by a constant times
$\sqrt{N_M}e_{m,M}^2$. Hence
\begin{equation}\label{e:feasible-root-remainder}
Z_{m,M}=\beta_{m,M}+\xi_{m,M}+R_{m,M},\qquad\text{with}\qquad
|R_{m,M}|\leq C\{\sqrt{N_M}e_{m,M}^2
                    +|\beta_{m,M}|\,|e_{m,M}|\}.
\end{equation}
For $m\in\mathcal A_M\setminus\widehat{\mathcal A}_M$, define
$R_{m,M}\coloneqq 0$. The bound on $|R_{m,M}|$ then holds for every
$m\in\mathcal A_M$.
Using the second and fourth share-error moments from
Lemma~\ref{l:binomialmoments},
\begin{align*}
\E(R_{m,M}^2\mid\mathcal F_M)
&\leq C\{N_M\E(e_{m,M}^4\mid\mathcal F_M)
          +\beta_{m,M}^2\E(e_{m,M}^2\mid\mathcal F_M)\}\\
&\leq C(1+\beta_{m,M}^2)/N_M.
\end{align*}

Let
\[
B_M\coloneqq \sum_{m\in\mathcal A_M}R_{m,M}^2,\qquad
V_M\coloneqq 2\sqrt{T_M^\circ B_M}+B_M.
\]
On the event $\widehat{\mathcal A}_M=\mathcal A_M$,
the implemented statistic sums
$(\beta_{m,M}+\xi_{m,M}+R_{m,M})^2$ over exactly the
same indices as $T_M^\circ$. Expanding those squares and applying
Cauchy--Schwarz gives
\[
|T_M(\theta_M^*)-T_M^\circ|
\leq2\sqrt{T_M^\circ B_M}+B_M=V_M.
\]
The conditional expectations satisfy
\[
\E(T_M^\circ\mid\mathcal F_M)=k_M+\nu_M,\qquad
\E(B_M\mid\mathcal F_M)\leq C(k_M+\nu_M)/N_M.
\]
Another application of Cauchy--Schwarz, this time to the conditional
expectation, gives
\begin{align*}
\E(V_M\mid\mathcal F_M)
&\leq2\sqrt{\E(T_M^\circ\mid\mathcal F_M)
                 \E(B_M\mid\mathcal F_M)}
        +\E(B_M\mid\mathcal F_M)\\
&\leq C(k_M+\nu_M)/\sqrt{N_M}.
\end{align*}
Here $k_M\leq M$, and the deterministic bound on $\nu_M$ in Lemma~\ref{l:sievebias}
is $o(\sqrt M)$. Taking expectations therefore yields
\[
\E V_M\leq C\{M+o(\sqrt M)\}/\sqrt{N_M}.
\]

This error must be small compared with the statistic's scale
$\sqrt{d_M}$, rather than small in absolute value.
For any $\varepsilon>0$, bound separately the probability that the
trimming rules disagree and the probability that too few markets
satisfy the population trimming conditions:
\begin{align*}
\PP\left\{
\frac{|T_M(\theta_M^*)-T_M^\circ|}{\sqrt{d_M}}>\varepsilon
\right\}
&\leq\PP(\widehat{\mathcal A}_M\ne\mathcal A_M)
       +\PP(k_M<p_*M/2)\\
&\quad+\frac{\E V_M}{\varepsilon\sqrt{p_*M/2}}.
\end{align*}
The first two terms vanish by Lemmas~\ref{l:retainednumber} and \ref{l:trimmingagreement}.
The third is $O(\sqrt{M/N_M})+o(N_M^{-1/2})$ and vanishes
because F1 implies $M/N_M\to0$.
Consequently,
\[
\frac{T_M(\theta_M^*)-T_M^\circ}{\sqrt{d_M}}\to_p0.
\qedhere
\]

\end{proof}

\begin{proof}[Proof of Theorem~\ref{t:coverage}]
Lemmas~\ref{l:populationstatlimit} and
\ref{l:estimatedstandardization} give
\[
\frac{T_M(\theta_M^*)-d_M}{\sqrt{2d_M}}\dto N(0,1).
\]
I first use this limit to obtain
$\PP\{\theta_M^*\in\mathcal C^s_{M,\gamma}\}$.
F5 then shows that, conditional on this acceptance,
$[L_M,U_M]$ contains $\phi(\theta_0)$ with
probability approaching one.
For deterministic integers $k\to\infty$, let $Z_1,\ldots,Z_k$ be
i.i.d.\ $N(0,1)$ and set $C_k\coloneqq \sum_{j=1}^kZ_j^2\sim\chi_k^2$. The squared variables $Z_j^2$ have mean one and variance two, so the ordinary central limit theorem
\citep[Theorem~27.1]{billingsley} gives
$(C_k-k)/\sqrt{2k}\Rightarrow N(0,1)$. Since the standard normal
CDF is continuous and strictly increasing, the corresponding
quantiles satisfy
\[
\frac{\chi^2_{1-\gamma,k}-k}{\sqrt{2k}}
\longrightarrow z_{1-\gamma},
\]
where $z_{1-\gamma}$ is the standard normal $(1-\gamma)$ quantile.
To justify using the random integer $d_M$, write
$c_k\coloneqq (\chi^2_{1-\gamma,k}-k)/\sqrt{2k}$.
For every $\varepsilon>0$, choose $K$ such that
$|c_k-z_{1-\gamma}|<\varepsilon$ for all $k\geq K$.
Then
\[
\PP\{|c_{d_M}-z_{1-\gamma}|\geq\varepsilon\}
\leq\PP(d_M<K)\longrightarrow0,
\]
because $d_M\to_p\infty$. This proves
\[
\frac{\chi^2_{1-\gamma,d_M}-d_M}{\sqrt{2d_M}}
\to_p z_{1-\gamma}.
\]

Define the acceptance event
$E_M\coloneqq \{\theta_M^*\in\mathcal C^s_{M,\gamma}\}$.
On $\{\widehat{\mathcal A}_M=\mathcal A_M,\ k_M=d_M\}$,
the test's actual count $\widehat k_M$ equals $d_M$, so acceptance
is equivalently
\[
\frac{T_M(\theta_M^*)-d_M}{\sqrt{2d_M}}
\leq
\frac{\chi^2_{1-\gamma,d_M}-d_M}{\sqrt{2d_M}}.
\]
The left side converges in distribution to a standard normal and
the right side converges in probability to $z_{1-\gamma}$.
Slutsky's theorem therefore gives
\[
\frac{T_M(\theta_M^*)-\chi^2_{1-\gamma,d_M}}{\sqrt{2d_M}}
\dto N(-z_{1-\gamma},1).
\]
Since this limiting distribution assigns probability zero to zero,
the probability that the displayed difference is nonpositive tends
to $1-\gamma$.
The event excluded in making the comparison has probability
tending to zero. Hence
\begin{equation}\label{e:approximant-acceptance-proof}
\PP(E_M)=\PP\{\theta_M^*\in\mathcal C^s_{M,\gamma}\}
\longrightarrow1-\gamma.
\end{equation}
The test still accepts $\theta_M^*$ when
$\widehat{\mathcal A}_M=\varnothing$.
Lemmas~\ref{l:retainednumber} and \ref{l:trimmingagreement} show that
the probability of this event tends to zero.

Write $\phi_M^*\coloneqq \phi(\theta_M^*)$ and define the interval-inclusion
event
\[
\mathcal H_M\coloneqq 
\{[\phi_M^*-\varepsilon_M,\phi_M^*+\varepsilon_M]
      \subseteq[L_M,U_M]\}.
\]
F5 gives
\[
|\phi_M^*-\phi(\theta_0)|=o(\varepsilon_M),\qquad
\PP(\mathcal H_M\mid E_M)\to1.
\]
\Needspace{8\baselineskip}
The bound $|\phi_M^*-\phi(\theta_0)|=o(\varepsilon_M)$ means that,
for all sufficiently large $M$, $\phi(\theta_0)$ lies in
$[\phi_M^*-\varepsilon_M,\phi_M^*+\varepsilon_M]$.
On $E_M\cap\mathcal H_M$, that interval is contained in the
interval $[L_M,U_M]$. Thus, for all sufficiently large $M$,
\begin{align*}
\PP\{\phi(\theta_0)\in[L_M,U_M]\}
&\geq\PP(E_M\cap\mathcal H_M)\\
&=\PP(E_M)\PP(\mathcal H_M\mid E_M).
\end{align*}
Using \eqref{e:approximant-acceptance-proof} and F5 proves
\[
\liminf_{M\to\infty}
\PP\{\phi(\theta_0)\in[L_M,U_M]\}\geq1-\gamma.
\]
\Needspace{7\baselineskip}
If $\mathcal C^s_{M,\gamma}=\varnothing$, the interval $[L_M,U_M]$
is taken to be empty. On $E_M$, $\theta_M^*\in\mathcal C^s_{M,\gamma}$,
so this accepted set is nonempty.
\end{proof}

\subsection{Proof of Theorem~4}\label{a:multipleproof}
I use the model and notation in Appendix~\ref{a:multipleidentification}.
The payoff of alternative $0$ is normalized to zero.
Write $\alpha\coloneqq (\alpha_1,\ldots,\alpha_J)'$ and
$\delta\coloneqq (\delta_1,\ldots,\delta_J)'$. The two parameter values
$\theta\coloneqq (\alpha,\delta,\lambda,G)$ and
$\widetilde\theta\coloneqq (\widetilde\alpha,\widetilde\delta,\widetilde\lambda,\widetilde G)$
compared in the proof must each satisfy A1--A6.
If $\lambda(x)=0$ or $\widetilde\lambda(x)=0$, A6(a) requires
$\mathcal Z_x=\R^J$.
Write $p_j(z_j;x)$ for the choice probability of alternative $j$
in the binary RI problem on $\{0,j\}$, using the same marginal distribution
of $\Omega_j$ under $G$ and unit cost $\lambda(x)$.
This probability describes choice when only alternatives $0$ and $j$
are available. I aim to recover $p_j(z_j;x)$ for every $z_j\in\R$,
including values outside $\supp(Z_j\mid X=x)$.
When $\lambda(x)>0$, $p_j(z_j;x)$ maximizes
\[
\phi_j(p;z_j,x)\coloneqq \lambda(x)\E_G
\log\left\{1-p+p e^{(\alpha_j+\delta_jz_j+\Omega_j)/\lambda(x)}\right\}
\quad(p\in[0,1]).
\]
When $\lambda(x)=0$, $p_j(z_j;x)=\PP_G\{\alpha_j+\delta_jz_j+\Omega_j>0\}$.
Assumption~A6(b) gives probability zero to equality of the two payoffs.

The following lemmas establish properties of the choice probabilities,
recover the functions $p_j(\cdot;x)$, and identify the joint taste-shock distribution.
The last lemma treats the case $\lambda(x)=0$ for every $x\in\mathcal X$.

\begin{lemma}[Choice probabilities and full consideration]\label{l:multshares}
Suppose A2 holds. Fix $x\in\mathcal X_+$ and write $\lambda\coloneqq \lambda(x)$.
For $a\in\R^J$, let $\bar b(a;x)=(\bar b^0(a;x),\ldots,\bar b^J(a;x))'$
maximize $q\mapsto\Phi(q;a,x)$ over $\Delta_J$, as in
\eqref{e:appendix-multshare}. This maximizer is unique, and the map
$a\mapsto\bar b(a;x)$ is continuous. In covariate coordinates,
$b(z;x)=\bar b(a(z);x)$, where $a(z)\coloneqq (\alpha_j+\delta_jz_j)_{j=1}^J$.
The full-consideration region
\[
\mathcal W_x\coloneqq \{a\in\R^J:\bar b^j(a;x)>0\text{ for every }j=0,\ldots,J\}
\]
is connected and open. Each function $a\mapsto\bar b^j(a;x)$ is real
analytic on $\mathcal W_x$. The map
\[
\begin{aligned}
c:\mathcal W_x&\longrightarrow\R^J,\\
a&\longmapsto c(a)\coloneqq (c_1(a),\ldots,c_J(a))',\\
c_j(a)&\coloneqq -a_j+\lambda\log\frac{\bar b^0(a;x)}{\bar b^j(a;x)}
\quad(j=1,\ldots,J)
\end{aligned}
\]
is a real-analytic diffeomorphism from $\mathcal W_x$ onto $\R^J$.
For every $j=1,\ldots,J$ and $a\in\mathcal W_x$,
$\partial\bar b^j(a;x)/\partial a_j>0$.
\end{lemma}
\Needspace{8\baselineskip}
\begin{proof}
The finite second moments of $G$ stated in Appendix~\ref{a:multipleidentification} imply
\[
\E_G\|\Omega\|_1
=\sum_{j=1}^J\E_G|\Omega_j|
\leq\sum_{j=1}^J(\E_G\Omega_j^2)^{1/2}<\infty.
\]
Lemma~\ref{l:shareobjectivecontinuity} therefore makes
$(q,a)\mapsto\Phi(q;a,x)$ finite and continuous on
$\Delta_J\times\R^J$. A2 and Lemma~\ref{l:shareobjectiveunique}
give a unique maximizer $\bar b(a;x)$ and continuity of
$a\mapsto\bar b(a;x)$, including values of $a$ at which the maximizing
choice-probability vector lies on the boundary of $\Delta_J$.

Lemmas~\ref{l:sharepayoffbijection} and \ref{l:sharevaluegradients}
show that the map
\[
B_x:\mathcal W_x\longrightarrow\mathcal D_J^\circ,\qquad
a\longmapsto(\bar b^1(a;x),\ldots,\bar b^J(a;x))'
\]
is a bijection with a real-analytic inverse $a^*(\cdot;x)$.
In particular, $\bar b^j(a;x)$ is real analytic on $\mathcal W_x$.
The set $\mathcal D_J^\circ$, consisting of vectors $(q_1,\ldots,q_J)'$ with $q_j>0$ for every $j$ and $\sum_jq_j<1$, is convex and hence connected. Its continuous image under
$a^*(\cdot;x)$ is $\mathcal W_x$, so $\mathcal W_x$ is connected
\citep[Theorem~23.5]{munkres}. It is open by continuity of $\bar b$.

I next verify the asserted inverse for the threshold map $c$.
For any threshold vector $c\in\R^J$, define
\[
p_0(\omega;c)\coloneqq \frac1{1+\sum_j e^{(\omega_j-c_j)/\lambda}},\qquad
p_j(\omega;c)\coloneqq e^{(\omega_j-c_j)/\lambda}p_0(\omega;c),
\qquad q_j(c)\coloneqq \E_G p_j(\Omega;c).
\]
Write $q(c)\coloneqq (q_0(c),\ldots,q_J(c))'$; every component $q_j(c)$ is positive.
Define the map $a_x:\R^J\to\R^J$, $c\mapsto a_x(c)$, by
\begin{equation}\label{e:mult-index-from-threshold}
a_{x,j}(c)\coloneqq -c_j+\lambda\log\{q_0(c)/q_j(c)\}.
\end{equation}
At $a_x(c)$, the expression inside the logarithm in $\Phi$ satisfies
\[
q_0(c)+\sum_{j=1}^J q_j(c)e^{(a_{x,j}(c)+\Omega_j)/\lambda}
=q_0(c)\left\{1+\sum_{j=1}^J e^{(\Omega_j-c_j)/\lambda}\right\}.
\]
The conditional probabilities in the share first-order conditions
are therefore precisely $p_j(\Omega;c)$.
Their expectations equal $q_j(c)$ by definition. As shown in Lemma~\ref{l:sharepayoffbijection}, after substituting $q_0=1-\sum_{k=1}^Jq_k$ in $\Phi$, these equalities make its derivatives with respect to $q_j$, $j=1,\ldots,J$, zero. Strict concavity then gives
\[
\bar b(a_x(c);x)=q(c),\qquad\text{so}\qquad a_x(c)\in\mathcal W_x.
\]
Substitution into the definition of $c(a)$ now gives
\[
c_j(a_x(c))=-a_{x,j}(c)+\lambda\log\{q_0(c)/q_j(c)\}=c_j.
\]
Thus $c(a_x(c))=c$ for every $c\in\R^J$.

For the other inverse identity, fix $a\in\mathcal W_x$ and write
$\bar q\coloneqq \bar b(a;x)$. The definition of $c_j(a)$ implies
\[
e^{(\Omega_j-c_j(a))/\lambda}
=\frac{\bar q_j}{\bar q_0}e^{(a_j+\Omega_j)/\lambda}.
\]
Hence $p_j(\Omega;c(a))$ is the conditional probability associated
with the interior optimal vector $\bar q$ at $a$.
Lemma~\ref{l:sharepayoffbijection} gives
$\E_Gp_j(\Omega;c(a))=\bar q_j$ for every $j=0,\ldots,J$.
Consequently $q(c(a))=\bar q$, and
\[
a_{x,j}(c(a))=-c_j(a)+\lambda\log(\bar q_0/\bar q_j)=a_j.
\]
This proves $a_x(c(a))=a$ and establishes the bijection between
$\mathcal W_x$ and $\R^J$.

The functions $q_j(c)$ are real analytic by
\citet[Theorem~1(a)]{iariawang2025}. Thus $c\mapsto a_x(c)$ is real
analytic by \eqref{e:mult-index-from-threshold}. Its inverse
$a\mapsto c(a)$ is real analytic because the shares are analytic
and strictly positive on $\mathcal W_x$. These explicit inverse maps
prove the asserted analytic diffeomorphism.
For completeness, with $\mathbf q(c)\coloneqq (q_1(c),\ldots,q_J(c))'$,
differentiating $B_x(a_x(c))=\mathbf q(c)$ gives
\begin{equation}\label{e:mult-threshold-jacobian}
D_ca_x(c)=\{D_aB_x(a_x(c))\}^{-1}D_c\mathbf q(c).
\end{equation}
The first factor is nonsingular by Lemma~\ref{l:sharevaluegradients}.
To verify the second, put
\[
\mathbf p(\Omega;c)\coloneqq (p_1(\Omega;c),\ldots,p_J(\Omega;c))'.
\]
Differentiating the probabilities under the expectation gives
\[
D_c\mathbf q(c)=-\frac1\lambda\E_G\{
\operatorname{diag}(\mathbf p(\Omega;c))-
\mathbf p(\Omega;c)\mathbf p(\Omega;c)'\}.
\]
For any nonzero $u=(u_1,\ldots,u_J)'\in\R^J$,
$-u'D_c\mathbf q(c)u$ is $1/\lambda$ times the expected variance
of $0,u_1,\ldots,u_J$ under the positive probabilities
$p_0(\Omega;c),\ldots,p_J(\Omega;c)$.
That variance is positive by the argument in
Lemma~\ref{l:sharepayoffbijection}. Thus $D_c\mathbf q(c)$ is
negative definite and nonsingular.

Finally, Lemma~\ref{l:sharevaluegradients} gives a positive-definite
share Jacobian $D_aB_x(a)$. For the coordinate vector $e_j\in\R^J$,
\[
\frac{\partial\bar b^j(a;x)}{\partial a_j}
=e_j'D_aB_x(a)e_j>0.
\]
Thus increasing $a_j$ while holding the other payoff indices fixed
locally increases $\bar b^j(a;x)$ at every $a\in\mathcal W_x$.
\end{proof}

\Needspace{8\baselineskip}
The next lemma recovers the probability $p_j(z_j;x)$ for choice from
$\{0,j\}$ using probabilities from the model with all alternatives
available. At positive information cost, fix $u\in\R$ with
$p\coloneqq p_j(u;x)\in(0,1)$. I construct covariate vectors $z_n$
with $z_{n,j}=u$ for every $n$ and $b^i(z_n;x)>0$ for every
$i=0,\ldots,J$, such that $b^j(z_n;x)\to p$,
$b^0(z_n;x)\to1-p$, and the probabilities of all other alternatives
tend to zero. The comparison uses equality of
$b(z_n;x)$ and $\widetilde b(z_n;x)$ only at points where every
alternative has positive probability.
\begin{lemma}[Choice probabilities when only alternatives 0 and $j$ are available]\label{l:identifiedbinarymargins}
Two parameter values satisfying A1--A6 and generating the same
$\mathcal B_{\mathrm{obs}}$ have the same functions $p_j(\cdot;x)$:
\[
p_j(z_j;x)=\widetilde p_j(z_j;x)
\quad\text{for every }z_j\in\R,\ j=1,\ldots,J,\ x\in\mathcal X.
\]
\end{lemma}
\begin{proof}
I first show that $\delta_j$ and $\widetilde\delta_j$ have the same
sign for each $j=1,\ldots,J$.
Suppose first that at least one parameter value has positive cost at
some $x\in\mathcal X$. Interchanging the two parameter values if
necessary, choose $x^\circ$ with $\lambda(x^\circ)>0$.
Assumption~A3 supplies a covariate vector
\[
z^\circ\in\operatorname{int}(\mathcal Z_{x^\circ}),\qquad
\text{such that}\qquad
b^i(z^\circ;x^\circ)>0\quad(i=0,\ldots,J).
\]
Since $z^\circ$ lies inside the conditional support, the choice
probabilities are identified in a neighborhood of $z^\circ$.
Continuity keeps every probability positive in a sufficiently small
such neighborhood. Holding $z_k$ fixed for $k\ne j$,
Lemma~\ref{l:multshares} and the chain rule give
\[
\frac{\partial b^j(z;x^\circ)}{\partial z_j}
=\delta_j\frac{\partial\bar b^j(a(z);x^\circ)}{\partial a_j},
\qquad
\frac{\partial\bar b^j(a(z);x^\circ)}{\partial a_j}>0.
\]
Thus $b^j(z;x^\circ)$ strictly increases with $z_j$ when $\delta_j>0$
and strictly decreases when $\delta_j<0$ in this neighborhood.

The other parameter value gives the same observed probabilities there.
If $\widetilde\lambda(x^\circ)>0$, the same derivative calculation
applies to $\widetilde b^j$ and $\widetilde\delta_j$.
If $\widetilde\lambda(x^\circ)=0$, increasing $z_j$ changes only
alternative $j$'s payoff, by $\widetilde\delta_j$ times the increase
in $z_j$. Hence $\widetilde b^j$ is nondecreasing in $z_j$ when
$\widetilde\delta_j>0$ and nonincreasing when
$\widetilde\delta_j<0$. It cannot agree with $b^j$ on this
neighborhood if the two coefficients have opposite signs.
Therefore the two coefficients have the same sign whenever at least
one of the two cost functions takes a positive value.

If instead $\lambda(x)=\widetilde\lambda(x)=0$ for every
$x\in\mathcal X$, fix any $x^*\in\mathcal X$ and hold $z_k=0$
for $k\ne j$. A6(a) gives $\mathcal Z_{x^*}=\R^J$. Write
$e_j\in\R^J$ for the $j$th coordinate vector. When $\delta_j>0$,
for every realization of the taste-shock vector, alternative $j$ eventually has the
largest payoff as $t\to+\infty$ at $z=te_j$, and eventually loses to
alternative 0 as $t\to-\infty$. Bounded convergence gives
\[
\lim_{t\to+\infty}b^j(te_j;x^*)=1,\qquad
\lim_{t\to-\infty}b^j(te_j;x^*)=0.
\]
When $\delta_j<0$, the two limits are reversed. The same argument
applies to $\widetilde\delta_j$. Equality of the observed probabilities
at every $te_j$ therefore rules out opposite signs in this case as well.
Thus, in either case,
\[
\operatorname{sign}(\widetilde\delta_j)=\operatorname{sign}(\delta_j)
\quad(j=1,\ldots,J).
\]
Replacing $Z_j$ by $-Z_j$ whenever this common sign is negative makes
both $\delta_j$ and $\widetilde\delta_j$ positive. I use these shifters
throughout the rest of the proof.

Fix $x$ with $\lambda(x)>0$ and $\widetilde\lambda(x)>0$. Define
\[
\begin{aligned}
W&\coloneqq \{z\in\R^J:b^j(z;x)>0\text{ for every }j=0,\ldots,J\},\\
\widetilde W&\coloneqq \{z\in\R^J:\widetilde b^j(z;x)>0\text{ for every }j=0,\ldots,J\}.
\end{aligned}
\]
By A3, there is $z_x^\circ\in\operatorname{int}(\mathcal Z_x)$ at which
every $b^j(z_x^\circ;x)$ is positive. The observed equality
$b(z_x^\circ;x)=\widetilde b(z_x^\circ;x)$ makes every candidate
probability positive as well. Continuity therefore supplies an open
ball $D\subseteq\operatorname{int}(\mathcal Z_x)\cap W\cap\widetilde W$
around $z_x^\circ$. Since $D$ lies in the observed support,
$b(z;x)=\widetilde b(z;x)$ for every $z\in D$.
The regions $W$ and $\widetilde W$
are open and connected, and $z\mapsto b(z;x)$ and
$z\mapsto\widetilde b(z;x)$ are real analytic there, by Lemma~\ref{l:multshares}.
Let $V$ be the connected component of $W\cap\widetilde W$ containing
$D$. Since an open Euclidean set is locally connected, its components
are open \citep[Theorem~25.3]{munkres}; thus $V$ is open.

I now show that $b(z;x)=\widetilde b(z;x)$ for every $z\in V$.
For a fixed $j\in\{0,\ldots,J\}$, put
$f(z)\coloneqq b^j(z;x)-\widetilde b^j(z;x)$ for $z\in V$, and define
\[
E_f\coloneqq \{z\in V:D^\nu f(z)=0\text{ for every }\nu\in\mathbb N_0^J\},
\]
where $D^\nu$ is the partial derivative with multi-index $\nu$.
For each multi-index $\nu$, the set
$\{z\in V:D^\nu f(z)=0\}$ is closed relative to $V$ because $D^\nu f$
is continuous. Their intersection $E_f$ is therefore closed relative
to $V$. If $z\in E_f$, every coefficient in the Taylor series of $f$
at $z$ is zero. Real analyticity makes $f$ identically zero on a
neighborhood of $z$; every derivative is then zero on that neighborhood,
so it lies in $E_f$. Thus $E_f$ is open relative to $V$ as well.
Since $f=0$ on the open ball $D$, all its derivatives vanish there,
and $D\subseteq E_f$. Connectedness of $V$ now gives $E_f=V$.
In particular, the derivative with multi-index zero is $f$ itself,
so $b^j(z;x)=\widetilde b^j(z;x)$ for every $z\in V$.
Applying this to all $j$ gives $b(z;x)=\widetilde b(z;x)$ on $V$. This is the same
open-and-closed argument used for the one-dimensional identity result
in \citet[Corollary 1.2.6]{krantzparks}; the connectedness criterion is
given in \citet[Section~23]{munkres}.

This component is also closed relative to $W$. Indeed, if $z_n\in V$
and $z_n\to z\in W$, continuity gives
$\widetilde b(z;x)=b(z;x)$ with every coordinate positive.
Thus $z\in W\cap\widetilde W$. Because this intersection is open,
choose $r>0$ with $B(z,r)\subseteq W\cap\widetilde W$.
For sufficiently large $n$, $z_n\in B(z,r)\cap V$.
The ball and $V$ are connected and intersect, so their union is
connected. Maximality of the connected component $V$ implies
$B(z,r)\subseteq V$, and in particular $z\in V$.
This proves that $V$ is closed relative to $W$.
Because $V$ is nonempty and both open and closed relative to the
connected set $W$, $V=W$ \citep[Section~23]{munkres}; interchanging the two parameter values gives
$V=\widetilde W$. Consequently
\begin{equation}\label{e:common-fullconsideration}
W=\widetilde W,\qquad
b(z;x)=\widetilde b(z;x)\quad(z\in W).
\end{equation}
The equality in \eqref{e:common-fullconsideration} has been proved
only for $z\in W$, where every $b^j(z;x)$ is positive.

Now fix $j\in\{1,\ldots,J\}$ and $u\in\R$ with
$0<p_j(u;x)<1$. Write
\[
p\coloneqq p_j(u;x),\qquad q_0\coloneqq1-p,\qquad
a_j^*\coloneqq\alpha_j+\delta_ju,\qquad
\lambda\coloneqq\lambda(x),\qquad
c^*\coloneqq\lambda\log(q_0/p)-a_j^*.
\]
I will keep $z_{n,j}=u$ by choosing a threshold near $c^*$ at each
$n$. For $s\in\R$ and $0\leq\varepsilon\leq1$, define
\[
R_k\coloneqq e^{\Omega_k/\lambda},\qquad
L_\varepsilon(\Omega;s)\coloneqq
1+e^{(\Omega_j-s)/\lambda}
+\varepsilon\sum_{k\notin\{0,j\}}R_k,
\]
and
\[
\begin{aligned}
Q_0(s,\varepsilon)&\coloneqq\E_G\{1/L_\varepsilon(\Omega;s)\},\\
Q_j(s,\varepsilon)&\coloneqq
\E_G\{e^{(\Omega_j-s)/\lambda}/L_\varepsilon(\Omega;s)\},\\
Q_k(s,\varepsilon)&\coloneqq
\E_G\{\varepsilon R_k/L_\varepsilon(\Omega;s)\}
\quad(k\notin\{0,j\}).
\end{aligned}
\]
Each integrand lies in $[0,1]$. The probabilities $Q_0(s,\varepsilon)$
and $Q_j(s,\varepsilon)$ are strictly positive, and when
$\varepsilon>0$ so is every $Q_k(s,\varepsilon)$.
These probabilities sum to one. The payoff index for alternative $j$
reconstructed from these probabilities and the threshold $s$ is
\[
f_\varepsilon(s)\coloneqq-s+
\lambda\log\frac{Q_0(s,\varepsilon)}{Q_j(s,\varepsilon)}.
\]
Bounded convergence makes $f_\varepsilon$ continuous in $s$ and gives
$f_\varepsilon(s)\to f_0(s)$ as $\varepsilon\downarrow0$ for each
$s\in\R$.

The binary first-order condition at $u$ gives
\[
p=\E_G\frac{p e^{(a_j^*+\Omega_j)/\lambda}}
{q_0+p e^{(a_j^*+\Omega_j)/\lambda}}.
\]
Since $e^{(\Omega_j-c^*)/\lambda}
=(p/q_0)e^{(a_j^*+\Omega_j)/\lambda}$, this equation and its
complement imply
\[
Q_j(c^*,0)=p\qquad\text{and}\qquad Q_0(c^*,0)=q_0,
\qquad\text{so}\qquad f_0(c^*)=a_j^*.
\]
To solve $f_\varepsilon(s)=a_j^*$ for small positive $\varepsilon$,
I first show that $f_0$ is strictly decreasing. Put
\[
P_s(\omega)\coloneqq
\frac{e^{(\omega-s)/\lambda}}{1+e^{(\omega-s)/\lambda}}.
\]
The derivative of $P_s(\omega)$ with respect to $s$ is bounded in
absolute value by $1/(4\lambda)$. Differentiating under the expectation
and using $Q_0(s,0)=1-Q_j(s,0)$ therefore gives
\[
\begin{aligned}
f_0'(s)
&=-1+\frac{\E_G[P_s(\Omega_j)\{1-P_s(\Omega_j)\}]}
{Q_0(s,0)Q_j(s,0)}\\
&=-\frac{\Var_G(P_s(\Omega_j))}{Q_0(s,0)Q_j(s,0)}<0.
\end{aligned}
\]
The strict inequality follows because A2 rules out a constant
$\Omega_j$ and $P_s$ is strictly increasing in its argument.

For each $n\geq1$, choose $0<\varepsilon_n<1/n$ small enough that
\[
f_{\varepsilon_n}(c^*-1/n)>a_j^*
>f_{\varepsilon_n}(c^*+1/n).
\]
Such a choice exists because $f_0$ is strictly decreasing,
$f_0(c^*)=a_j^*$, and $f_\varepsilon$ converges to $f_0$ at the
two displayed arguments. Continuity and the intermediate value theorem
then give $c_n\in(c^*-1/n,c^*+1/n)$ satisfying
\[
f_{\varepsilon_n}(c_n)=a_j^*,\qquad\text{and}\qquad c_n\to c^*.
\]
Define
\[
\begin{gathered}
Q_{i,n}\coloneqq Q_i(c_n,\varepsilon_n)\quad(i=0,\ldots,J),\qquad
Q_n\coloneqq(Q_{0,n},\ldots,Q_{J,n})',\\
c_{j,n}\coloneqq c_n,\qquad
c_{k,n}\coloneqq-\lambda\log\varepsilon_n
\quad(k\notin\{0,j\}),\\
a_{i,n}\coloneqq-c_{i,n}+\lambda\log(Q_{0,n}/Q_{i,n}),\qquad
z_{n,i}\coloneqq(a_{i,n}-\alpha_i)/\delta_i
\quad(i=1,\ldots,J).
\end{gathered}
\]
Lemma~\ref{l:multshares} gives $z_n\in W$ and $b(z_n;x)=Q_n$.
The defining equation for $c_n$ gives, for every $n$,
\[
a_{j,n}=f_{\varepsilon_n}(c_n)=a_j^*,\qquad\text{so}\qquad z_{n,j}=u.
\]
Since $c_n\to c^*$ and $\varepsilon_n\to0$, bounded convergence in
the formulas for $Q_i$ yields
\begin{equation}\label{e:binary-share-limit}
Q_{0,n}\to q_0,\qquad Q_{j,n}\to p,\qquad
Q_{k,n}\to0\quad(k\notin\{0,j\}).
\end{equation}

I also need an upper bound on the payoff indices $a_{k,n}$ for
$k\notin\{0,j\}$. Because $c_n>c^*-1$ and $\varepsilon_n\leq1$,
\[
\begin{gathered}
L_{\varepsilon_n}(\Omega;c_n)\leq\overline L(\Omega),\\
\text{where}\qquad
\overline L(\Omega)\coloneqq
1+e^{(\Omega_j-c^*+1)/\lambda}
+\sum_{k\notin\{0,j\}}R_k.
\end{gathered}
\]
For each such $k$, let
$m_k\coloneqq\E_G\{R_k/\overline L(\Omega)\}\in(0,1]$.
The definitions of $Q_{k,n}$ and $a_{k,n}$ then give
\begin{equation}\label{e:rival-index-bound}
\frac{Q_{k,n}}{\varepsilon_n}
=\E_G\frac{R_k}{L_{\varepsilon_n}(\Omega;c_n)}\geq m_k,
\qquad
a_{k,n}=\lambda\log\frac{Q_{0,n}}{Q_{k,n}/\varepsilon_n}
\leq-\lambda\log m_k.
\end{equation}

By \eqref{e:common-fullconsideration}, the same $Q_n$ maximizes the
objective under $\widetilde\theta$ at $z_n$. Write
$\widetilde\lambda\coloneqq\widetilde\lambda(x)$ and
$\widetilde\Omega\sim\widetilde G$.
The payoff index of alternative $j$ under $\widetilde\theta$ is
$\widetilde a_j\coloneqq\widetilde\alpha_j+\widetilde\delta_ju$
for every $n$. The other indices are bounded above because
\[
\widetilde a_{k,n}
=\frac{\widetilde\delta_k}{\delta_k}a_{k,n}
+\widetilde\alpha_k
-\frac{\widetilde\delta_k}{\delta_k}\alpha_k
\]
and the slope ratios are positive. Set
$\widetilde a_{j,n}\coloneqq\widetilde a_j$ and write
\[
\widetilde D_n\coloneqq Q_{0,n}+\sum_{i=1}^JQ_{i,n}
e^{(\widetilde a_{i,n}+\widetilde\Omega_i)/\widetilde\lambda},
\qquad
\widetilde\Phi_n(Q_n)\coloneqq
\widetilde\lambda\E_{\widetilde G}\log\widetilde D_n.
\]
For every realization of $\widetilde\Omega$,
\eqref{e:binary-share-limit} and the upper bounds on the indices give
\[
\widetilde D_n\longrightarrow
q_0+p e^{(\widetilde a_j+\widetilde\Omega_j)/\widetilde\lambda}.
\]
For all sufficiently large $n$,
$\widetilde\lambda\log\widetilde D_n
\geq\widetilde\lambda\log(q_0/2)$.
Choose constants $\bar a_i$ bounding $\widetilde a_{i,n}$ above.
Since $Q_n$ is a probability vector,
\[
\widetilde\lambda\log\widetilde D_n
\leq\max\{0,\max_{i\geq1}(\bar a_i+\widetilde\Omega_i)\}.
\]
Consequently
\[
\bigl|\widetilde\lambda\log\widetilde D_n\bigr|
\leq\bigl|\widetilde\lambda\log(q_0/2)\bigr|
+\max_i|\bar a_i|+\sum_{i=1}^J|\widetilde\Omega_i|.
\]
The right-hand side is integrable under $\widetilde G$ because each
$\widetilde\Omega_i$ has a finite first moment. Dominated convergence
therefore gives
\[
\widetilde\Phi_n(Q_n)\longrightarrow\widetilde\phi_j(p;u,x).
\]

For any $p'\in[0,1]$, the probability vector assigning $1-p'$ to
alternative 0, $p'$ to alternative $j$, and zero to every other
alternative is feasible in every problem under $\widetilde\theta$.
Because $z_{n,j}=u$ for every $n$, its objective is exactly
$\widetilde\phi_j(p';u,x)$. Optimality of $Q_n$ and the preceding
limit give
\[
\widetilde\Phi_n(Q_n)\geq\widetilde\phi_j(p';u,x)
\quad\Longrightarrow\quad
\widetilde\phi_j(p;u,x)\geq\widetilde\phi_j(p';u,x)
\quad(p'\in[0,1]).
\]
Assumption~A2 implies that $\widetilde\Omega_j$ is nondegenerate.
Lemma~\ref{l:shareobjectiveunique}, applied with $J=1$ to this marginal
distribution, makes $\widetilde p_j(u;x)$ the unique maximizer of
$\widetilde\phi_j(\cdot;u,x)$. Thus $p_j(u;x)=\widetilde p_j(u;x)$
whenever $0<p_j(u;x)<1$.
By Lemma~\ref{l:completecurve}, both functions are continuous,
nondecreasing CDFs with limits zero and one. Their equality on
$\{u:0<p_j(u;x)<1\}$ forces equality also where $p_j(u;x)$ equals
zero or one, by monotonicity. Therefore
$p_j(u;x)=\widetilde p_j(u;x)$ for every $u\in\R$.

If exactly one of $\lambda(x)$ and $\widetilde\lambda(x)$ is zero,
A6(a) gives $\mathcal Z_x=\R^J$. Thus the observed equality
$b(z;x)=\widetilde b(z;x)$ holds for every $z\in\R^J$.
Write $p_j^+(z_j;x)$ and $p_j^{\mathrm{zero}}(z_j;x)$ for the probabilities of choosing $j$ when only $j$ and 0 are available, under the parameter values with positive and zero information costs, respectively.
Fix $u\in\R$ with $p\coloneqq p_j^+(u;x)\in(0,1)$ and apply
the preceding construction to the parameter value with positive cost.
It gives $z_n$ and $Q_n$ satisfying \eqref{e:binary-share-limit},
with $z_{n,j}=u$ for every $n$.
Under the parameter value with zero cost, choosing $j$ when all
alternatives are available implies that its payoff exceeds
alternative 0's. If $j$ would be chosen from $\{0,j\}$ but is not
chosen when all alternatives are available, another alternative must
be chosen. Assumption~A6(b) excludes ties. Since the observed choice
probabilities agree at $z_n$,
\[
0\leq p_j^{\mathrm{zero}}(u;x)-Q_{j,n}
\leq\sum_{k\notin\{0,j\}}Q_{k,n}\longrightarrow0.
\]
Together with $Q_{j,n}\to p$, this gives
$p_j^{\mathrm{zero}}(u;x)=p_j^+(u;x)$.
Both functions are continuous, nondecreasing CDFs with limits zero
and one, by A6(b) and Lemma~\ref{l:completecurve}. Their equality
therefore extends to points where $p_j^+(u;x)$ equals zero or one.
Thus $p_j^{\mathrm{zero}}(u;x)=p_j^+(u;x)$ for every $u\in\R$.

If $\lambda(x)=\widetilde\lambda(x)=0$, their choice probabilities with
all alternatives available are identified on $\R^J$ by A6(a).
For $J=1$, these are already the probabilities for choice from
$\{0,1\}$, so the conclusion follows directly.
For $J\geq2$, fix $z_j$ and write $a_j\coloneqq \alpha_j+\delta_jz_j$.
Define the covariate vector $z_n$ by $z_{j,n}\coloneqq z_j$ and $z_{k,n}\coloneqq -n$ for
$k\in\{1,\ldots,J\}\setminus\{j\}$.
Under either parameter value, each rival's payoff tends to
$-\infty$ for every realization of the taste-shock vector because its slope is positive.
Assumption~A6(b) gives $\PP_G(a_j+\Omega_j=0)=0$.
\Needspace{6\baselineskip}
Outside this event,
\[
\ind\{a_j+\Omega_j>
\max(0,\max_{k\notin\{0,j\}}(\alpha_k-\delta_kn+\Omega_k))\}
\longrightarrow\ind\{a_j+\Omega_j>0\}.
\]
Bounded convergence gives $b^j(z_n;x)\to p_j(z_j;x)$.
Applying the same argument under $\widetilde\theta$ and using equality
of the identified choice probabilities gives
$p_j(z_j;x)=\widetilde p_j(z_j;x)$ for every \mbox{$z_j\in\R$}.
\end{proof}

\Needspace{8\baselineskip}
\begin{lemma}[The probability of choosing alternative 0 identifies the joint taste-shock distribution]\label{l:jointtastes}
Suppose the payoff coefficients $(\alpha,\delta)$ and the cost function
$\lambda:\mathcal X\to[0,\infty)$ are known and A2--A3 hold.
Fix $x\in\mathcal X_+$. On an observed nonempty open set
$D_x\subseteq\operatorname{int}(\mathcal Z_x)$ with $b^j(z;x)>0$ for
every $j=0,\ldots,J$, the choice-probability vector $z\mapsto b(z;x)$ identifies
the joint taste-shock distribution $G$ through the probability of choosing alternative 0
$z\mapsto b^0(z;x)$.
\end{lemma}
\begin{proof}
Fix such an $x$, write $\lambda\coloneqq \lambda(x)$, and take independent
standard Gumbel variables $\eta_0,\ldots,\eta_J$, independent of $\Omega$.
At $z$ with all $b^j(z;x)>0$, the conditional logit formula gives
the probability that alternative zero has the largest value among
\[
a_j(z)+\Omega_j+\lambda\log b^j(z;x)+\lambda\eta_j
\quad(j=0,\ldots,J),\qquad\text{where }a_0(z)=\Omega_0=0.
\]
Let $\zeta_j\coloneqq \eta_j-\eta_0$ and $H\coloneqq \Omega+\lambda\zeta$.
Let $D_x$ be the observed open set in the lemma, on which
$b^j(z;x)>0$ for every $j=0,\ldots,J$.
Assumption~A3 and continuity ensure that such a nonempty set exists.
Define the known map
\[
\begin{aligned}
c_x:D_x&\longrightarrow\R^J,\\
z&\longmapsto c_x(z)\coloneqq (c_{x,1}(z),\ldots,c_{x,J}(z))',\\
c_{x,j}(z)&\coloneqq -\alpha_j-\delta_jz_j+
\lambda\log\frac{b^0(z;x)}{b^j(z;x)}.
\end{aligned}
\]
The probability of choosing alternative 0 satisfies
\[
b^0(z;x)=\PP\{H_j\leq c_{x,j}(z)\text{ for every }j\}=F_H(c_x(z)).
\]
The two threshold maps satisfy $c_x(z)=c(a(z))$.
The affine map $z\mapsto a(z)$ has nonsingular Jacobian
$\operatorname{diag}(\delta_1,\ldots,\delta_J)$, and Lemma~\ref{l:multshares} makes $a\mapsto c(a)$ a diffeomorphism.
Thus $c_x(D_x)$ is a nonempty open subset of $\R^J$.
The values $b^0(z;x)$ therefore determine $F_H$ on $c_x(D_x)$.

By Lemma~\ref{l:gumbelconvolution}, $H$ has a real-analytic density $f_H$.
Taking one derivative of its CDF in each coordinate determines that
density $f_H$ on $c_x(D_x)$. If two taste-shock distributions $G$ and
$\widetilde G$ give the same
observed choice probabilities, their implied densities of $H$ agree
on this open set. The identity argument just used
in Lemma~\ref{l:identifiedbinarymargins} extends equality to the connected
domain $\R^J$. Thus the complete distribution
of $H$ is determined.
For $t\in\R^J$, write
$\varphi_G(t)\coloneqq \int e^{it'\omega}\,dG(\omega)$ and
$\varphi_H(t)\coloneqq \E e^{it'H}$ for the characteristic functions of
$\Omega$ and $H$. Independence and \eqref{e:aux-gumbel-cf} then give
\[
\varphi_H(t)=\varphi_G(t)\varphi_\zeta(\lambda t),\qquad
\varphi_G(t)=\frac{\varphi_H(t)}{\varphi_\zeta(\lambda t)}
\quad(t\in\R^J).
\]
The denominator never vanishes. Uniqueness of multivariate
characteristic functions \citep[Section~29, p.\ 382]{billingsley}
identifies the joint distribution $G$, including dependence among
$\Omega_1,\ldots,\Omega_J$.
\end{proof}

\Needspace{8\baselineskip}
\begin{lemma}[Payoff coefficients and the taste-shock distribution when information is free]\label{l:fullinformationmultinomial}
Suppose A1 and A6 hold, $\lambda(x)=0$ for every $x\in\mathcal X$,
$\E_G\Omega=0$, $\E_G\|\Omega\|^2<\infty$, and $\delta_j>0$ for
$j=1,\ldots,J$. The observed family $\mathcal B_{\mathrm{obs}}$
identifies $\alpha_j/\delta_j$ for $j=1,\ldots,J$ and
$\delta_j/\delta_1$ for $j=2,\ldots,J$.
Once $\delta_1$ is identified, it therefore identifies
$(\alpha,\delta,G)$.
\end{lemma}
\begin{proof}
Fix $x^*\in\mathcal X$. A6(a) gives $\mathcal Z_{x^*}=\R^J$, so
every vector $z\in\R^J$ lies in the observed conditional support. A6(b) excludes ties between any two payoffs.
For a fixed $z_j=v$, sending every rival shifter $z_k$, $k\ne j$,
to $-\infty$ gives the probability of choosing $j$ when only $j$ and 0 are available, by bounded convergence,
as in the zero-cost case of Lemma~\ref{l:identifiedbinarymargins}.
When $J=1$, no alternatives need to be excluded. The function $p_j(\cdot;x^*)$ is
the CDF of the switch point under full information:
\[
S^{(j)}\coloneqq \frac{-\alpha_j-\Omega_j}{\delta_j},\qquad\text{so}\qquad
p_j(v;x^*)=\PP_G\{S^{(j)}\leq v\},\qquad v\in\R.
\]
Its mean exists by the finite second-moment restriction, and
\[
\mu_j\coloneqq \int_{\R}v\,d p_j(v;x^*)
=\E_G S^{(j)}=-\frac{\alpha_j}{\delta_j}.
\]
Thus the functions $p_j(\cdot;x^*)$ determine $\alpha_j/\delta_j$ for each $j$.

To recover the slope ratios when $J\geq2$, fix $j\in\{2,\ldots,J\}$.
For $r>0$, $u\in\R$, and $t>0$, define the covariate vector
$z^{(j)}(t;r,u)\in\R^J$ by
\[
z^{(j)}_1(t;r,u)\coloneqq t,\qquad
z^{(j)}_j(t;r,u)\coloneqq rt+u,\qquad
z^{(j)}_k(t;r,u)\coloneqq 0\quad(k\notin\{1,j\}).
\]
The payoffs of alternatives $1$ and $j$ both tend to $+\infty$
for every realization of the taste-shock vector as $t\to\infty$. Every other payoff,
including alternative 0's, stays fixed along this sequence. Consequently,
eventually the chosen alternative is either $1$ or $j$, except when
their payoffs tie. Their payoff difference is
\[
v_j(\Omega;z^{(j)}(t;r,u))-v_1(\Omega;z^{(j)}(t;r,u))
=(\delta_jr-\delta_1)t+
\alpha_j-\alpha_1+\delta_ju+\Omega_j-\Omega_1.
\]
If $\delta_jr>\delta_1$, this difference tends to $+\infty$;
if $\delta_jr<\delta_1$, it tends to $-\infty$.
If $\delta_jr=\delta_1$, the difference does not depend on $t$, and
A6(b) gives probability zero to its being zero.
Bounded convergence therefore establishes the observable limit
\[
\begin{aligned}
L_j(r,u)&\coloneqq \lim_{t\to\infty}
b^j(z^{(j)}(t;r,u);x^*)\\
&=\begin{cases}
0,&r<\delta_1/\delta_j,\\
\PP_G\{\Omega_1-\Omega_j<\alpha_j-\alpha_1+\delta_ju\},
&r=\delta_1/\delta_j,\\
1,&r>\delta_1/\delta_j.
\end{cases}
\end{aligned}
\]
The distribution of $\Omega_1-\Omega_j$ is continuous by A6(b).
Its CDF has limits zero and one at the two ends of $\R$.
Because $\delta_j>0$, there is a value $u\in\R$ for which the middle probability is strictly between zero and one. At every other value of
$r$, the limiting probability is zero or one for every $u$.
Thus the observed limits determine the unique slope ratio through
\[
\{r>0:\exists u\in\R\text{ with }0<L_j(r,u)<1\}
=\{\delta_1/\delta_j\}.
\]
Once $\delta_1$ is known, these ratios give every $\delta_j$.
The already identified means give $\alpha_j=-\delta_j\mu_j$.
For $J=1$, no additional slope ratio is needed, and the same mean
formula identifies $\alpha_1$.

It remains to recover $G$. Alternative 0 is chosen when all other
payoffs are nonpositive; using weak rather than strict inequalities
does not affect this probability by A6(b). Therefore
\[
b^0(z;x^*)
=\PP_G\{\Omega_j\leq-\alpha_j-\delta_jz_j
\text{ for every }j=1,\ldots,J\}.
\]
For any $h=(h_1,\ldots,h_J)'\in\R^J$, set
$z_j(h)\coloneqq (-\alpha_j-h_j)/\delta_j$ and
$z(h)\coloneqq (z_1(h),\ldots,z_J(h))'$. This vector lies in the observed support, and
\[
G(h)=\PP_G\{\Omega_j\leq h_j\text{ for every }j\}
=b^0(z(h);x^*).
\]
Hence the probabilities of choosing alternative 0 identify the joint CDF on all of
$\R^J$, including dependence among the taste shocks.
\end{proof}

\begin{proof}[Proof of Theorem~\ref{t:multipleidentification}]
Compare $\theta\coloneqq (\alpha,\delta,\lambda,G)$ and
$\widetilde\theta\coloneqq (\widetilde\alpha,\widetilde\delta,\widetilde\lambda,\widetilde G)$
satisfying the assumptions and generating the same
$\mathcal B_{\mathrm{obs}}$.
Lemma~\ref{l:identifiedbinarymargins} gives
$p_j(z_j;x)=\widetilde p_j(z_j;x)$ for every $z_j\in\R$, $j$, and $x$.
For each $j$ and $x$, let $S_x^{(j)}$ be the switch point from Lemma~\ref{l:switchmoments}, with CDF $p_j(\cdot;x)$ given by the probability of choosing $j$ when only $j$ and 0 are available. At positive cost, this CDF is continuous by Lemma~\ref{l:completecurve}; at zero cost, A6(b), applied to $(j,0)$, gives continuity. The identified functions $p_j(\cdot;x)$ therefore determine
\[
\mu_j(x)\coloneqq \E S_x^{(j)},\qquad V_j(x)\coloneqq \Var(S_x^{(j)}),
\qquad C_j(x)\coloneqq \Cov\!\left(S_x^{(j)},\ell(p_j(S_x^{(j)};x))\right).
\]
If $\mathcal X_+\ne\varnothing$, A2 rules out a constant
$\Omega_j$: otherwise $1$ and $e^{\Omega_j/\lambda(x)}$ would be
linearly dependent at any $x\in\mathcal X_+$.
If $\mathcal X_+=\varnothing$, A6(b), applied to the pair $(j,0)$,
gives a continuous distribution for $\Omega_j$, which also rules out
a constant. Thus $\sigma_j^2\coloneqq \Var_G(\Omega_j)\in(0,\infty)$ in
either case. Lemma~\ref{l:switchmoments} gives
\begin{equation}\label{e:appendix-multmoments}
\mu_j(x)=-\frac{\alpha_j}{\delta_j},\qquad
\delta_j^2V_j(x)+2\delta_j\lambda(x)C_j(x)=\sigma_j^2,
\qquad C_j(x)>0.
\end{equation}

I first recover $\delta_1$, $\alpha_1$, and $\lambda$ using the
normalization $\sigma_1^2=\Var_G(\Omega_1)=1$.
A4 gives a sequence of cost-shifter values $x_M$ with
$\lambda(x_M)<1/M$. The cost may equal zero at some of these values.
The variance bound in Lemma~\ref{l:switchmoments} gives
$V_1(x_M)\to1/\delta_1^2$, while every variance is at most
$1/\delta_1^2$. Hence the functions $p_1(\cdot;x)$, $x\in\mathcal X$, identify
\[
\delta_1=(\sup_x V_1(x))^{-1/2},\qquad
\alpha_1=-\delta_1\mu_1(x),\qquad
\lambda(x)=\frac{1-\delta_1^2V_1(x)}{2\delta_1C_1(x)}.
\]
Applying the same formulas to the other parameter value gives $\widetilde\delta_1=\delta_1$, $\widetilde\alpha_1=\alpha_1$, and
$\widetilde\lambda(x)=\lambda(x)$ for every $x\in\mathcal X$,
including values at which $\lambda(x)=0$.

It remains to identify the remaining payoff coefficients and $G$.

First suppose $\mathcal X_+\ne\varnothing$ and choose
$\bar x\in\mathcal X_+$. For $j>1$, the variance $\sigma_j^2$ is not
normalized, but $\lambda$ is now known. At $\bar x$,
$V_j(\bar x)<\sigma_j^2/\delta_j^2$.
The same low-cost sequence satisfies
$V_j(x_M)\to\sigma_j^2/\delta_j^2$, so for some $x'$,
$V_j(x')>V_j(\bar x)$.
Subtracting \eqref{e:appendix-multmoments} at these two cost-shifter values
cancels the unknown taste variance:
\[
\delta_j^2\{V_j(\bar x)-V_j(x')\}
+2\delta_j\{\lambda(\bar x)C_j(\bar x)-\lambda(x')C_j(x')\}=0.
\]
Since the variance difference is nonzero and $\delta_j>0$,
\[
\delta_j=
\frac{2\{\lambda(x')C_j(x')-\lambda(\bar x)C_j(\bar x)\}}
{V_j(\bar x)-V_j(x')}.
\]
The mean identity then gives
\[
\alpha_j=-\delta_j\mu_j(x).
\]
These common moments give $\widetilde\alpha_j=\alpha_j$ and $\widetilde\delta_j=\delta_j$ for every $j>1$. Substitution also identifies each $\sigma_j^2$.

Only the joint taste-shock distribution $G$ remains to be identified.
Lemma~\ref{l:jointtastes}, applied at $\bar x$, recovers it from
$z\mapsto b(z;\bar x)$ on the observed open set
$D_{\bar x}\subseteq\operatorname{int}(\mathcal Z_{\bar x})$, where
$b^j(z;\bar x)>0$ for every $j=0,\ldots,J$.
The argument does not require the choice probabilities
at unobserved covariate values. Thus $G=\widetilde G$ and
$\theta=\widetilde\theta$ when $\mathcal X_+\ne\varnothing$.

If $\mathcal X_+=\varnothing$, the already identified common cost
function satisfies $\lambda(x)=\widetilde\lambda(x)=0$ for every
$x\in\mathcal X$. The value of $\delta_1$ is also already known.
Lemma~\ref{l:fullinformationmultinomial} then identifies every remaining
slope $\delta_j$, every intercept $\alpha_j$, and the joint distribution
$G$ from the same observed family $\mathcal B_{\mathrm{obs}}$.
Consequently $\theta=\widetilde\theta$ in this case as well.
The two cases exhaust the nonnegative cost functions allowed in the
model and prove the theorem.

\Needspace{8\baselineskip}
Since $\theta=\widetilde\theta$, the choice probabilities satisfy
\[
b(z;x)=\widetilde b(z;x)
\quad\text{for every }z\in\R^J,\ x\in\mathcal X.
\]
When $\lambda(x)>0$, both choice-probability vectors maximize the same strictly
concave objective $q\mapsto\Phi(q;a(z),x)$.
When $\lambda(x)=0$, both are generated by the same payoff
comparisons under the same distribution $G$; A6(b) excludes payoff
ties. This equality also holds at $z$ where $b^0(z;x)=0$.
\end{proof}

\end{document}